\documentclass [prb,preprint,superscriptaddress,floatfix,amsmath,amssymb] {revtex4-2}
\usepackage{amsmath}
\usepackage{graphicx}
\usepackage{hyperref}
\usepackage{color}
\usepackage{amssymb}
\usepackage{bm}

\newcommand{\beq} {\begin{equation}}
\newcommand{\eeq} {\end{equation}}
\newcommand{\bea} {\begin{eqnarray}}
\newcommand{\eea} {\end{eqnarray}}
\newcommand{\be} {\begin{equation}}
\newcommand{\ee} {\end{equation}}

\newcommand{\bo}{\bar \omega}

\newcommand{\sign}{\text{sign}}

\definecolor{darkgreen}{RGB}{0,170,0}

\begin{document}
\title{Interplay between superconductivity and non-Fermi liquid at a quantum-critical
point in a metal. VI. The $\gamma$ model and its phase diagram at $2 < \gamma <3$}
\author{Shang-Shun Zhang}
\affiliation{School of Physics and Astronomy and William I. Fine Theoretical Physics
Institute, University of Minnesota, Minneapolis, MN 55455, USA}
\author{Yi-Ming Wu}
\affiliation{School of Physics and Astronomy and William I. Fine Theoretical Physics
Institute, University of Minnesota, Minneapolis, MN 55455, USA}
\author{Artem Abanov}
\affiliation{Department of Physics, Texas A\&M University, College Station, USA}
\author{Andrey V. Chubukov}
\affiliation{School of Physics and Astronomy and William I. Fine Theoretical Physics
Institute, University of Minnesota, Minneapolis, MN 55455, USA}
\date{\today}
\begin{abstract}
In this paper, the sixth in series, we continue our analysis of the interplay between non-Fermi liquid and pairing in the effective low-energy model of fermions with singular dynamical interaction $V(\Omega_m) = {\bar g}^\gamma/|\Omega_m|^\gamma$
 (the $\gamma$ model).   The model describes low-energy physics of various quantum-critical metallic systems at the verge of an instability towards density or spin order,  pairing of fermions at the half-filled Landau level, color superconductivity, and pairing in SYK-type models. In previous Papers I-V we analyzed the $\gamma$ model for $\gamma \leq 2$ and argued that the ground state is an ordinary  superconductor for $\gamma <1$,  a peculiar one for $1<\gamma <2$, when the phase of the gap function winds up along real frequency axis due to emerging dynamical vortices in the upper half-plane of frequency, and that there is a quantum phase transition at $\gamma =2$, when the number of dynamical vortices becomes infinite.  In this paper we consider larger  $2< \gamma <3$ and address the issue what happens on the other side of this quantum transition. We argue that the system moves away from criticality in that the number of dynamical vortices becomes finite and decreases with increasing $\gamma$.  The ground state is again a superconductor, however a highly unconventional one with a non-integrable singularity in the density of states at the lower edge of the continuum.  This implies
   that the
      spectrum of excited states now contains a level with  a macroscopic degeneracy, proportional to the total number of states in the system. We argue that the
       phase diagram in variables $(T,\gamma)$ contains two distinct superconducting phases for $\gamma <2$ and $\gamma >2$, and an intermediate pseudogap state of preformed pairs.
\end{abstract}
\maketitle

\section{Introduction.}

This paper continues our studies of the interplay between non-Fermi liquid (NFL) and superconductivity for  itinerant fermions near a quantum-critical point (QCP) towards charge or spin order.   The key interaction between fermions in this situation
 is mediated by soft bosonic order parameter fluctuations.  When soft bosons are slow compared to electrons (e.g., when bosons are Landau-overdamped collective modes of fermions), the low-energy physics is described by an effective dynamical model  with 4-fermion interaction $V(\Omega) \propto 1/|\Omega|^\gamma$.
 At a QCP, when order parameter propagator is massless, this form holds down to $\Omega =0$.

 The model with $V(\Omega) \propto 1/|\Omega|^\gamma$ has been  nicknamed the $\gamma$-model.
 The exponent $\gamma$ has particular values for
a growing number of
specific microscopic realizations: $\gamma =0+$  for 3D QC-systems and for pairing of quarks, mediated by gluon exchange, $\gamma =1/3$ for a system near a nematic QCP and for fermions at a half-filled Landau level,  $\gamma =1/2$ near an antiferromagnetic QCP, $\gamma = 0.68$ for Sachdev-Ye-Kitaev (SYK) model of $N$ fermions coupled to equal number of bosons,  $\gamma =1$ for pairing by propagating bosons, $\gamma =2$ for phonon-mediated pairing at vanishing Debye frequency, etc.
 Microscopic models with varying $\gamma$ have also been proposed.   We listed and discussed
some  microscopic models in the first paper of the series (Paper I).
  In all cases, the same interaction, mediated by low-energy bosons, gives rise to fermionic self-energy, which accounts
   for NFL behavior in the normal state, and at the same time serves as glue that binds fermions into pairs.
    The two tendencies (NFL and SC) are intertwined as they come from the same interaction,  and compete with each other: a  fermionic self-energy makes fermions incoherent and reduces the tendency to pairing, while if
     bound  pairs develop, they provide a feedback  on the self-energy, which at lowest frequencies recovers the Fermi liquid form, i.e., fermions become propagating rather than diffusive excitations.

 For non-SYK systems, in each  case SC emerges in
   a particular momentum channel, e.g, in a $d-$wave channel near an antiferromagnetic QCP.  However, once the pairing symmetry is incorporated and momentum integration in the formulas for the fermionic self-energy  and the pairing vertex is carried out, the effective low-energy model  for different microscopic realizations becomes the same one, specified only by  the value of $\gamma$. The sign of $V(\Omega)$ is attractive, i.e., if fermions were free, the ground state would
necessarily be a superconductor.

 In previous papers (Refs.~\cite{paper_1,paper_2,paper_3,paper_4,paper_5}), which we refer to as Papers I-V, we treated $\gamma$ as a parameter and  analyzed the interplay between NFL and pairing for $\gamma \leq 2$. In this paper we consider $\gamma >2$.
 For convenience of a reader, we list some results of previous works, which form the base for the analysis in this paper.

\begin{figure}
\centering
\includegraphics[scale=0.6]{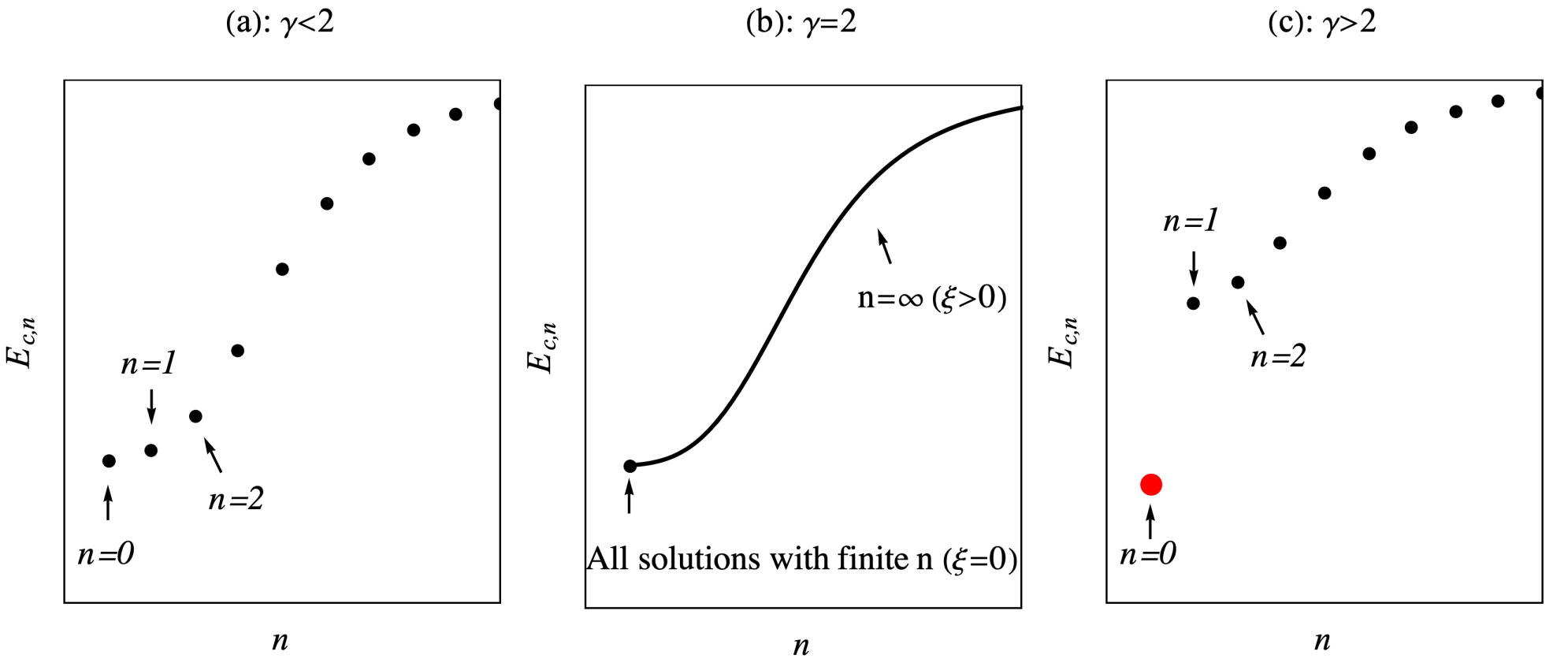}\caption{Condensation energy ($E_{c,n}$) of each topologically distinct solution labeled by integer $n=0,1,2,...$, where (a) $\gamma<2$, (b) $\gamma=2$, and (c) $\gamma>2$.}
\label{fig:Ec}
\end{figure}

 \begin{itemize}
 \item
 For any $\gamma >0$, the ground state is a superconductor, i.e., superconductivity wins the competition with a NFL.
 However, in  distinction to the pairing of coherent fermions in a Fermi liquid, the pairing of incoherent fermions is a threshold phenomenon, and in an extended $\gamma$ model with different magnitudes of $V(\Omega)$ in the particle-hole and particle-particle channels ($V(\Omega)$ and $V(\Omega)/N$, respectively), there exists a $\gamma-$dependent critical $N_{cr} >1$ separating a SC state for $N < N_{cr}$ (including the original model with $N=1$) and a NFL ground state for $N > N_{cr}$.
 \item
 In another crucial distinction from pairing in a Fermi liquid,  the gap equation at a QCP at $T=0$  has an infinite set of solutions  $\Delta_n (\omega)$, where $n$ runs between $0$ and $\infty$. At zero frequency, $\Delta_n (0) \sim {\bar g}  e^{-An}$ are all finite (${\bar g}$ is electron-boson coupling and $A$ is a $\gamma-$dependent number).  However, the $n$-th solution changes sign $n$ times along the the Matsubara axis,  $\omega \equiv \omega_m$. The solutions are then topologically distinct as each zero of $\Delta_n (\omega_m)$ is a center of a dynamical vortex on the upper complex plane of frequency.
 The $n=0$ solution is
 sign-preserving 
  and its structure along the Matsubara axis
  is similar to a conventional gap function in a Fermi liquid with attraction. The $n =\infty$ solution has an infinitesimally small magnitude and is the solution of the linearized gap equation.  We presented the exact proof that the solution of the linearized gap equation exists  along with the solutions of the non-linear gap equation.
  Away from a QCP,  only a finite number of solution remains, and above a certain deviation from a QCP only the conventional $n=0$ solution survives.
  \item
   Each solution from the infinite set at  QCP evolves with $T$ and vanishes at a separate $T_{c,n}$.
    The largest $T_{c,0} \sim {\bar g}$,  At large $n$,
   $T_{c,n} \propto e^{-An}$.  We presented strong numerical evidence for the existence of the set of critical temperatures and showed that the corresponding eigenfunctions change sign $n$ times along the Matsubara axis.
 \item
 For $\gamma <2$, the set is discrete, and the largest condensation energy at $T=0$  and the highest $T_c$ is for the $n=0$ solution. In this respect, the ground state is still a  ``conventional'' superconductor in the sense that $\Delta_0 (\omega_m)$ is a regular, sign-preserving function of the Matsubara frequency.
   Phase fluctuations of $\Delta_0 (\omega_m)$ are weak in the same parameter by which soft bosons are slow modes compared to fermions. However, as  $\gamma$ increases towards $2$, the other solutions become progressively more relevant.  Namely,  the spectrum of the condensation energy $E_{c,n}$ becomes more dense and $E_{c,n}$ with $n >0$ come closer to $E_{c,0}$. Simultaneously,  the frequency range, where $\Delta_n (\omega_m)$ changes sign $n$ times,
 shifts to  progressively smaller
  $\omega_m \propto (2-\gamma)$, while at larger  frequencies all $\Delta_n (\omega_m)$ nearly coincide with  $\Delta_0 (\omega_m)$.
 \item
  At $\gamma =2$, a critical behavior emerges: all $\Delta_n (\omega_m)$ with
   finite $n$  become undistinguishable from $\Delta_0 (\omega_m)$ at any $\omega_m >0$, while the solutions with $n \to \infty$ form a
    continuum spectrum  $\Delta_\xi (\omega_m)$.  A continuous $\xi$ is the product of $n$ and $2-\gamma$, and its value is determined by how the double limit $n \to \infty$ and $\gamma \to 2$ is taken. This is similar to how a continuous phonon spectrum emerges in the thermodynamic limit from a discrete set of energy levels. The condensation energy $E_{c, \xi}$ also becomes a continuous function of $\xi$. A visual picture is that an infinite set of $E_{c,n}$ approaches $E_{c,0}$ at $\gamma \to 2$ and touches it at $\gamma =2-0$ (Fig.~\ref{fig:Ec}(a) and \ref{fig:Ec}(b)).
This creates a branch of gapless ``longitudinal'' fluctuations. We argued that these fluctuations destroy phase coherence at any $T>0$ and give rise to pseudogap behavior at $0 < T < T_{p}$, where $T_p \sim {\bar g}$ is a would be transition temperature if the solutions with $n >0$ didn't exist.
 Away from a QCP, when a pairing boson has a gap $\omega_D$,   $T_c \propto \omega_D$.  This last result applies to electron-phonon pairing at small $\omega_D$.
 \item
 Extra information about the critical behavior emerges at $\gamma \to 2$, comes from the analysis of the gap equation on the real axis.  Here,  $V(\Omega) \propto e^{i \pi \gamma/2}$ is complex and hence $\Delta_0 (\omega)$ is  also complex. For $\gamma <1$, Re$V(\Omega) \propto \cos{\pi \gamma/2}$ is positive (attractive), and  Re$\Delta_0 (\omega)$ is a regular,  sign-preserving  function of $\omega$.  The corresponding density of states (DOS) vanishes at $\omega <\Delta$ and is non-zero for larger frequencies, as is expected on general grounds for the case when the pairing boson is massless. For $\gamma >1$, Re$V(\Omega)$ changes sign.  We found that in this situation there appears a finite
   frequency range where the phase $\eta_0 (\omega)$ of $\Delta_0 (\omega) = |\Delta_0 (\omega)| e^{i\eta_0 (\omega)}$  winds up by $2\pi m$, where $m$ is an  integer. The value of $m$ increases
     in increments of one
     at $\gamma >1$, and the
     increase accelerates as $\gamma$ approaches $2$.
     As the consequence, the DOS
      develops a set of maxima and minima in the range where the phase winds up.
    We extended $\Delta_0 (z)$ to complex $z$ in the upper half-plane and traced the phase winding $2\pi m$ to the emergence of $m$  vortices at complex $z$; each vortex moves from the lower to the upper frequency half-plane as $\gamma$ increases, leaving a $2\pi$ phase winding along the real axis. At $\gamma =2$, the number of vortices becomes infinite and the frequency range,
   where $\eta_0 (\omega)$ winds up, extends to an infinity, where $\Delta_0 (z)$ develops an essential singularity. Its presence is a must as otherwise an extension from an infinite set of vortex points  would give $\Delta_0 (z) =0$.
     In explicit form, the gap function along the real frequency axis at $\gamma =2$ is  $\Delta_0 (\omega) \sim \omega/\sin{\phi (\omega +i\delta)}$, where $\phi (x)$ is an increasing function of the argument~\cite{combescot,Karakozov_91,Marsiglio_91}.
     The DOS for such $\Delta_0 (\omega)$  consists of a set of $\delta$-functional peaks at frequencies where $\sin{\phi (\omega)} = \pm 1$. This is qualitatively different from a continuum DOS for $\gamma <2$.  Away from a QCP (i.e., for a non-zero $\omega_D$), the continuum remains, but with sharp maxima and nearly zero DOS between the maxima.
       The other  $\Delta_\xi (z)$ from a continuum set at $\gamma =2$ also have an infinite number of vortices, likely at the same $z$,  and essential singularity at $z = \infty$.   This clearly indicates that (i) the $\gamma =2$ model is indeed  critical and (ii) there is an ultimate connection between criticality and topology.
       \end{itemize}

In this paper, we  show that the new phase develops on the other side of the critical point, at $\gamma >2$.
We present evidence for this from calculations on the Matsubara axis and on the real axis.
On the Matsubara axis, we  find the spectrum of condensation energies $E_{c,n}$ again becomes a discrete one at $\gamma >2$  and $E_{c,0}$ is the largest. In simple words,
condensation energies $E_{c,n}$ with $n >0$ approach $E_{c,0}$ as $\gamma$ increases towards $2$, merge with $E_{c,0}$ and form a continuous spectrum at $\gamma=2$,
 and then bounce back
  at  $\gamma >2$, re-creating the gap between $E_{c,0}$  and other $E_{c,n}$ (Fig.~\ref{fig:Ec} (c)).
At a first glance it looks that the system  behavior at $\gamma >2$ is a mirror copy of that at  $\gamma <2$.
   However, we show that there is one crucial difference: for  $\gamma >2$   the condensation energy $E_{c,0}$
    behaves differently from other $E_{c,n}$. This can be seen most explicitly in the extend $\gamma$  model with reduced interaction in the pairing channel relative to the one in the particle-hole channel.
      For $\gamma <2$,
      all $E_{c,n}$,  including the one for $n=0$, vanish simultaneously once the pairing interaction reduces below
      a certain  threshold.
      For $\gamma >2$,  $E_{c,n}$  with $n >0$ all vanish at the threshold, while $E_{c,0}$ remains finite, i.e., the  $n=0$ solution of the gap equation  exists
        below the threshold for all other $n$.
      We see the this last  behavior in the original model: $E_{c,n}$ with $n >0$ vanish at $\gamma_{cr} \approx 2.81$,  while $E_{c,0}$ remains finite.

We present another evidence for the decoupling between the $n=0$ solution and the solutions with $n >0$, this time from the analysis of the gap function $\Delta_n (z)$ in the upper half-plane of complex frequency, $z = \omega' + i \omega^{''}$.   We recall that for $\gamma <2$, $\Delta_n(z)$ with all $n$ have the same number of zeros (centra of $2\pi$ vortices) away from the
 Matsubara axis.  The number
of zeros, $m$,
is finite and increases with $\gamma$. The analysis of the exact solution for  $\Delta_{\infty} (z)$ shows that these $m$ vortices are part of an infinite set of vortices, which crosses into the
lower frequency half-plane at larger $z$  and eventually approaches the ``critical'' line in the lower half-plane, at the angle $\pi(1/2- 1/\gamma)$,
counted from the real axis.
We show this in detail in Fig.\ref{fig:vortex}.
There is an essential singularity at the end of this line. At $\gamma =2$, the critical line is along the real axis, what causes the special behavior of $\Delta_n (\omega)$ with all $n$ and essential singularity at $\omega = \infty$.  It turns out that at $\gamma >2$, the critical line for all $n >0$ gradually continues into the upper half-plane, and each $\Delta_n (z)$ possesses an infinite number of zeros (vortex points), whose positions approach this line at $z \to \infty$.  For $n=0$, the critical line bounces back into the lower half-plane, and, as a result,  $\Delta_0 (z)$ has a finite number of zeros in the upper half-plane and show regular behavior at $z \to \infty$ along any direction in the upper half-plane.

We next focus on the $n=0$  solution and search for qualitative  changes in the physical properties of the system between $\gamma <2$ and $\gamma >2$.  For this we analyze the form of $\Delta_0 (\omega)$ on the real axis and use it to obtain the DOS. We recall that for $\gamma <2$, the DOS, $N(\omega)$ is a continuous function of $\omega$ at frequencies
 above the gap.  We show that for $\gamma >2$, the DOS again forms a gapped continuum,  but
  there is a non-integrable singularity (an ``infinite'' peak) at the lower end of the continuum.  The
  prefactor for this singular term increases with $\gamma$, initially as $\gamma-2$, i.e.,
   the weight of the ``infinite peak'' increases with $\gamma$.

We extend the analysis of $\Delta_0 (\omega)$ and $N(\omega)$ to
the case when a boson has a finite mass, which we label  $\omega_D$ by analogy with the phonon case. We show that the ``infinite'' peak survives in a finite range of $\omega_D$,
 i.e., the new structure is stable against small perturbations and
  occupies a finite region in the phase diagram.
   This  state is a  superconductor with a non-zero  superfluid stiffness $\rho_s$ at $T=0$, as we explicitly show,
    yet it is qualitatively different from a superconductor at $\gamma <2$. In essence the total area of the peak, divided by the total number of states,  can be regarded as the ``order parameter'' of the new state.

\begin{figure}
\centering
\includegraphics[scale=0.7]{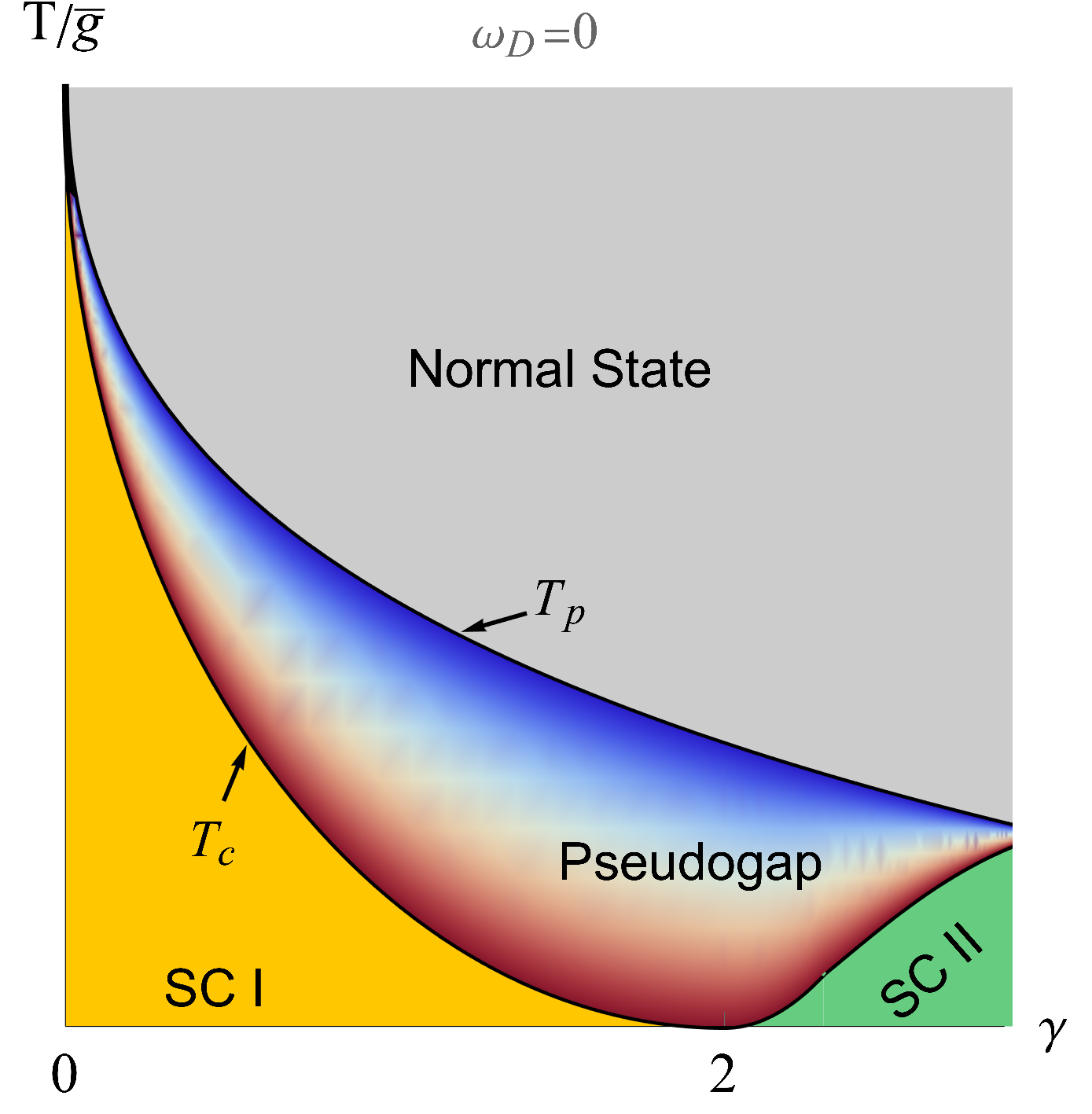}\caption{Phase diagram on $T-\gamma$ plane at a QCP ($\omega_D=0$), where the model parameter $0\leq \gamma < 3$.}
\label{fig:pd1}
\end{figure}

\begin{figure}
\centering
\includegraphics[scale=0.7]{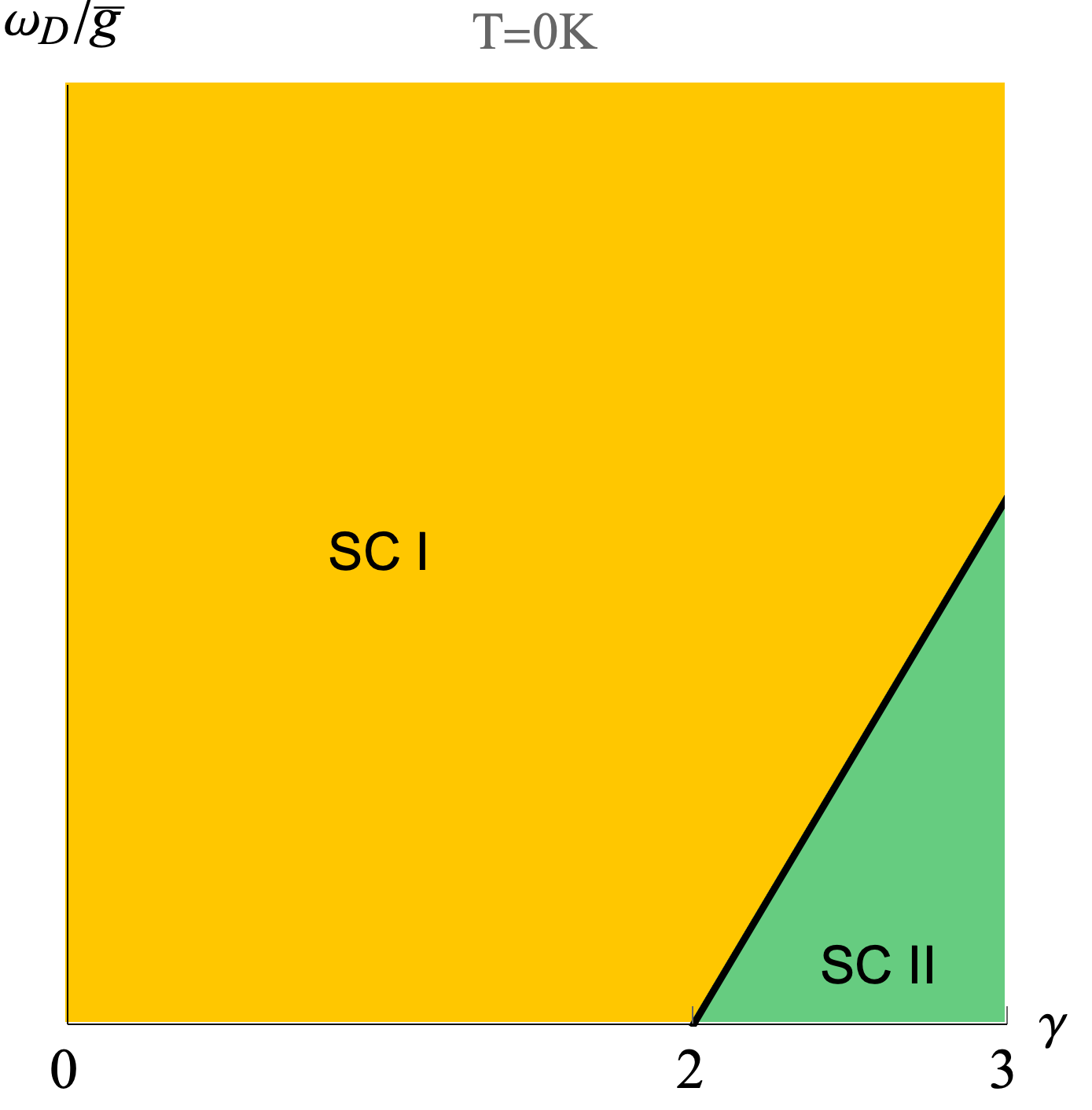}\caption{Phase diagram on $\omega_D-\gamma$ plane at $T=0K$, where the model parameter $0\leq \gamma < 3$.}
\label{fig:pd2}
\end{figure}

The phase diagrams for the $\gamma$ model in variables $(T, \gamma)$ at $\omega_D =0$ and in $(\omega_D, \gamma)$ at $T=0$ are shown in Figs.~\ref{fig:pd1} and \ref{fig:pd2}.
To obtain these phase diagrams, we combined the results for $\gamma >2$ with the results of previous papers from the series, Refs.~\cite{paper_1,paper_2,paper_3,paper_4,paper_5}.
At $T=0$, the two superconducting  phases
SC I and SC II
 merge at the critical $\gamma =2$.
SC I is a superconducting phase with conventional properties, and SC II is the new state, which we discuss  in this paper, with an
 ``infinite'' peak in the DOS.
  At a finite temperature, there is an intermediate regime between the two ordered phases, where long range superconducting order is destroyed by  ``longitudinal'' gap fluctuations, associated with the presence of an infinite set of low-lying states  with $n >0$.
 In this regime, fermions form bound pairs, which, however, remain incoherent and do not superconduct. The observables in this regime display pseudogap behavior, e.g., fermionic spectral function has a  peak at the gap value, but the spectral weight below the gap remains finite.

The structure of the paper is the following.
In the next section, we briefly review the $\gamma$ model and present the gap equations along the Matsubara and the real frequency axis.
In Sec.~\ref{sec:matsubara}, we analyze the gap equation along the Matsubara axis  and show that for $\gamma >2$ it still  has an
  infinite number of topologically distinct solutions, $\Delta_n (\omega_m)$, with $n =0,1,2..$, like for smaller $\gamma$.  We present the exact solution of the linearized gap equation, $\Delta_\infty (\omega_m)$ and then use it to obtain discrete solutions of the non-linear gap equation, $\Delta_n (\omega_m)$ (Sec.~\ref{sec:Mats_expansion}).
 In Sec.~\ref{sec:Mats_sign_preserving}, we discuss the structure of the $n=0$ solution, $\Delta_0 (\omega_m)$.
 In Sec.~\ref{sec:decouple}, we extend the model to non-equal interactions in  particle-hole and particle-particle channels, taking special care
to avoid introducing unphysical divergencies. We show that for $\gamma <2$, all $\Delta_n (\omega_m)$ disappear once the pairing interaction drops below a certain threshold, while for $\gamma >2$,  the  solutions with all $n >0$ disappear at the threshold, while the $n=0$ solution survives.   In Sec.~\ref{sec:vortex},  we extend the gap equation from Matsubara axis into the upper half-plane of frequency and show that  the distinction between $n=0$ and $n >0$ can be seen by analyzing the structure of dynamical vortices.
In Sec.~\ref{sec:real}, we analyze the gap function $\Delta_0 (\omega)$ along the real axis, particularly its form near the frequency $\omega_0$,
where $\Delta_0 (\omega) = \omega$. We first present, in Sec.\ref{sec:real_2}, an approximate treatment, in which we replace the integral gap equation by the differential one and keep only the lowest derivatives of $\Delta (\omega)$.  We show that at $\omega \sim  \omega_0$, $\Delta_0 (\omega)$ is entirely real and
$\Delta_0 (\omega)/\omega -1 $ scales as $(\omega_0 - \omega -i \delta)^4$.
In Sec.~\ref{sec:dos_2}, we obtain the DOS for this $\Delta_0 (\omega)$ and show that it has an infinite peak (a non-integrable singularity) at $\omega_0$.  In Sec.~\ref{sec:real_sub1}, we present more accurate treatment, in which we include higher-order derivatives of $\Delta (\omega)$. We show that
  the form of $\Delta (\omega)$ near $\omega_0$ get modified, yet  the DOS still has an infinite peak.
  In Sec.~\ref{sec:integral}, we show that this non-integrable singularity can be extracted directly from the integral gap equation.
    In Sec.~\ref{sec:omega_D}, we extend the analysis to
  finite mass of a boson and show that the infinite peak survives in a finite range of the mass.
   We summarize our results in Sec.~\ref{sec:gamma}, combine them with earlier results  for smaller $\gamma$, and
   present the phase diagram of the $\gamma$-model. The phase diagram in $(T,\gamma)$ plane contains two different superconducting phases and intermediate regime of preformed pairs with pseudogap behavior of observables.

Some technical details of calculations are moved to the Appendices.  Throughout the paper we use $\omega_m$ for  fermionic frequency along the Matsubara axis (a continuous variable at $T=0$ and a discrete  one at a finite $T$, $\omega_m = \pi T(2m+1)$),
$\omega$ for fermionic frequency along the real axis, and $z = \omega' + i \omega^{''}$, $\omega^{''} >0$,  for complex frequency in the upper half-plane.

\section{Model and Eliashberg equations}
\label{sec:Eli}

The $\gamma$-model is an effective model that describes low-energy fermions with dynamical interaction $V(\Omega_m) \propto 1/|\Omega_m|^\gamma$.  This model is obtained from an underlying
 model of itinerant dispersion-full fermions with interaction mediated by a soft boson near a charge or spin QCP, after one integrates over momenta in the expressions for the fermionic self-energy and the pairing vertex. When collective bosons are slow modes compared to fermions (e.g., when they are Landau overdamped by fermions), the momentum integration factorizes between the one transverse to the Fermi surface, which involves only fermionic propagators, and the one along the Fermi surface, which involves the bosonic propagator between points on the Fermi surface and converts it into the local propagator. At a QCP, the local bosonic propagator is massless, and its frequency dependence is singular, $1/|\Omega_m|^\gamma$.  The dimensionless interaction, mediated by this boson, is then $V(\Omega_m) = \bar{g}^\gamma/|\Omega_m|^\gamma$, where $\bar{g}$ is
 the effective fermion-boson coupling constant.
  The exponent $\gamma$ is determined by the type of the underlying microscopic model. We refer a reader to Paper I for the list of specific examples~\cite{paper_1}.

 The interaction $V(\Omega_m)$ is sign-preserving on the Matsubara axis and singular at $\Omega_m \to 0$. It gives rise to two competing effects: (i) a NFL behavior in the normal state and (ii) an attraction in one or more pairing channels (chosen within the original model with momentum and frequency-dependent interaction). The two trends are described by coupled equations for the fermionic self-energy $\Sigma (\omega_m)$ and the pairing vertex $\Phi (\omega_m)$ (see Papers I-IV for the exact forms of these equations).  One can replace these two equations by the equation for the pairing gap $\Delta (\omega_m) = \Phi (\omega_m)/(1 + \Sigma (\omega_m)/\omega_m)$  and the inverse quasiparticle residue $Z(\omega_m) = 1 + \Sigma (\omega_m)/\omega_m$. One advantage of using $\Delta$ instead of $\Phi$ is that the equation for $\Delta (\omega_m)$ can be expressed solely in terms of $\Delta (\omega_{m'})$.  In explicit form, the non-linear gap equation is
\begin{equation}
\Delta(\omega_{m})=\bar{g}^{\gamma}\pi T\sum_{\omega_{m^\prime}}\frac{\Delta(\omega_{m^\prime})-
\Delta(\omega_{m})\frac{\omega_{m^\prime}}{\omega_{m}}}{\sqrt{(\omega_{m^\prime})^{2}+\Delta^{2}(\omega_{m^\prime})}}
\frac{1}{\rvert\omega_{m^\prime}-\omega_{m}\rvert^{\gamma}},
\label{eq:gap1}
\end{equation}
Another advantage of using $\Delta$ instead of $\Phi$ is that a potentially singular contribution from $V(\Omega_m \to 0)$, i.e., from $\omega_{m'} \to    \omega_{m}$,  is eliminated by
 vanishing numerator.
The cancellation holds both at a finite $T$ and at $T=0$. At a finite $T$, the would be divergent contribution
 comes from the term with $m' = m$ in the summation over discrete $m'$.  It vanishes, because the numerator vanishes exactly at $m=m'$, and this holds even if we keep a small mass in the bosonic propagator in intermediate calculations. We note in passing that the term with $V(0)$ describes thermal fluctuations, whose role for the pairing parallels that of
   non-magnetic impurities. The cancellation of the thermal contribution can  then be viewed as a realization of the Anderson theorem. At $T=0$, the integral $\int d \omega'/|\omega - \omega'|^\gamma$ is singular for $\gamma >1$, but the singular behavior is eliminated as the expansion of the numerator yields compensating  $(\omega - \omega')^{2}$.
The frequency integral then remains convergent as long as $\gamma <3$, which we consider here.

At the onset of the pairing, when $\Delta (\omega_m)$  is infinitesimally small,
 the gap equation reduces to
\begin{equation}
\Delta(\omega_{m})=\bar{g}^{\gamma}\pi T\sum_{\omega_{m}^{\prime}}\left(\frac{\Delta(\omega_{m}^{\prime})}{\omega_{m}^{\prime}}-\frac{\Delta(\omega_{m})}
{\omega_{m}}\right)\frac{\text{sgn}(\omega_{m})}{\rvert\omega_{m}^{\prime}-\omega_{m}\rvert^{\gamma}}.\label{eq:gap2}
\end{equation}
At zero temperature, one can replace the sum over $\omega'_m$ in (\ref{eq:gap1}) and (\ref{eq:gap2})
by the integral $\pi T\sum_{\omega_{m}^{\prime}}\rightarrow(1/2)\int d\omega_{m}^{\prime}$.

The gap equation on the real axis is obtained by applying spectral representation to Eq.~(\ref{eq:gap1})  [see Refs.~\cite{Marsiglio_88,Karakozov_91,combescot} and Papers I, IV and V for details]. It takes the form
\begin{equation}\label{el8}
\Delta (\omega) B(\omega)=A(\omega)+C(\omega),
\end{equation}
where $D(\omega)=\Delta(\omega)/\omega$, the functions $A(\omega)$, $B(\omega)$ and $C(\omega)$ are given by Eqs.~(\ref{last_1}), (\ref{real_a_3_1}) in Sec.~\ref{sec:real}.
\begin{figure}
\centering
\includegraphics[scale=0.7]{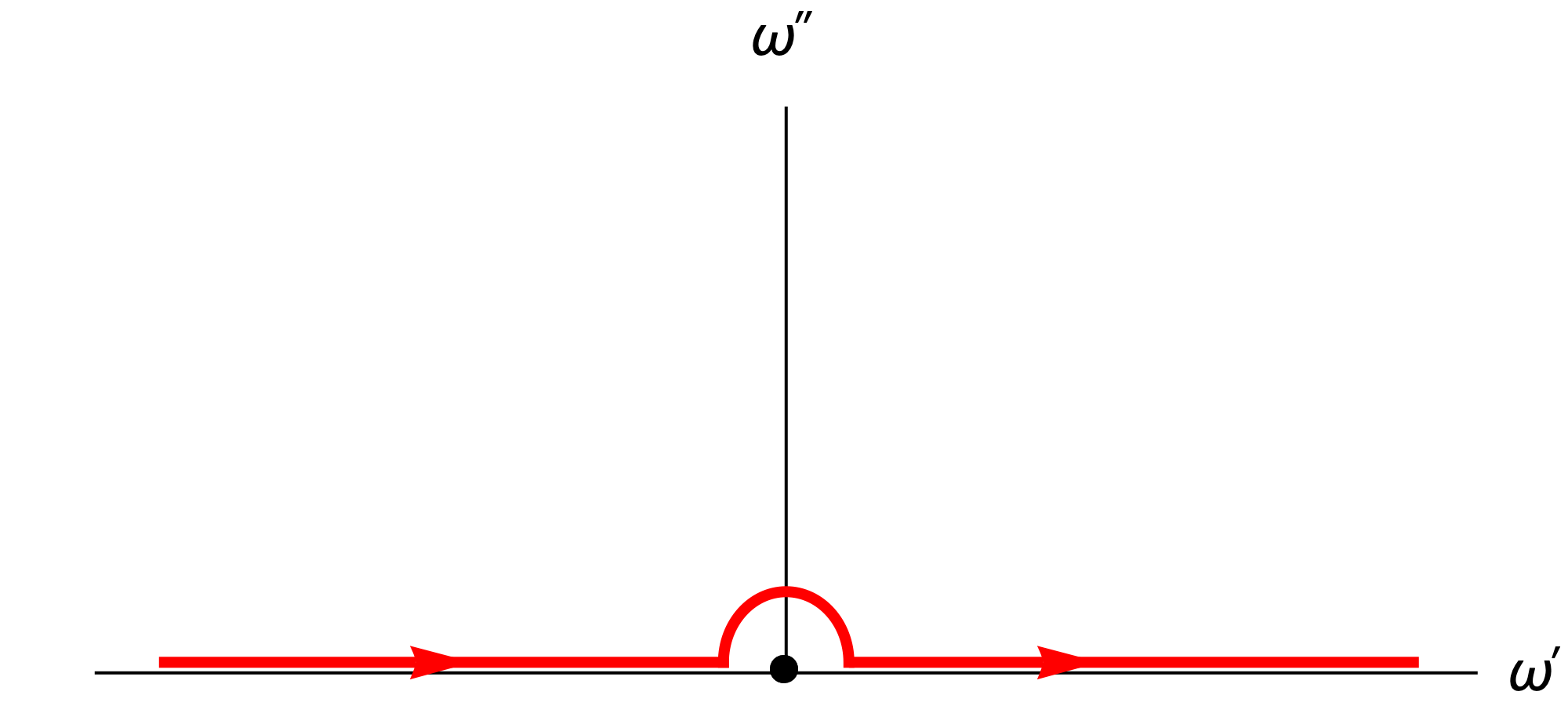}\caption{The
integration contour for $C(\omega)$  in Eq. (\ref{real_a_3_1}). The contour 
 bypasses the point $\Omega_m=0$, where the interaction $V(\omega)$ is singular and $\int d \Omega \Omega {\text {Im}} V (\omega)$ diverges.} \label{fig:contour}
\end{figure}

\section{Solution of the gap equation along the Matsubara axis}
\label{sec:matsubara}

In this Section, we present two sets of results. First, we show that  at $T=0$, there exists an infinite number of topologically distinct solutions of the non-linear gap equation. We label these solutions as  $\Delta_n (\omega_m)$, where an integer $n$ indicates how many  times $\Delta_n (\omega_m)$  changes sign along the positive Matsubara axis.
 We recall that we previously found that an infinite discrete set of solutions exists for $0<\gamma <2$ (Papers I-IV)  and
   becomes continuous at $\gamma =2$ (paper V).
 Here we show that the set again  becomes a discrete one for $\gamma >2$. In simple words, condensation energies $E_{c,n}$  with $n >0$ come closer to $E_{c,0}$  as $\gamma$ approaches $2$, ``touch'' it $\gamma =2$, where the condensation energy becomes a continuous function,
   and then pull back at larger $\gamma$, leaving $E_{c,0}$ the largest and separated by the gap from other $E_{c,n}$.
  Second, we show that the behavior of $\Delta_0 (\omega_m)$ before and after  ``touching''
 is qualitatively different. Namely,  for $\gamma <2$, $\Delta_0 (\omega)$  disappears  simultaneously with other $\Delta_n (\omega_m)$  once the pairing interaction drops below some critical value.
  For $\gamma >2$,  $\Delta_0 (\omega)$  remains non-zero when all other $\Delta_n (\omega_m)$ vanish.  To demonstrate this explicitly, we extend the model and introduce a parameter $M$, which distinguishes between the strength of the interaction in the particle-particle and the particle-hole channel ($M=1$ in the original model).
    For $\gamma <2$,
    $\Delta_n (\omega_m)$ with all $n$, including $n=0$, vanish at $M < M_{cr} (\gamma)$. For $\gamma >2$,  $\Delta_n (\omega_m)$  with $n >0$ still vanish at $M < M_{cr} (\gamma)$, but $\Delta_0 (\omega_m)$ remains finite down to $M=0$ and vanishes there in a highly non-trivial manner.
    Later, in Sec.~\ref{sec:real}, we analyze the gap function  on the real axis and show that the   $n=0$  solution does change qualitatively compared to that for $\gamma <2$ and yields qualitatively different structure of the density of states.

\subsection{Discrete set of $\Delta_n (\omega_m)$ for $\gamma >2$}
\label{sec:discrete_set}

\subsubsection{Solution of the linearized gap equation}
\label{sec:Mats_linearized}

\begin{figure}
\centering
\includegraphics[scale=0.5]{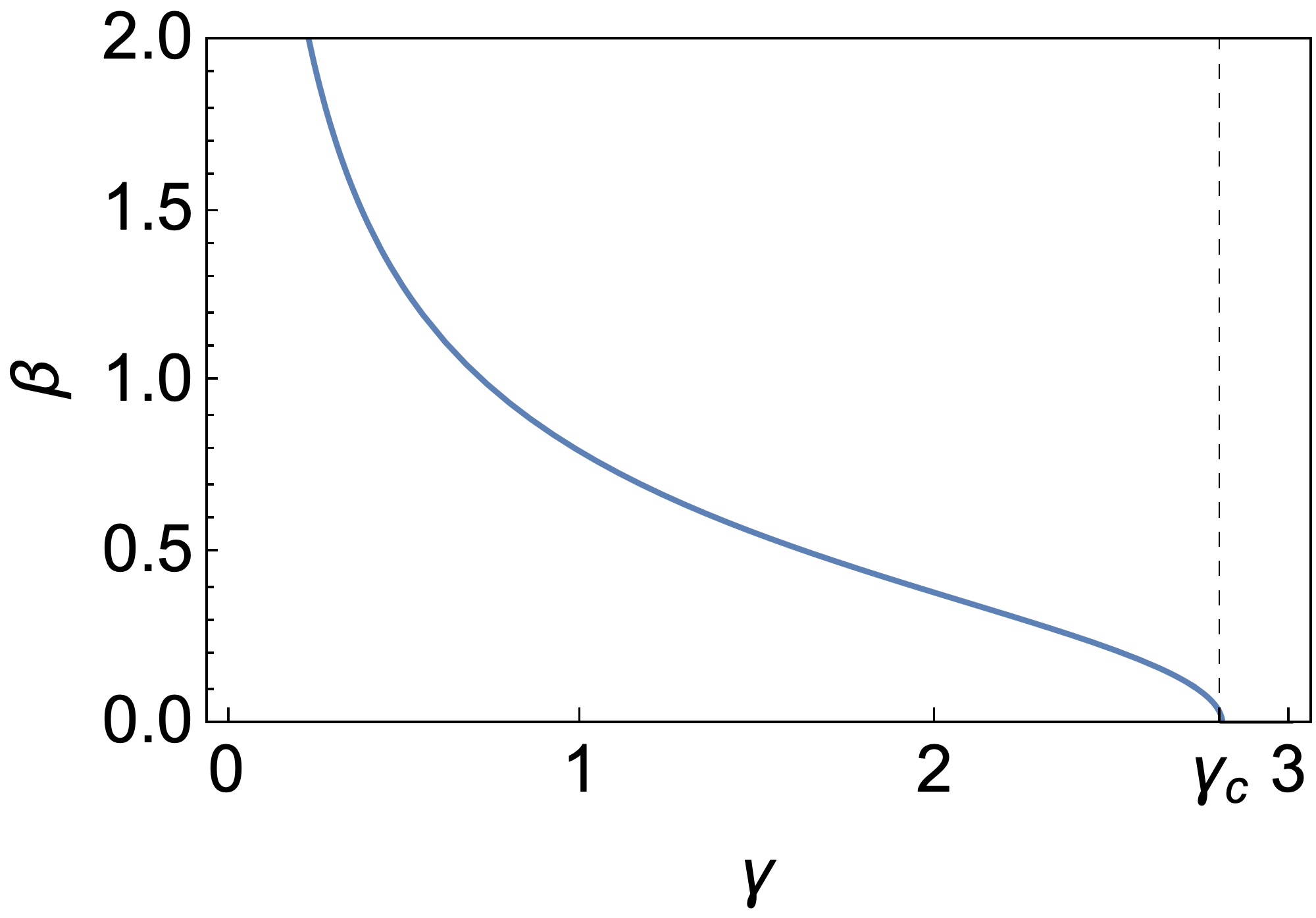}\caption{The parameter $\beta$, which sets the periodicity of the logarithmic
oscillation of $\Delta_{\infty}(\omega_{m})$ at the smallest $\omega_{m}$
regime, as a function of $\gamma$, where $\gamma_{c}\simeq2.81$.}
\label{fig:beta}
\end{figure}

We begin by showing that the solution of the linearized gap equation at $T=0$ still exists for $\gamma >2$, like for smaller $\gamma$.  We label this solution $\Delta_{\infty} (\omega_m)$ as the corresponding gap function changes sign an infinite number of times as a function of $\omega_m$.

\begin{figure}
\centering
\includegraphics[scale=0.7]{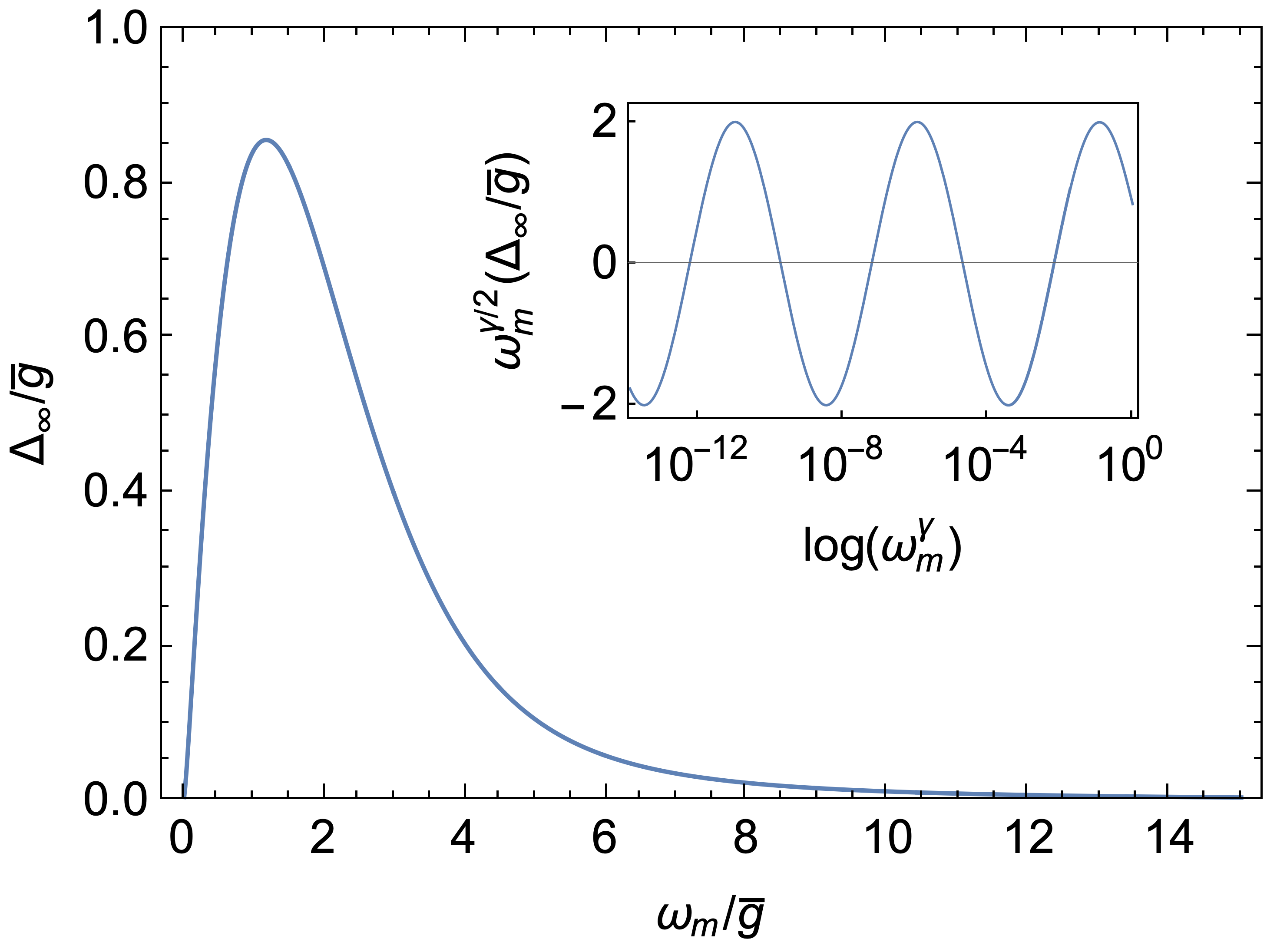}\caption{The gap function $\Delta_{\infty}(\omega_m)$ along the Matsubara axis for $\gamma=2.5$.
The inset shows the logarithmic oscillation in the infrared limit.}
\label{fig:delta_mats}
\end{figure}

 At $T=0$, the linearized gap equation (\ref{eq:gap2}) reads
\begin{equation}
\Delta_{\infty}(\omega_{m})=\frac{\bar{g}^{\gamma}}{2}\int_{-\infty}^{\infty}d\omega_{m^\prime}
\left(\frac{\Delta_{\infty}(\omega_{m^\prime})}{\omega_{m^\prime}}-\frac{\Delta_{\infty}(\omega_{m})}{\omega_{m}}
\right)\frac{\text{sgn}(\omega_{m^\prime})}{\rvert\omega_{m^\prime}-\omega_{m}\rvert^{\gamma}}.\label{eq:gap3_a}
\end{equation}
 Candidate solutions of this equation can be identified analytically at frequencies much larger and much smaller than ${\bar g}$.
 At large $\omega_m \gg {\bar g}$, one can pull out $1/\rvert\omega_{m}\rvert^{\gamma}$  from the integral and obtain
 $\Delta_{\infty}(\omega_{m}) \propto 1/\rvert\omega_{m}\rvert^\gamma$.  At small $\omega_m \ll {\bar g}$, the solution  is a combination of two power-laws $\Delta  (\omega_m) \propto |\omega_m|^{a_{1,2}}$. Substituting this form into
  (\ref{eq:gap3_a}) we find the condition on $a$:
 \begin{equation}
\int_{-\infty}^{\infty}dx\frac{\rvert x\rvert^{a}-\text{sgn}(x)}{\rvert x-1\rvert^{\gamma}}=0.
\end{equation}
For  $\gamma\leq\gamma_{cr} \simeq 2.81$, $a_{1,2}$ are complex-conjugated numbers,
$ \gamma/2 \pm i \gamma \beta$, where $\beta$ is determined from
\begin{equation}
\frac{1-\gamma}{2}\frac{\Gamma(\frac{\gamma}{2}+i\beta\gamma)\Gamma(\frac{\gamma}{2}-i\beta\gamma)}{\Gamma(\gamma)}
\left(1+\frac{\cosh(\pi\gamma\beta)}{\cos(\pi\gamma/2)}\right) =1.
\label{beta}
\end{equation}
($\gamma_{cr}$ is the solution of this equation for $\beta =0$).
We plot $\beta = \beta (\gamma)$ in Fig.~\ref{fig:beta}.   The gap function $\Delta_{\infty} (\omega_m) = |\omega_m|^{\gamma/2} \left(C |\omega|^{i\gamma \beta} + C^* |\omega|^{-i\gamma \beta}\right)$ oscillates
 as a function of $\log{|\omega_m|}$ as ($C = |C| e^{i \phi}$)
 \begin{equation}
\Delta_{\infty}(\omega_{m}\ll\bar{g})= |C| \rvert\omega_{m}\rvert^{\gamma/2}\cos(\beta\log\rvert\omega_{m}\rvert+\phi),\label{eq:smallw_mats}
\end{equation}
where $\phi$ is a free phase factor in this approximation.
The
infrared  
behavior is the same as we previously found for  smaller $\gamma$.
It is tempting to use $\phi$ as a tool that allows one to smoothly connect the limits of large and small $\omega_m$.
 There is no guarantee that this is possible as the gap equation is integral rather than differential.
  In Papers I-V we went a step further and obtained the exact solution of the linearized gap equation at $T=0$.  It reproduces
   $1/|\omega_m|^\gamma$  behavior at large $\omega_m$ and log-oscillations at small $\omega_m$ with some particular
    $\phi$.  This eventually allowed us to obtain a discrete set of solutions of the non-linear gap equation, $\Delta_n (\omega_m)$, in which
  $\Delta_{\infty} (\omega_m)$ is the smallest member. Here, we borrowed computational technique  from Papers I-V and obtained the exact solution $\Delta_{\infty} (\omega_m)$ for $\gamma >2$ (up to $\gamma_{cr} =2.81$). The exact solution again matches with analytical high-frequency and small-frequency forms, with some $\gamma-$dependent parameter $\phi$.
 We show $\Delta_{\infty} (\omega_m)$ for representative $\gamma =2.5$ in Fig. \ref{fig:delta_mats}.
  Note that because log-oscillations extend down to $\omega_m =0$,  $\Delta_{\infty} (\omega_m)$
   changes sign an infinite number of times, what justifies labeling it as $n=\infty$ solution.

\subsubsection{Sign-preserving solution.}

\label{sec:Mats_sign_preserving}

\begin{figure}
\centering
\includegraphics[scale=0.5]{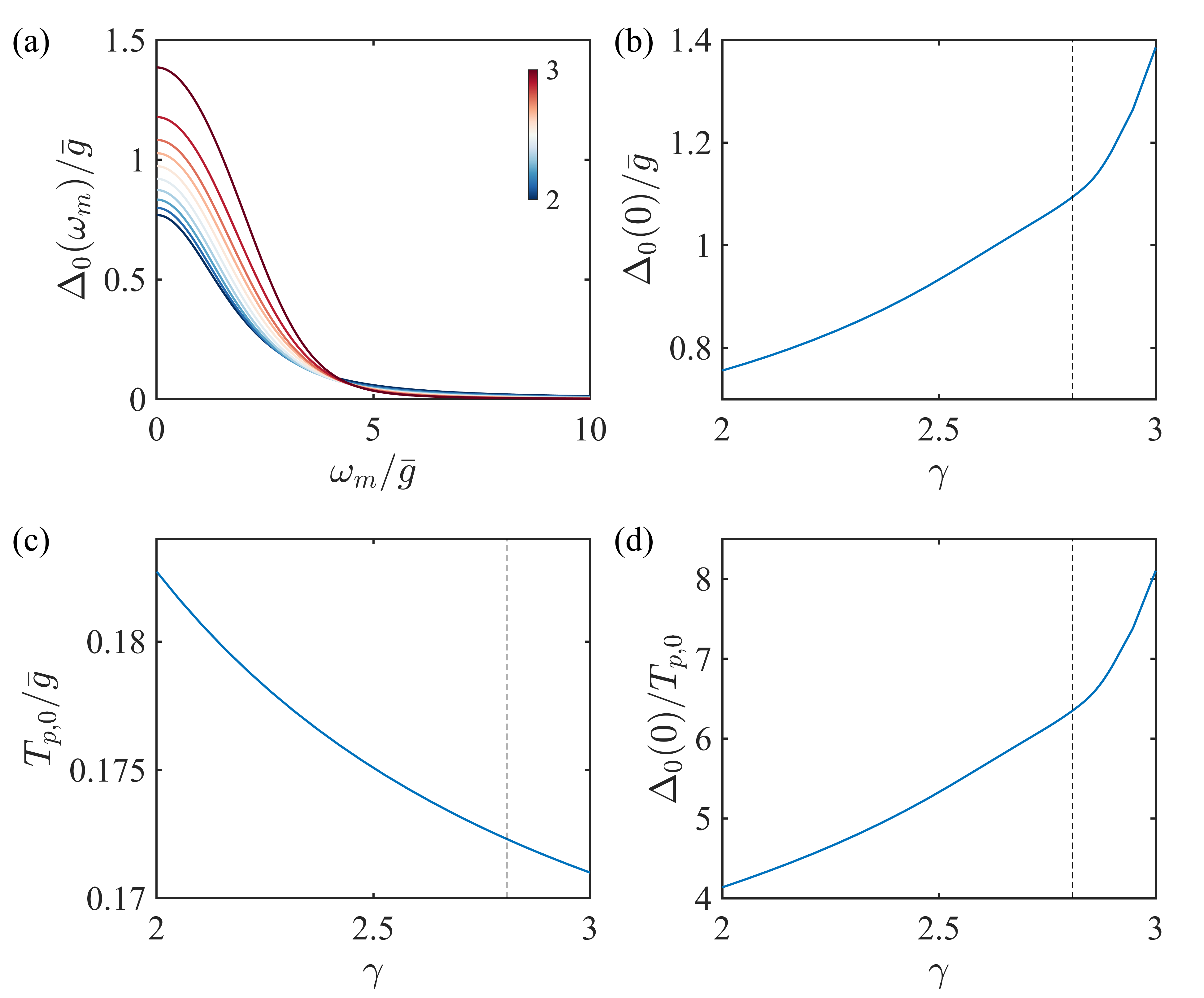}\caption{(a) The numerical solution of the gap function $\Delta_{0}(\omega_{m})$
for $2<\gamma<3$ at temperature $T=10^{-3}\bar{g}\ll T_{p,0}$. (b)
The amplitude of $\Delta_{0}(0)$, (c) the onset temperature $T_{p,0}$
and (d) the ratio $\Delta_{0}(0)/T_{p,0}$ as a function of $\gamma\in(2,3)$.
The dashed vertical line indicates the critical $\gamma_{c}$ above
which the $n\protect\geq1$ solutions do not exist. Correspondingly,
it appears as a kink in the curve of $\Delta_{0}(0)$ and the ratio
$\Delta_{0}(0)/T_{p,0}$. }
\label{fig:mats_n=0}
\end{figure}

\begin{figure}
\centering
\includegraphics[scale=0.5]{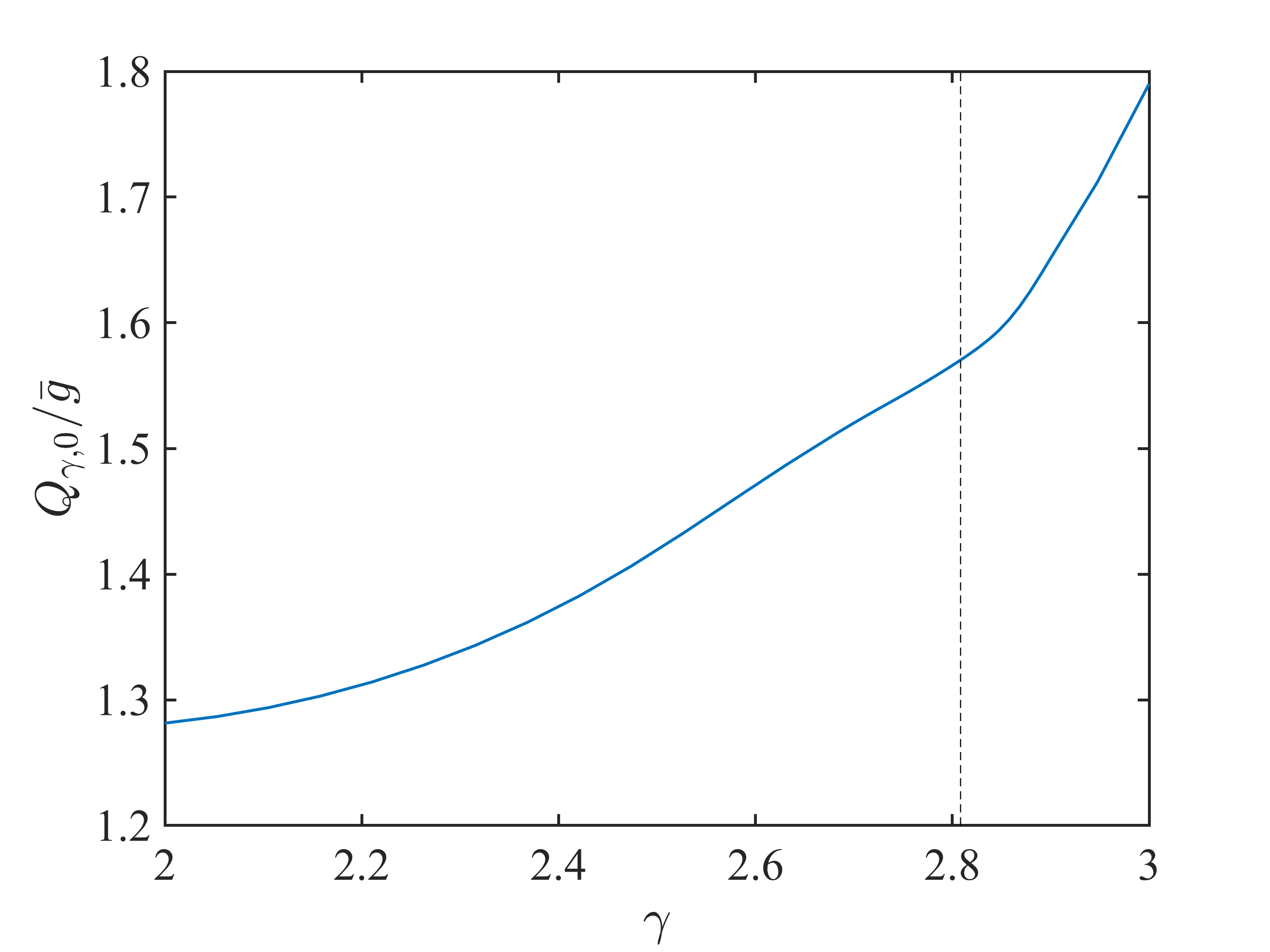}\caption{$Q_{\gamma,0}$, defined in Eq. (\ref{eq:Qgamma}), as a function
of $\gamma\in(2,3)$. It is obtained from the numerical solution of
the non-linear gap equation shown in Fig. \ref{fig:mats_n=0}.
The dashed vertical line has the same meaning as in Fig. \ref{fig:mats_n=0}.
A kink in this curve also appears around this critical $\gamma$.}
\label{fig:Qgamma}
\end{figure}

We now consider the opposite limit  -- the sign-preserving, $n=0$ solution of the non-linear gap equation.
We obtained this solution numerically and show the results in Fig.~\ref{fig:mats_n=0}.  In Fig.  \ref{fig:mats_n=0} (a), we show $\Delta_0 (\omega_m)$ for several representative $2 <\gamma <3$.   We see that the $\Delta_0 (\omega_m)$ has a finite value at $\omega_m =0$ and  monotonically decreases with increasing $\omega_m$.  This is similar to the behavior of $\Delta_0 (\omega_m)$ at smaller $\gamma$.  In Fig. \ref{fig:mats_n=0} (b), we show  $\Delta_0 (0)$ vs $\gamma$.  For a generic $\gamma$ between $2$ and $3$, $\Delta_0 (0) \sim {\bar g}$.
 At $\gamma \to 3$, $\Delta_0 (0)$ diverges logarithmically (Ref. \cite{wu2019pairing}).
 For completeness, in  Fig.~\ref{fig:mats_n=0} (c) and (d) we show the corresponding
onset temperature for the pairing $T_{p,0}$ and the ratio $\Delta_0 (0)/T_{p,0}$. The results are consistent with what has been reported earlier~\cite{wu2019pairing,lee2018pairing}.
 At large frequencies, $\Delta_0 (\omega_m)$ scales as $1/|\omega_m|^\gamma$. This form can be straightforwardly extracted from the gap equation in the same way as for the $n = \infty$ solution, by pulling out  $1/|\omega_m|^\gamma$ from the integrand.  For the $n=0$ solution, this gives
 \begin{equation}
\Delta_{0}(\omega_{m})=Q_{\gamma,0}\left(\frac{\bar{g}}{|\omega_{m}|}\right)^{\gamma},
\end{equation}
where
\begin{equation}
Q_{\gamma,0}=\int_{0}^{\infty}\frac{d\omega_{m}^{\prime}\Delta_{0}(\omega_{m}^{\prime})}{\sqrt{\Delta_{0}^{2}(
\omega_{m}^{\prime})+(\omega_{m}^{\prime})^{2}}}.\label{eq:Qgamma}
\end{equation}
Substituting $\Delta_0 (\omega_m) \propto 1/|\omega_m|^\gamma$, we find that the integral is ultra-violet convergent, what justifies pulling out $1/|\omega_m|^\gamma$. For a generic $\gamma$ between $2$ and $3$, the frequency integral in (\ref{eq:Qgamma})  converges at $\omega_{m}^\prime  \sim \Delta_0 (0) \sim {\bar g}$, hence  $Q_{\gamma,0}$ is of order ${\bar g}$. We show $Q_{\gamma,0}$ in Fig. \ref{fig:Qgamma}. We see that it is indeed of order ${\bar g}$.

\begin{figure}
\centering
\includegraphics[scale=0.22]{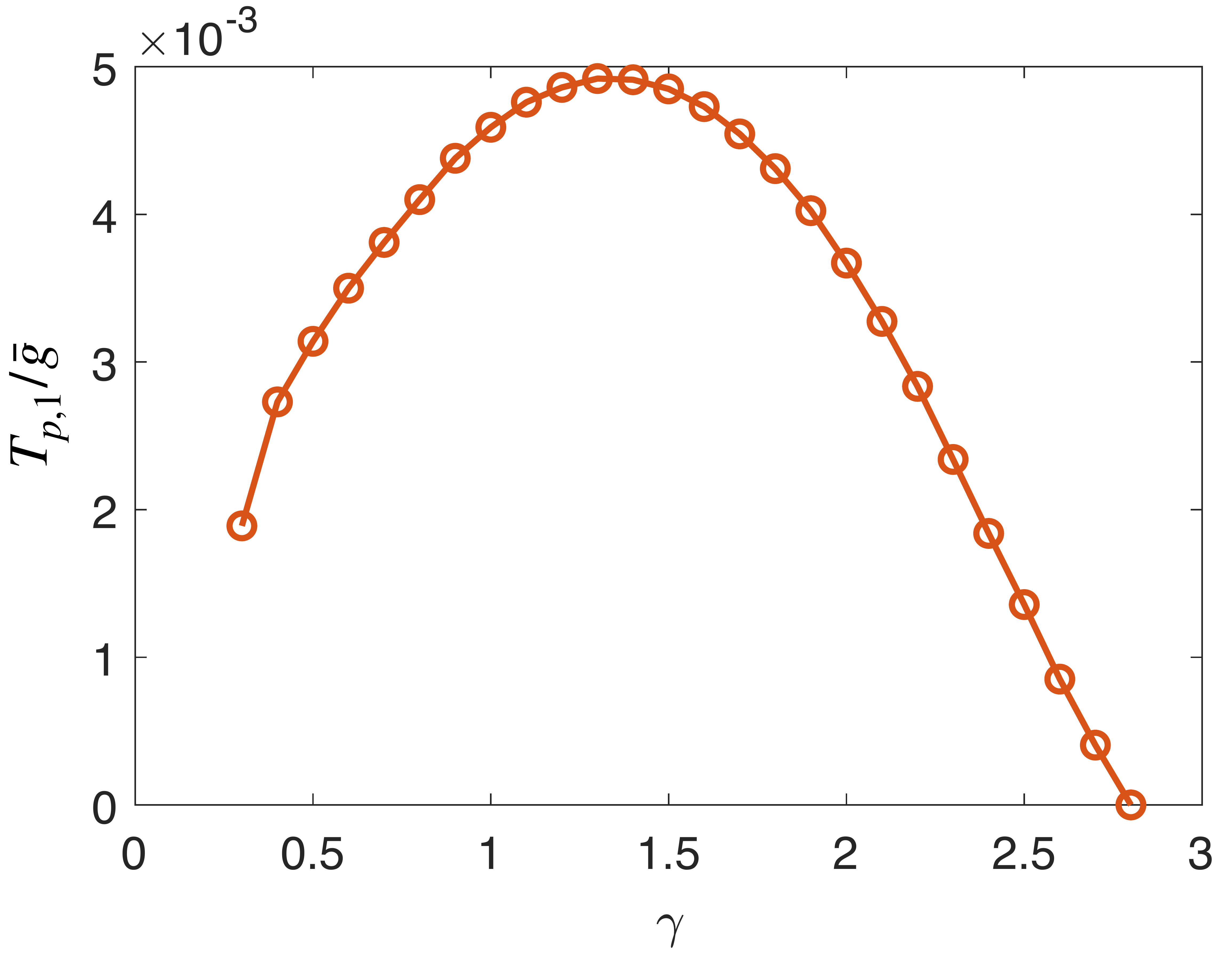}
\caption{The onset temperature for the pairing $T_{p,1}$  as a function of $\gamma$. It is a non-monotonic function of $\gamma$ and vanishes at $\gamma=\gamma_{cr}=2.81$.}
\label{fig:Tp1}
\end{figure}

\subsubsection{Discrete set of solutions.}
\label{sec:Mats_expansion}

For $\gamma <2$, we demonstrated in Papers I-IV that  $\Delta_{\infty}(\omega_{m})$ and $\Delta_{0}(\omega_{m})$ are the two end points of an infinite discrete set of solutions $\Delta_n (\omega_m)$. A gap function labeled by $n$ changes sign $n$ times along the positive Matsubara axis.
The set becomes continuous at $\gamma =2$ (Paper V).  Here we show that an infinite set of  $\Delta_n (\omega_m)$ exists also for $\gamma >2$, but again becomes discrete.

To demonstrate this, we search for the solution of the non-linear gap equation by expanding to infinite order in
 $\Delta(\omega_{m'})$ in Eq.~(\ref{eq:gap1}). This yields
 \begin{equation}
\Delta(\omega_{m})=\sum_{j=0}^{\infty} \epsilon^{2j+1} \Delta^{(2j+1)}(\omega_{m}),\label{exx1_a}
\end{equation}
where  $\Delta^{(1)}(\omega_{m}) = \Delta_{\infty} (\omega_m)$
 from Eq.~\eqref{eq:smallw_mats},
 $\epsilon$ is a parameter, which we adjust to get a solution.
The two limits we considered earlier correspond to an infinitesimally small $\epsilon$, when $\Delta(\omega_{m}) = \epsilon  \Delta_{\infty} (\omega_m)$, and to some finite $\epsilon = \epsilon_{0}$ for the $n=0$ solution.

In general, the conditions on $\epsilon$ are obtained by substituting $\Delta (\omega_m)$ from Eq.~(\ref{exx1_a})  into Eq.~(\ref{eq:gap1}), solving iteratively for $\Delta^{(2j+1)}$ in terms of $\Delta^{(2j'+1)}$ and $j' <j$, and requiring that the series converge.
    For a BCS superconductor, the solution exists only for a single value of $\epsilon$.
   For the $\gamma$-model with $\gamma \leq 2$,  the solutions exist for a discrete set of $\epsilon_n$ for $\gamma <2$ and for arbitrary $0<\epsilon < \epsilon_{max}$ for $\gamma =2$.

 For $\gamma >2$, we find that the solutions exist for a discrete set of $\epsilon_n$, of which $\epsilon_0$ is the largest.  This is very similar to the case $\gamma <2$. The details of the calculations are rather involved and we moved them to Appendix~\ref{app:discrete_set}.

We also compute the condensation energy for different solutions using the expression for the free energy in the $\gamma$-model in Paper I. The set of condensation energies $E_{c,n}$ is discrete, and, as one could expect, the largest condensation energy is for the $n=0$ solution. This again is very similar to what we previously found for $\gamma <2$. We illustrate this in Fig.~\ref{fig:Ec} (c).

\subsection{Decoupling of the $n=0$  solution from the set}
\label{sec:decouple}

So far, our results for $\gamma>2$ agree with those for $\gamma <2$.  In both cases, there exists a discrete set of  $\Delta_n (\omega_m)$ with integer $n$, ranging from $0$ to $\infty$, and  the condensation energy $E_{c,0}$ is the largest.

 We now show that the analogy is only partially correct, and there is one crucial feature on which the
 two cases differ qualitatively.
  Namely, we argue that for $\gamma <2$ the solutions with all $n$ behave as one set, while for $\gamma >2$,
   the $n=0$ solution decouples from the set and  behaves differently from the other solutions with $n >0$.
   What we mean here is that for smaller $\gamma$, all $\Delta_n (0)$ disappear simultaneously once we extend the model and reduce the strength of the pairing interaction below a certain value (more on this below).  For $\gamma >2$, the solutions with $n >0$ disappear under the same conditions, but the one with $n=0$ survives. This distinction can be seen already in the original $\gamma$ model.  As we said before, the solution with $n = \infty$ exists only up to $\gamma_{cr} =2.81$~\footnote{We note in passing that at $\gamma=\gamma_{cr}$, the two power-law solutions merge into a single $|\omega|^{\gamma/2}$,  but at this point another solution $\Delta_\infty (\omega_m) \propto \rvert\omega_{m}\rvert^{\gamma/2}\log\rvert\omega_{m}\rvert$ emerges, as can be verified by using he identity
$\int_{-\infty}^{\infty}dx  \rvert x\rvert^{\gamma/2-1}\log\rvert x\rvert /\rvert x-1\rvert^{\gamma}=0$.
 As a result, the low-frequency $\Delta_{\infty}(\omega_{m}) \propto \rvert\omega_{m}\rvert^{\gamma/2}\log{\rvert\omega_{m}\rvert/\omega_{*}}$ still contains a free parameter $\omega^*$ that allows one to match this low-frequency form with   $\Delta_{\infty}(\omega_{m}) \propto 1/ \rvert\omega_{m}\rvert^{\gamma}$ at high frequencies.}.
 If the solutions form a single set, the solutions with non-infinite $n$ should disappear at the same $\gamma_{cr}$. This can be verified by computing the corresponding onset pairing temperatures $T_{p,n}$.
 In Fig.~\ref{fig:Tp1} we plot $T_{p,1}$ as a function of $\gamma$. We see that it vanishes at $\gamma_{cr}$, as we anticipated.
 We verified that $T_{p,2}$ vanishes as well, this leaves little doubt that all $T_{p,n}$ with $n >0$ vanish
 at $\gamma_{cr}$. Then, at $T=0$, all $\Delta_n (\omega_m)$ with $n >0$  vanish simultaneously at $\gamma_{cr}$.
 However, we see from Fig.~\ref{fig:delta_mats} (b) and (c) that $T_{p,0}$ and the gap function $\Delta_0 (\omega_m)$  at $T=0$  remain finite at this $\gamma$, the only signature of $\gamma_{cr}$ in these figures is a kink in the $\gamma$ dependence of $T_{p,0}$ and of $\Delta_0 (0)$.   Clearly then, the $n=0$ solution decouples from the set of $\Delta_n (\omega_m)$  with $n >0$.

\subsubsection{Extended $\gamma$ model}

To see this more clearly and also to understand the difference between $\gamma <2$ and $\gamma >2$, we extend the $\gamma$ model in the same way as in Papers IV and V, by introducing a parameter $M \neq 1$, which separates the pairing interaction and the one in the particle-hole channel. The original $\gamma$-model, in which both interactions are  $V(\Omega_m)$, corresponds to $M=1$.
We introduce $M \neq 1$ in such  a way that the pairing interaction gets weaker at $M <1$.
 The extension has to be done carefully to avoid emerging singularities from $\int d \omega_{m'}/|\omega_m-\omega_{m'}|^\gamma$, which cancel out in the gap equation at $M=1$ (see Eq.~(\ref{eq:gap1})).

  We already used this extension for different purposes in Papers IV and V. There, we
   derived the modified gap equation:
   \bea
&&D (\bo_m) \left(\bo_m + \frac{1-M}{2} \int ~\frac{d \bo'_{m}}{|\bo_m - \bo'_{m}|^{\gamma}}  \left(\frac{\mbox{\sign} \bo_{m}}{\sqrt{1 + D^2 (\bo_{m})}} - \frac{\mbox{\sign} \bo'_{m}}{\sqrt{1 + D^2 (\bo'_{m})}}
\right)\right) = \nonumber \\
&& \frac{1}{2} \int ~\frac{d \bo'_{m}}{|\bo_m - \bo'_{m}|^{\gamma}}  \frac{D(\bo'_{m})-D(\bo_m)}{\sqrt{1 + D^2 (\bo'_{m})}} \mbox{\sign} \bo'_{m}
\label{3_12b}
\eea
where $D({\bo_m})=\Delta(\omega_m)/\omega_m$,  $\bo_m=\omega_m/{\bar g}_M$ and ${\bar g}_M = {\bar g}/M^{1/\gamma}$.
At $M=1$, Eq. (\ref{3_12b}) reduces to Eq. (\ref{eq:gap1})
\begin{figure}
\centering
\includegraphics[scale=0.8]{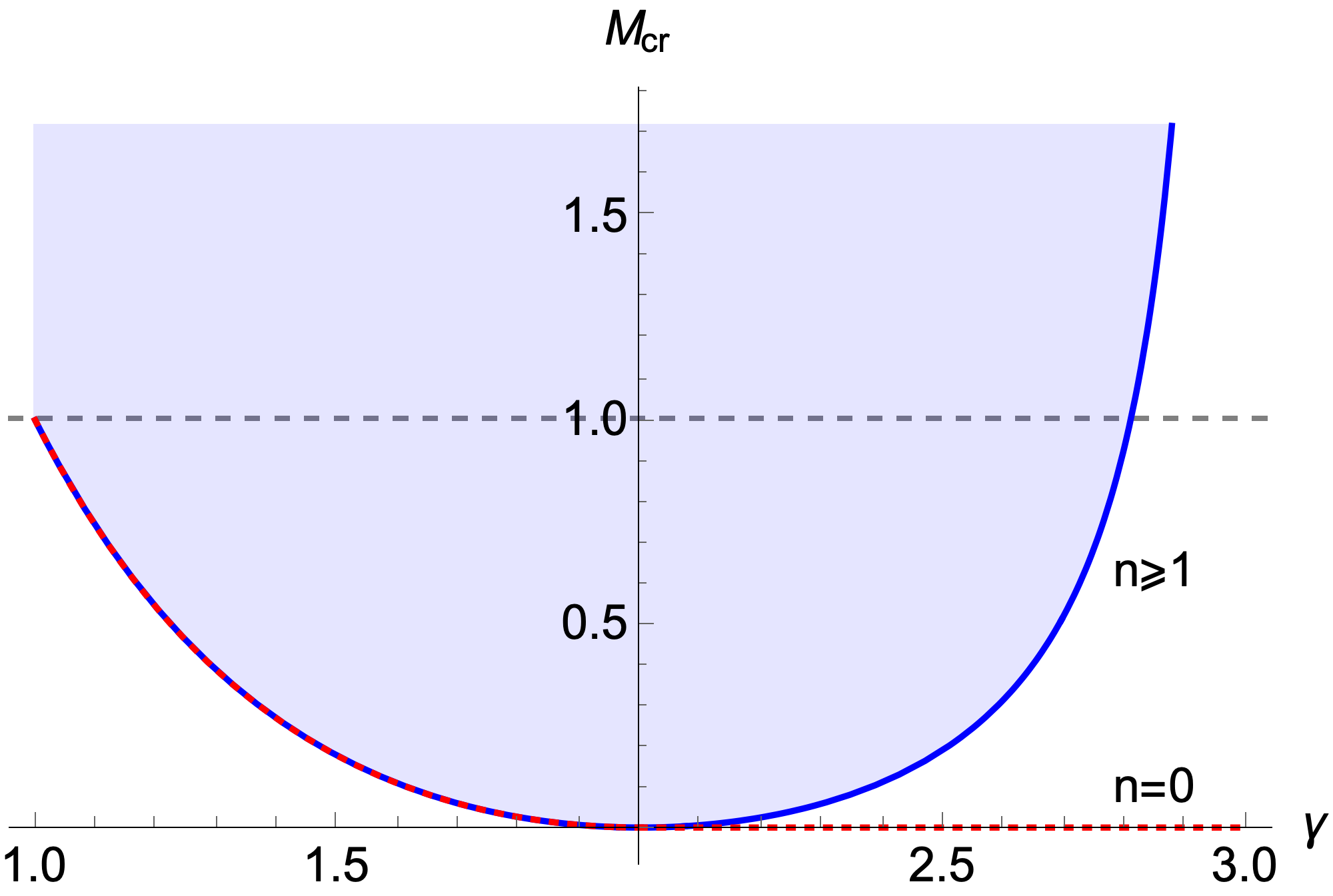}
\caption{The critical value of the parameter $M$ (see text)  as a function of $\gamma$. At $\gamma\leq 2$, $M_{\text{cr}}$ is non-zero and the same for all solutions
 with $n\geq 0$.
At $\gamma>2$, critical $M$ is zero for the $n =0$ solution (red dots) and is finite (and the same) for all other solutions with $n\geq 1$.}
\label{fig:Mcr}
\end{figure}

The extended model has the same structure of solutions as the original one: there is a discrete set of solutions $\Delta_n (\omega_m)$ for $\gamma <2$ and $\gamma >2$ and a continuous set for $\gamma =2$.  The end point, $\Delta_{\infty} (\omega_m)$ is
the solution of the linearized gap equation.
 Like for the original model, at small $\omega_m$, $\Delta_{\infty} (\omega_m) \propto |\omega_m|^{\gamma/2} \cos ({\beta \log{|{\bar \omega}_m|^\gamma} + \phi})$.
The parameter $\beta$ must be real, which restricts $M$ to  $M\geq M_{cr}(\gamma)$. The critical value is
\beq
M_{cr}(\gamma)=\frac{1-\gamma}{2}\frac{\Gamma^2(\frac{\gamma}{2})}{\Gamma(\gamma)}\left(1+\frac{1}{\cos(\pi\gamma/2)}\right).
\eeq
We plot $M_{cr}$ vs $\gamma$ in Fig. ~\ref{fig:Mcr}. The solution with $n = \infty$ exists in the blue area in this figure.
 The boundary crosses $M=1$ at $\gamma_{cr} =2.81$, as we found earlier.

\begin{figure}
\centering
\includegraphics[scale=0.6]{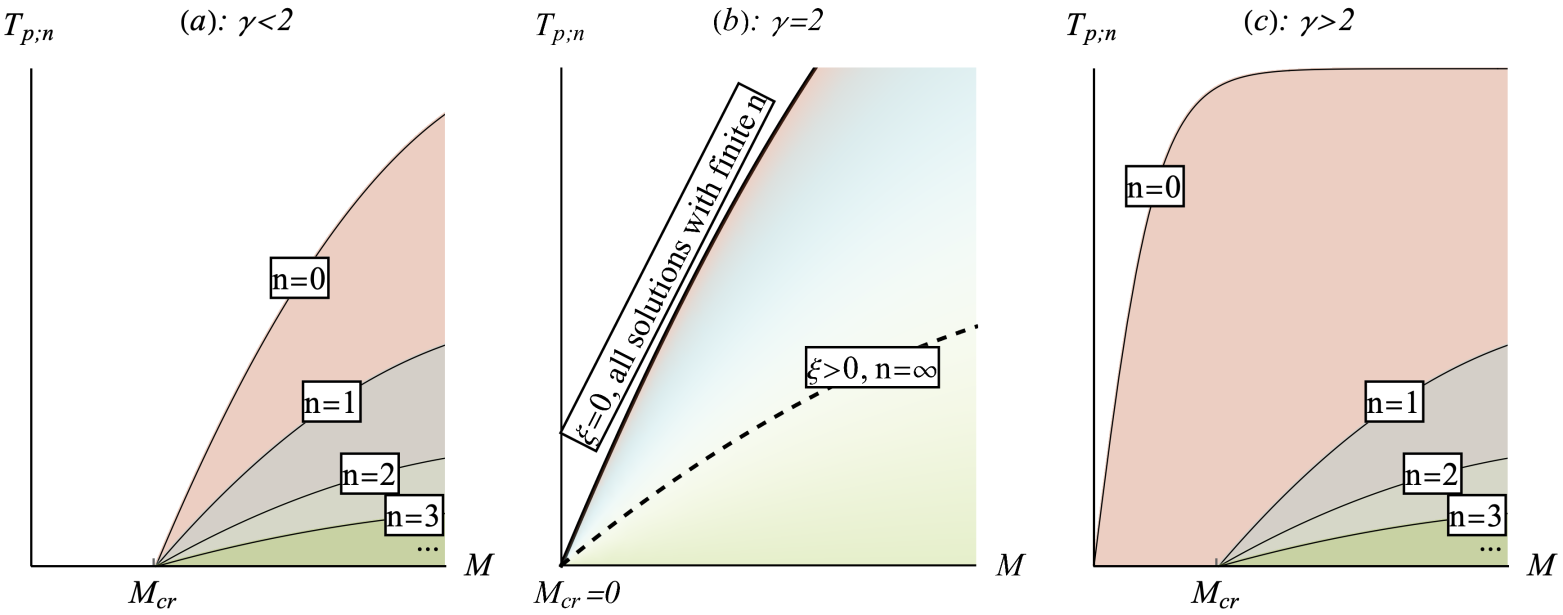}\caption{Onset pairing temperature ($T_{p;n}$) of each topologically distinct solution labeled by integer $n=0,1,2,...$, where (a) $\gamma<2$, (b) $\gamma=2$, and (c) $\gamma>2$.}
\label{fig:Tpn}
\end{figure}

\begin{figure}
\centering
\includegraphics[scale=0.4]{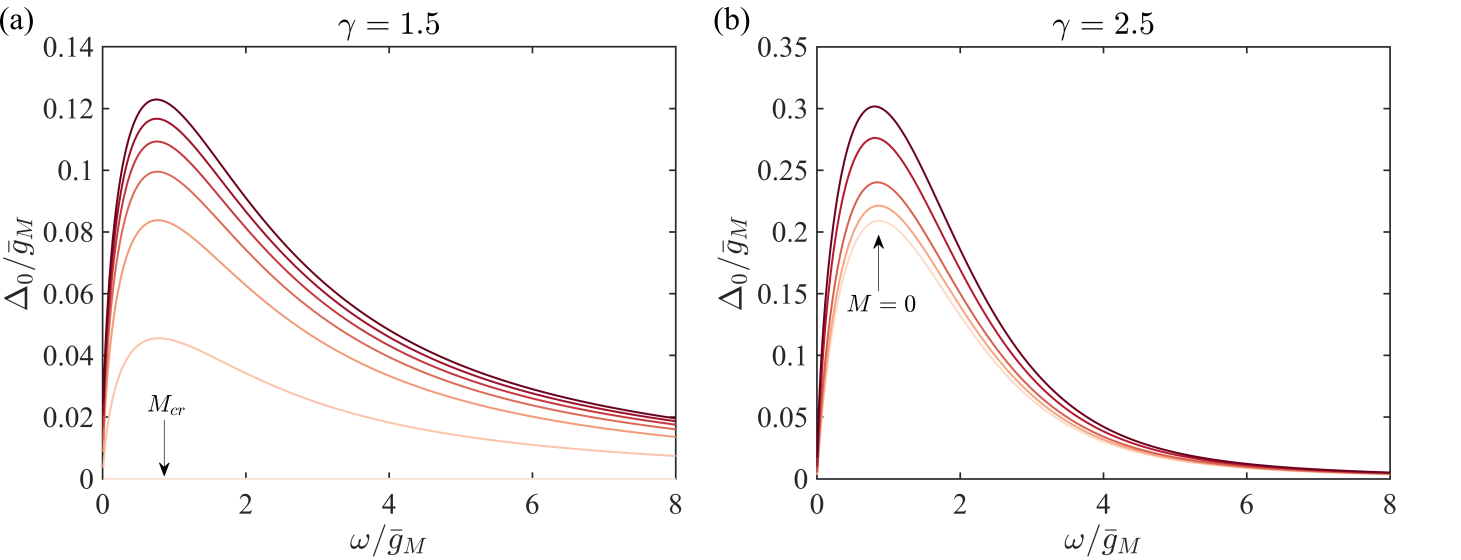}
\caption{Solution of the extended model, Eq.~(\ref{3_12b}), for various values of $M$ in the vicinity of the critical point, where (a) $\gamma=1.5$ and (b) $\gamma=2.5$.}
\label{fig:nleqM}
\end{figure}

We obtained numerically the onset temperatures for the pairing $T_{p,n} (M)$. For $\gamma \leq 2$ we found that all $T_{p,n}$  vanish at the same $M=M_{cr}$.  This implies at $T=0$, $\Delta_n (\omega_m)$ with all $n$, including $n=0$, vanish upon approaching the critical line $M_{cr} (\gamma)$ from above.  We show the behavior of $\Delta_0 (\omega_m)$ in Fig.~\ref{fig:nleqM} (a) and illustrate this result in Fig.~\ref{fig:Tpn} (a).
For $\gamma =2$, $M_{cr} =0$. The set is continuous, and all gap functions from the set vanish upon approaching $M_{cr} =0$ from above (see Fig.~\ref{fig:Tpn} (b)).

For $\gamma >2$, the result is different.
The onset temperatures $T_{p,n}$ with $n >0$ still vanish at $M_{cr} >0$, along with the corresponding $\Delta_n (\omega_m)$ at $T=0$.  However, $T_{p,0}$ and $\Delta_0 (\omega_m)$ remain finite at $M_{cr}$ (see Fig.~\ref{fig:Tpn} (c) for illustration).
We show the numerical results for $\Delta_0 (\omega_m)$ at different $M$ in Fig. ~\ref{fig:nleqM} (a) for representative $\gamma =2.5$ ($M_{cr} = 0.192$). This clearly shows that for $\gamma >2$  the solution with $n=0$  decouples from the set of solutions with $n\geq 1$. A non-zero $\Delta_0 (\omega_m)$  exists down to $M=0$ (the red dashed line in Fig.~\ref{fig:Mcr}), where it vanishes in a rather peculiar way: $\Delta_0 (0)$ gradually tends to zero at $M \to 0$, while the full function $\Delta_0 (\omega_m)$ remains finite (see Fig.~\ref{fig:nleqM} (b)) and at $M =0+$ becomes the end point of a continuum of solutions
(see Appendix~\ref{sec:app_extended_model} for details).

\subsection{Disparity between gap functions with $n=0$ and $n >0$ in the upper frequency half-plane}
\label{sec:vortex}

We now present complimentary evidence for qualitative difference between the gap functions with $n=0$ and $n >0$, by
 extending $\Delta_{n} (\omega_m)$ from the Matsubara axis into the upper half-plane of frequency.  To obtain $\Delta_n (z)$, where $z = \omega' + i \omega^{''}$ and $\omega^{''} >0$,  we  first obtained $\Delta_n (\omega)$ along the real axis by solving the gap equation (\ref{el8}) and then used Cauchy relation
 \begin{equation}
\Delta_n(z)=\frac{2}{\pi}\int_{0}^{\infty}d\omega\frac{\omega\Delta_n^{''}(\omega)}{\omega^{2}-z^{2}},\label{eq:cauchy}
\end{equation}
We will discuss the gap function along the real axis in the next section.
Here, we focus on zeros of $\Delta_n (\omega)$ away from the Matsubara axis, i.e., at the dynamical vortices at complex $z$.

\begin{figure}
\centering
\includegraphics[scale=0.3]{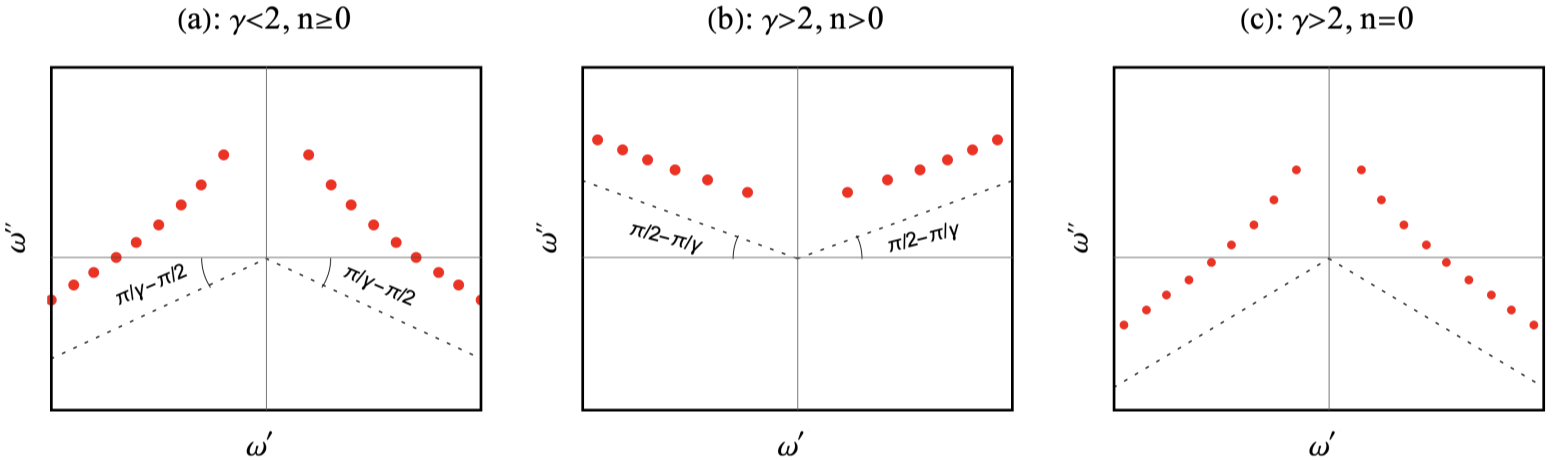}\caption{Schematic plot of the locations of dynamical vortices on the complex frequency plane for $\Delta_n (z)$:
 (a) $\gamma <2$, all $n \geq 0$,
 (b) $\gamma >2$, $n>0$, 
 and (c), $\gamma >2$, $n=0$.
Vortices along the Matsubara axis for the gap functions with $n>0$  are not shown.}
\label{fig:vortex}
\end{figure}

In Paper IV we showed that the vortices appear at $\gamma >1$. For $\gamma <2$, the number of vortices is finite and the same for all $n$, including $n =0$,  which is another evidence that  gap functions  with all $n$  are members of
the same set.  The structure of vortices can be understood by extending the exact solution for $n=\infty$ to complex $z$.
  This analysis shows (see Appendix \ref{sec:app_exact} for details) that vortices are located above the line in the  lower frequency half-plane, at the angle
$\pi/2  -\pi/\gamma$,
   counted from the real axis, see Fig. ~\ref{fig:vortex} (a).
   (to obtain this, we allowed $z$ to move into the lower frequency half-plane).  As $\gamma$ approaches $2$ from below,
    more vortices cross from the lower to the upper frequency half-plane.  At $\gamma =2$, the line coincides with the real axis, and the number of vortices in the upper frequency half-plane becomes infinite. Again, this behavior holds for all $n \geq 0$, and our numerical analysis in Paper V shows that even the locations of vortices are the same for all $n$. The set of vortices ends up at an essential singularity at $z =\infty$. Its presence is crucial as otherwise the extension from an infinite set of vortex
    points would give rise to zero gap function everywhere in the upper half-plane, including the Matsubara axis.

\begin{figure}
\centering
\includegraphics[scale=0.4]{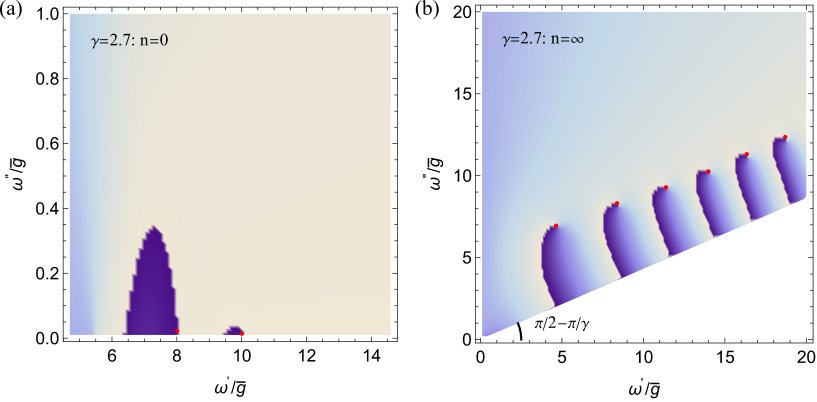}
\caption{The plot of the phase of the gap function $\Delta_n (z) = |\Delta_n (z)| e^{i\eta_n (z)}$ in the upper half-plane of complex frequency $z = \omega' + i \omega^{''}$ for representative $\gamma =2.7$. The positions of the vortices are marked by red dots.  Left panel, $n=0$ (obtained by using the approximated approach based on Eq.~(\ref{eq:phi})).  There is a finite number of vortices in the upper half-plane. Right panel - $n=\infty$.  The number of vortices is infinite; all vortices are located above the direction specified by the angle $\pi/2 (\gamma-2)/\gamma$ counted from the real axis. }
\label{fig:vortex_1}
\end{figure}

    For  $\gamma >2$, our numerical analysis shows  different behavior, schematically shown in panels (b) and (c) in Fig. ~\ref{fig:vortex}.   Namely,  $\Delta_n (z)$ with $n >0$  still possess an infinite number of vortices above the direction in the upper half-plane, specified by the angle
    $\pi/2  -\pi/\gamma$,
    see  Fig.~\ref{fig:vortex} (b).  However, for the $n=0$ solution, the corresponding axis bounds back into the lower half-plane, and as the consequence, the number of vortices in the upper half-plane of frequency becomes finite, Fig.~\ref{fig:vortex} (c).  In Fig.~\ref{fig:vortex_1}, we present the numerical results for $\Delta_0 (z)$ and $\Delta_\infty (z)$  which show this behavior for representative
    $\gamma =2.7$.
    The result for $\Delta_\infty (z)$ was obtained by analytical continuation of the exact solution on the Matsubara axis to complex $z$ in the upper frequency half-plane. We clearly see that
      for $\gamma >2$, the solution with $n=0$ decouples from other solutions with $n >0$,  like we found above in the Matsubara axis analysis. We also note that for $\gamma >2$, $\Delta_n (z)$ with $n >0$ are non-zero because of essential singularity at the end point of the set of vortices at $z = \infty$. Without it, an analytic continuation from the infinite set of vortex points into the upper half-plane would give $\Delta_{n>0} (z) =0$.
       That all non-zero  $\Delta_{n>0} (z)$ emerge at $\gamma >2$ as a multi-valued extension from the same essential singularity is fully consistent with our  earlier results that these solutions form the set with the same qualitative behavior of all members.
       The $n=0$ solution, on the other hand, does not come from an essential singularity and therefore is not a part of the set (at large $z$, $\Delta_0 (z) \sim 1/|z|^\gamma$ everywhere in the upper half-plane).

 \section{Gap equation along the real frequency axis}
\label{sec:real}

We now address the issue of whether there are any qualitative differences in the behavior of observables at $T=0$
 between $\gamma <2$ and $\gamma >2$.  In both cases the condensation energy is the largest for the $n=0$ solution, so we focus on  the form of $\Delta_0 (z)$.  For definiteness, we focus on the original $\gamma$ model with $M=1$.   We found earlier in this paper that the forms of $\Delta_0 (\omega_m)$ on the Matsubara axis at $\gamma <2$ and $\gamma >2$ are very similar. In both cases, the gap has a finite value at zero frequency, decreases monotonically with $\omega_m$, and scales as $1/|\omega_m|^\gamma$ at the largest $\omega_m$. Vortex structure at complex $z$ is also similar  -- in both cases there is a finite number of vortices in the upper frequency half-plane.
 Below we analyze the gap function along the real axis.  We show that its form changes {\it qualitatively} between $\gamma <2$ and $\gamma >2$, and the change in $\Delta_0 (\omega)$ leads to new feature in the DOS
  for $\gamma >2$:  the appearance of a bound state inside the gap with degeneracy proportional to the total number of particles in the system.   This bound state  shows up as a $\delta$-functional peak in the DOS  with an infinite weight in the thermodynamic limit.  We show later that this feature is robust against weak perturbations, in particular it survives when a pairing boson is massive, as long as the mass value is below a finite threshold.

We now analyze the gap function $\Delta_0 (\omega)$. At the smallest and the largest frequencies, $\Delta_0 (\omega)$  can be obtained by a direct rotation
from the Matsubara axis, i.e., by replacing
 $i\omega_m$ by $\omega + i 0^+$.  This yields
 $\Delta_{0}(\omega)\approx \Delta_0(0) \left( 1- O\left((\omega/{\bar g})^2\right) +...\right) $ at $|\omega|\ll {\bar g}$ and $\Delta_{0}(\omega>0)\approx Q_{0,\gamma}e^{i\pi\gamma/2} /(\omega+i0^+)^{\gamma}$ 
 at $|\omega|\gg {\bar g}$
respectively.  However, to obtain the form of $\Delta_0 (\omega)$ at intermediate $\omega \geq  {\bar g}$ one has to solve the non-linear gap equation in real frequencies, Eq. (\ref{el8}).

This equation contains three functions of frequency, $A(\omega)$, $B(\omega)$, and $C(\omega)$. The functions $A(\omega)$ and $B(\omega)$ can be expressed in terms of the gap function on the Matsubara axis:
 	\begin{equation}
		\begin{aligned}
			A(\omega)&=\pi T\sum_{\omega_m>0}\frac{D(\omega_m)}{\sqrt{1+D^2(\omega_m)}}\left(\frac{\bar{g}^\gamma}{(\omega_m+i\omega)^\gamma} + \frac{\bar{g}^\gamma}{(\omega_m-i\omega)^\gamma}\right), \\
			B(\omega)&=1+\frac{i\pi}{\omega} T\sum_{\omega_m>0}\frac{1}{\sqrt{1+D^2(\omega_m)}}\left(
\frac{\bar{g}^\gamma}{(\omega_m+i\omega)^\gamma} - \frac{\bar{g}^\gamma}{(\omega_m-i\omega)^\gamma}\right), \\
		\end{aligned}
\label{last_1}
	\end{equation}
 where, we remind,
 $D(\omega_m) = \Delta (\omega_m)/\omega_m$.
 For the $n=0$ solution, $\Delta (\omega_m) = \Delta_0 (\omega_m)$.   Using the fact that $\Delta_0 (\omega_m)$ is monotonically decreasing function of frequency, one can
 verify that at $\omega \geq {\bar g}$, $A(\omega)$ and $B(\omega)$ can be well approximated by
 $A(\omega) \simeq Q_{\gamma,0} (  \bar{g} / {\rvert\omega\rvert} )^{\gamma}\cos\frac{\pi\gamma}{2}$ and $B(\omega) \simeq 1$. The function $C(\omega)$ on the other hand is not expressed in terms of $\Delta (\omega_m)$.
 The function $C(\omega)$ depends on the gap function
\beq
C(\omega)  =i\bar{g}^{\gamma}\sin(\frac{\pi\gamma}{2})\int_{0^+}^{\rvert\omega\rvert}\frac{d\Omega}{\Omega^{\gamma}}
\frac{D(\rvert\omega\rvert-\Omega)-D(\rvert\omega\rvert)}{\sqrt{1-D^{2}(\rvert\omega\rvert-\Omega)}}.
\label{real_a_3_1}
 \eeq
The $0^+$ in the lower limit of the integral in  (\ref{real_a_3_1}) implies that a special care is needed to properly treat the limit $\Omega \to 0$, as   the integrand in (\ref{real_a_3_1}) is of order $1/\Omega^{\gamma-1}$ at small $\Omega$, and
$\int d\Omega/\Omega^{\gamma-1}$ is infra-red divergent.    The divergence is eliminated  by slightly shifting the integration contour into the
   upper half-plane of frequency, as shown in Fig.~\ref{fig:contour}.
    Such shift is necessary to satisfy the Kramers-Kronig (KK) relation for the interaction on the real axis: $V (\omega) = (1/\pi) \int d x V^{''} (x)/(x- \omega - i\delta)$ (see   Appendix~\ref{sec:KK}).
    For practical purposes, the correct result for $C(\omega)$ is obtained by integrating in (\ref{real_a_3_1}) along the real axis down to an infinitesimally small but finite $\epsilon$ and subtracting from the integral
    \beq
    \frac{-1}{\gamma-2} \frac{1}{\epsilon^{\gamma-2}} \frac{\frac{dD(\omega)}{d\omega}}{\sqrt{1-D^2 (\omega)}}
\eeq

The gap equation with $C(\omega)$ given by (\ref{real_a_3_1}) is an integral equation, even with the analytic expressions for $A(\omega)$ and $B(\omega)$.   In Papers IV and V we converted this equation into an approximate differential equation by Taylor expanding the integrand in (\ref{real_a_3_1}).  We use the same approach here.
 We present the results in two steps. First, we restrict with the lowest-order derivatives, like we did in Paper IV for $\gamma <2$ and  show that a qualitatively new behavior emerges for $\gamma >2$.  Then we present the results
  for the gap function at $\gamma >2$, obtained by expanding to an infinite order in derivatives. We show that the new feature, detected in the first procedure,  remains.

  Like in Papers IV and V, we follow Refs.~\cite{Karakozov_91,combescot} and express the gap function as $\Delta_0(\omega) = \omega/\sin [\phi_0 (\omega)]$,
   where
   $\phi (\omega)$ is in general a complex function of frequency.
    At $\gamma =2$, $C(\omega)$ contains only the  term with ${\dot \phi}_0$. The equation on $\phi_0$  then reduces to
    ${\dot \phi}_0 = (2/\pi {\bar g}^2) (\omega + Q_{2,0} ({\bar g}/\omega)^2 \sin{\phi_0})$, where $Q_{2,0}$ is given by
    Eq. (\ref{eq:Qgamma}).
     The solution of this equation is a monotonically increasing {\it real} function $\phi_0 (\omega)$  (Ref. \cite{combescot})
     This
     leads to rather peculiar behavior of $\Delta_0 (\omega)$ and the DOS consisting of a set of $\delta$-functions at $\omega_p$ where $\phi_0 (\omega_p) = \pi/2 + p \pi$ ($p$ is an integer).  For
     $\gamma \neq 2$,
      there appear an infinite number of other terms with the derivatives of ${\phi_0}$, all with coefficients $O(\gamma -2)$.
       To get some physics insight, below we first consider the toy model, in which
    keep only one of the leading new term - the one with ${\dot \phi}^2_0$. This toy model already shows that the system behavior at $\gamma >2$ is qualitatively different from that at $\gamma <2$.
    Then we consider the actual gap equation and
    sum up series of terms with higher-order derivatives and higher powers of ${\dot \phi}_0$.  We show that
     the structure of the gap function changes a bit, compared with the toy model, but qualitative difference between $\gamma <2$ and $\gamma >2$ holds.

\subsection{Expansion to order ${\dot \phi}^2_0 (\omega)$}
\label{sec:real_2}

For convenience, we keep $\gamma$ close to 2 and expand to first order in $\gamma -2$.
Expanding in the integrand for  $C(\omega)$  to order ${\dot \phi}^2_0$,
we express the gap equation as
\begin{align}
\dot{\phi}_0+
 \omega \delta \dot{\phi}_0^{2}\tan\phi_0 & =\frac{2}{\pi\bar{g}^{\gamma}}\left(\omega^{\gamma-1}-Q_{\gamma,0}\frac{\bar{g}^{\gamma}}{\omega^{2}}e^{i\pi\gamma/2}\sin\phi_0 \right),\label{eq:phi}
\end{align}
where
$\delta=(\gamma-2)/2$.

\begin{figure}
\centering
\includegraphics[scale=0.6]{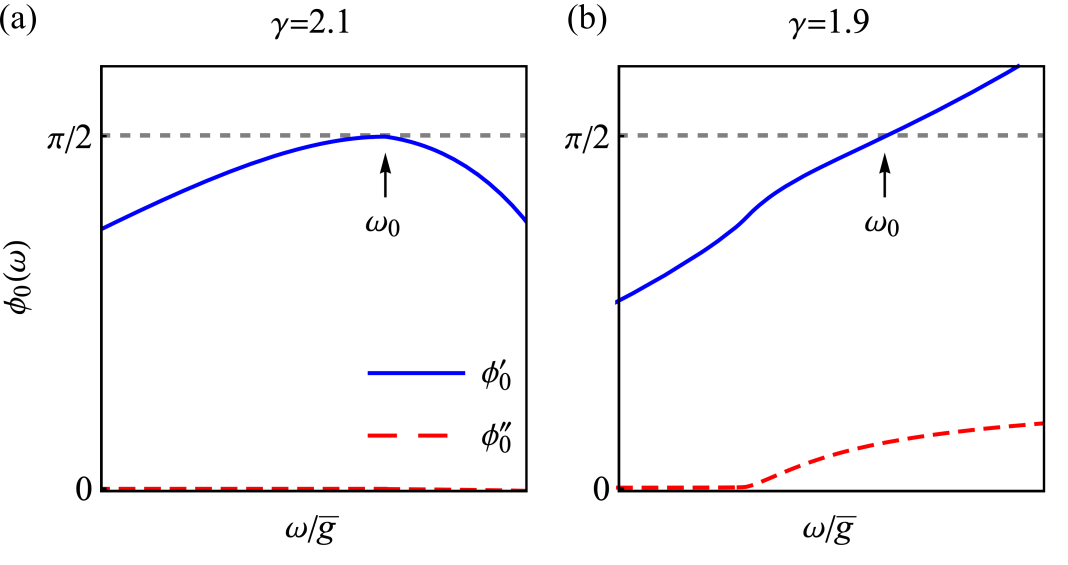}\caption{Distinct behaviors of the solution $\phi_0(\omega)$ around $\omega=\omega_0$ (vertical blue line) for (a) $\gamma=2.1$ and (b) $\gamma=1.9$, which are obtained by solving Eq.~(\ref{eq:phi}) numerically.}
\label{fig:real_compare}
\end{figure}

We are interested in  the behavior of $\phi_0 (\omega)$ 
at $\omega \geq {\bar g}$, where our approximations for $A(\omega)$ and $B(\omega)$ are valid.  The boundary condition for (\ref{eq:phi}) can be set at some initial $\phi_0 <\pi/2$ at, e.g., $\omega= {\bar g}$.   Extending from this to larger $\omega$, we see  that $\phi_0 (\omega)$ increases with $\omega$ and remains real as long as $\phi_0$ remains smaller than $\pi/2$.
At these frequencies, the
$Q_{\gamma,0}$ term is smaller than  $\omega$
 and can be safely neglected.
 Eq. (\ref{eq:phi}) then becomes the quadratic equation on ${\dot \phi}_0$. Solving it and choosing the solution that matches the boundary condition, we obtain
\begin{equation}
\dot{\phi}_0=-\frac{1}{2 \delta \omega\tan\phi_0}\left[1 -\sqrt{1+\frac{8 \delta \tan\phi_0}{\pi}\left(\frac{\omega}{\bar{g}}\right)^{\gamma}}\right].\label{eq:sol_quadratic}
\end{equation}
A simple analysis of this equation shows that, as we anticipated, the behavior of $\phi_0 (\omega)$ at $\gamma <2$ and at $\gamma >2$ is qualitatively different. Indeed, at $\gamma <2$, when $\delta <0$,  $\phi_0 (\omega)$ becomes complex at
 $(8/\pi) |\delta| \tan{\phi_0} (\omega/{\bar g})^\gamma =1$, before $\phi_0$ reaches $\pi/2$. For a complex $\phi_0$, $\tan{\phi_0}$ is non-singular, and $\phi_0 (\omega)$ evolves smoothly with $\omega$ -- its real part increases up to some value and then saturates when $Q_{\gamma,0}$ term becomes relevant, while Im $\phi_0 (\omega)$ increases logarithmically at large $\omega$, such that $Q_{\gamma,0} ({\bar g}^\gamma/\omega^2) \sin{\phi_0} \approx \omega^{\gamma-1}$. The high-frequency behavior yields $\Delta_0 (\omega) \propto 1/\omega^\gamma$.

  For $\gamma >2$, $\delta >0$.  Now $\phi_0$ remains real all the way up to a frequency, $\omega_0$, where $\phi_0 = \pi/2$ and $\tan{\phi_0}$ diverges. An elementary analysis of (\ref{eq:sol_quadratic}) shows that
  ${\dot \phi}_0$ then vanishes upon approaching this point.  Expanding  at $\omega \leq \omega_0$, we obtain from (\ref{eq:sol_quadratic})
\begin{equation}
\phi_0=\frac{\pi}{2}-\frac{1}{2\pi \delta}\frac{\omega_{0}^{\gamma-2}}{\bar{g}^{\gamma}}(\omega-\omega_{0})^{2}-
 \frac{1}{8\pi \delta^2} \frac{\omega_{0}^{\gamma-3}}{\bar{g}^{\gamma}}(\omega-\omega_{0})^{3}+....\label{eq:w0}
\end{equation}
We see that $\phi_0$ now approaches $\pi/2$ horizontally.  One can verify that Eq. (\ref{eq:w0}) also holds for $\omega \geq \omega_0$, this requires one to  choose another branch of the solution of the quadratic equation on ${\dot \phi}_0$.

We see that $\phi_0$ initially increases with $\omega$ and  approaches $\pi/2$ quadratically, and then bends back to smaller values. We verified this result by solving the full Eq.~(\ref{eq:phi}) numerically. In Fig.~\ref{fig:real_compare}, we show numerical results for $\gamma =1.9$ and $\gamma =2.1$. We see that, in the first case, Re $\phi_0 (\omega)$ increases monotonically and Im $\phi_0 (\omega)$ emerges before Re $\phi_0 (\omega)$ reaches $\pi/2$. In the second case, $\phi_0 (\omega)$ remains real and varies quadratically near $\omega_0$, where $\phi_0 (\omega_0) = \pi/2$.

\begin{figure}
\centering
\includegraphics[scale=0.5]{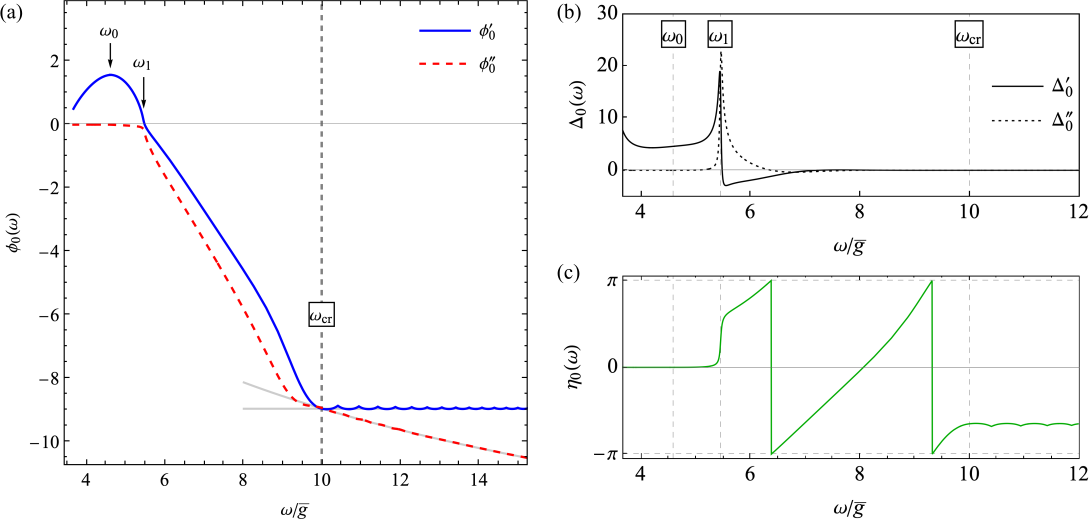}\caption{(a) Solution of the differential equation Eq.~(\ref{eq:phi}) at $\gamma=2.7$, where the $Q_{0,\gamma}$ term has been considered.
Gray lines are asymptotic behaviors at large frequency given in Eq.~(\ref{eq:largeW}).
(b) shows the gap function $\Delta_0(\omega)=\omega/\sin\phi_0(\omega)$ and (c) the phase $\eta_0(\omega)=\text{Arg}[\Delta_0(\omega)]$.}
\label{fig:real_axis_1}
\end{figure}

At $\omega>\omega_{0}$,  $\phi_0 (\omega)$ decreases and remains real, until it reaches $\phi_0 =0$ at some $\omega_1 > \omega_0$.  At around this frequency,
\begin{equation}
\phi_0 \simeq \sqrt{\frac{2}{\delta \omega_{1}}(\omega_{1}-\omega-i0^{+})}+\frac{4}{3\pi}\frac{\omega_{1}^{\gamma-1}}{\bar{g}^{\gamma}}
(\omega_{1}-\omega)+....\label{eq:w1}
\end{equation}
We again verified this behavior by solving numerically the full differential equation ~(\ref{eq:phi}). We show the result in Fig.~\ref{fig:real_axis_1} (a).
We clearly see that Im $\phi_0 (\omega)$ emerges at a frequency $\omega_1 > \omega_0$.
As $\omega$ increases above $\omega_1$, Im $\phi_0$ increases in amplitude. When it becomes large enough, $\tan \phi_0$ approaches $-i$, and the solution of Eq.(\ref{eq:phi}) without the $Q_{\gamma,0}$  term becomes Re $\phi_0 \sim$ Im $\phi_0 \sim -\sqrt{1/(\pi \delta\bar{g}^{\gamma})} \omega^{\gamma/2}$. This behavior is clearly reproduced in  Fig.~\ref{fig:real_axis_1} (a).
As  $\omega$ increases further,
$|$Im $\phi_0|$ continue 
increasing, hence $\sin{\phi_0}$ increases and above a certain frequency, the
 $Q_{\gamma,0}$ term becomes comparable to the $\omega^{\gamma-1}$ term.
  At even larger frequencies, the balance between these two terms holds, and we obtain
\bea \label{eq:largeW}
\text{ Re} \phi_0  &=& -2m\pi-
(\gamma-1) {\pi \over 2}, \nonumber \\
\text{ Im} \phi_0   &= & -\log {2{\bar g}\over Q_{0,\gamma}} - (1+\gamma) \log {\omega \over {\bar g}}.
\eea
where $m$ is the number of the additional vortices in the first quadrant.
 This yields  $\Delta_0 (\omega) \propto 1/\omega^{\gamma}$, as it should be.

We plot Re$\Delta_0 (\omega)$ and Im$\Delta_0 (\omega)$ in Fig. \ref{fig:real_axis_1} (b) and the phase $\eta_0 (\omega)$ of $\Delta_0 (\omega) = |\Delta_0 (\omega)| e^{i\eta_0 (\omega)}$ in Fig. ~\ref{fig:real_axis_1} (c).
  The phase undergoes two slips by $2\pi$ at positive $\omega$, consistent with the presence of two vortices in the first quadrant of the complex plane of frequency (see Fig.~\ref{fig:vortex_1} (a)).

\subsubsection{Density of states}
\label{sec:dos_2}

The density of single-electron states is  defined as $N(\omega)=
(-N_0/\pi) {\text{Im}}G_{l}(\omega)$,
where $N_{0}$ is the DOS in the normal state and $G_{l}(\omega)$
is the (retarded) single-electron Green's function, integrated
over the dispersion:
\begin{equation}
G_{l}(\omega)=-i\pi\sqrt{\frac{\omega^{2}}{\omega^{2}-\Delta_{0}^{2}(\omega)}},
\end{equation}
In terms of $\phi_0 (\omega)$, $N(\omega) = N_0 \text {Re} \sqrt{- \tan^2{\phi_0} }$.

\begin{figure}
\centering
\includegraphics[scale=0.6]{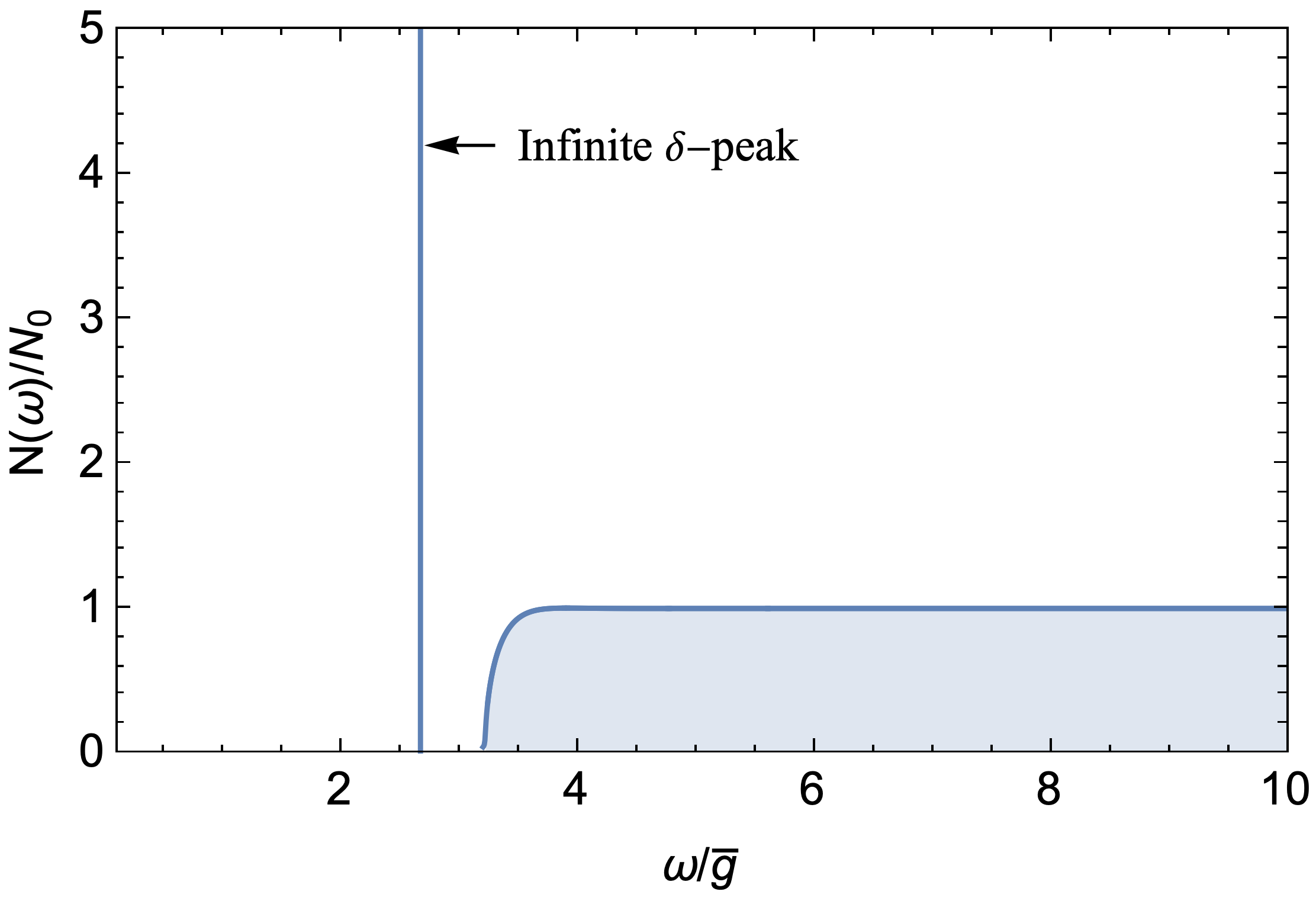}\caption{The density of states $N(\omega)$ in the approximate
approach, described by Eq.~(\ref{eq:phi}). We set $\gamma=2.1$.}
\label{fig:dos_1}
\end{figure}

 One can easily verify that the DOS vanishes at small frequencies, as  expected for a superconductor with a finite gap, and is non-zero at  frequencies $\omega > \omega_1$, where Im $\phi_0 (\omega)$ is finite.
  It is tempting to call $\omega_1$ a spectral gap, by analogy with a BCS/Eliashberg superconductor.
 For $\gamma <2$, there are no other features in the DOS, although there is a structure
 inside
 the continuum.
   For $\gamma >2$, there is also a continuum above $\omega_1$, but in addition, there is a level inside the continuum, at $\omega = \omega_0$, where $\tan{\phi_0}$ diverges and an imaginary part appears once we shift $\omega$  in the upper frequency half-plane by an infinitesimally small amount. Moreover, because
   $ N(\omega) \sim 1/(\omega - \omega_0 + i0^+)^2$, 
   the integral of the DOS over a narrow range around $\omega_0$ diverges.  The prefactor for the divergent term scales as
    $\delta$, the capacity of the level is proportional to $\gamma -2$.
     We show the result of numerical evaluation of the DOS for representative $\gamma =2.1$  in Fig.~\ref{fig:dos_1}.
     We clearly see that the DOS has a continuum, which starts at $\omega_1$, and an in-gap state at $\omega_0 < \omega_1$ with the
     ``infinite'' weight, comparable to the total weight of the continuum.
      We will see below that the DOS for the actual model also contains an ``infinite'' peak, but it is located at the
       lower end of the continuum and shows up as a non-integrable singularity.

\subsection{Equation  for $\phi_0 (\omega)$ with derivatives to all orders}
\label{sec:real_sub1}

We now analyze whether the results from the previous section survive if we add higher-order derivative.  These higher-order derivatives appear in combination with higher powers of $\tan{\phi_0}$, which diverges at $\omega = \omega_0$.  It is then a'priori unclear whether the macroscopically degenerate level at $\omega_0$ survives once we
 include higher-order terms.  We show below that it does survive.

 \begin{figure}
\centering
\includegraphics[scale=0.5]{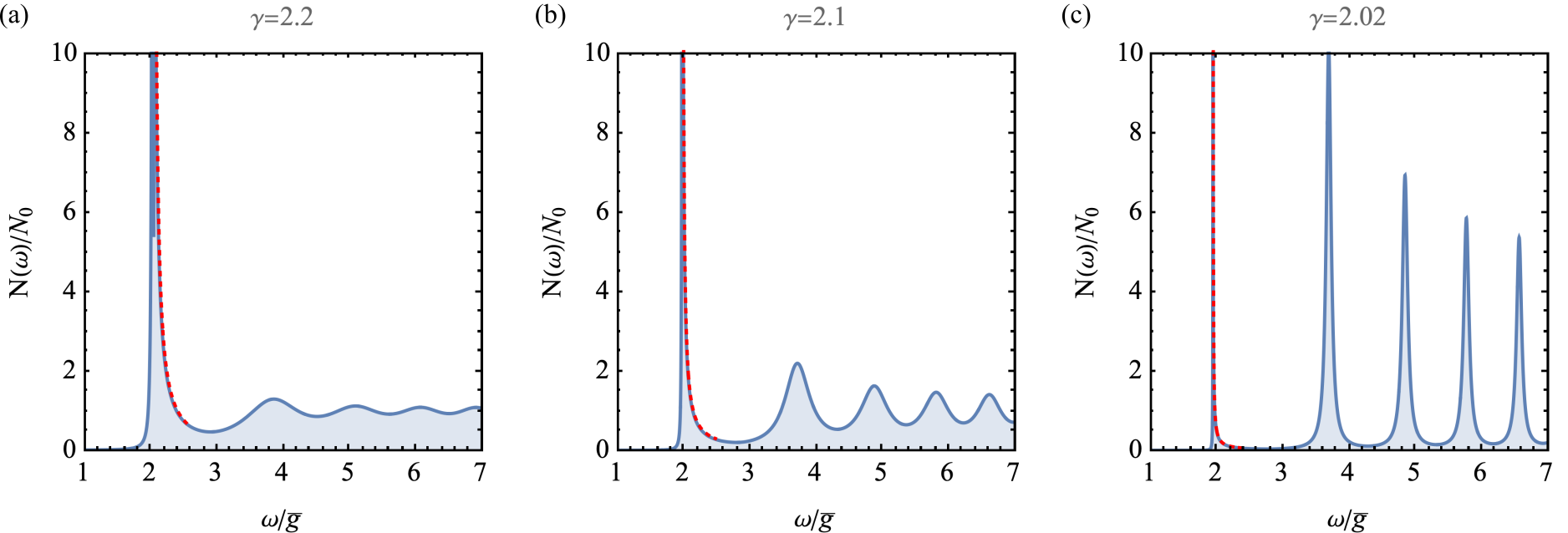}\caption{Evolution of DOS obtained by solving Eq.~(\ref{eq:phi21}) for different $\gamma$
approaching $2$ from above. The red dotted line is the analytic result, Eq.~(\ref{eq:dos}).}
\label{fig:dos}
\end{figure}

The analysis is rather involved and we present the details in Appendix \ref{sec:app_expand_C}.
 There are two types of terms in the expansion of $C(\omega)$ in the derivatives of $\phi_0 (\omega)$:
  terms with higher powers of ${\dot \phi_0}$, combined with higher powers of $\tan{\phi_0}$, and terms with
    higher derivatives of $\phi_0$, see Eq.
  (\ref{eq:C2}). We argue in Appendix \ref{sec:app_expand_C}
  that the terms with higher derivatives are irrelevant, but the terms with higher powers of ${\dot \phi_0}$ must be kept.  These last terms
   form series in $X=\omega \dot{\phi}_0 \tan \phi_0$ in the form
  \bea
  C(\omega) &=& \frac{{\bar g}^\gamma}{\omega^{\gamma-2}} \frac{\sin{\frac{\pi \gamma}{2}}}{2-\gamma} D(\omega) {\dot \phi}_0 \left[1 +  \frac{\gamma-2}{2(3-\gamma)} X -\frac{\gamma-2}{2(4-\gamma)} X^2 + \frac{\gamma-2}{2(5-\gamma)} X^3 +...\right].
  \label{eq:nn_11}
\eea
The series in Eq.~(\ref{eq:nn_11})
sum up into Hypergeometric function $_2 F_1 (1,2-\gamma, 3-\gamma,-X)$. Substituting into the gap equation and again neglecting the $Q_{\gamma,0}$ term, we obtain the differential equation on $\phi (\omega)$ in the form
\bea\label{eq:phi21}
{1\over 2}\dot{\phi}_0 \bigg[ 1 +  _2 F_1 (1,2-\gamma, 3-\gamma,-X)  \bigg] = \frac{2}{\pi\bar{g}^{\gamma}} \omega^{\gamma-1}. 
\eea
At large $X$, the asymptotic expansion of a Hypergeometric function  yields $_2 F_1 (1,2-\gamma, 3-\gamma,-X) \approx X^{\gamma-2} \Gamma(3-\gamma) \Gamma (\gamma-1)$. Substituting into (\ref{eq:phi21}) and solving for
$\phi_0 (\omega)$ near $\omega_0$, we obtain
\begin{align}
\phi_0(\omega) &\simeq {\pi\over 2} - \frac{4}{\pi} \left(\frac{\omega_0}{\bar g}\right)^\gamma
{B_{\gamma}\left(1- \omega/\omega_0 - i 0^+\right)^{\gamma -1} \over 1+ B_{\gamma} \left(1- \omega/\omega_0 - i 0^+\right)^{\gamma -2}},\label{eq:solution_phi0} \\
B_{\gamma}&=  \frac{1}{(\gamma-1)^{\gamma-1} \Gamma(3-\gamma) \Gamma (\gamma-1)}.
\end{align}
We see that $\phi_0 (\omega)$  approaches $\pi/2$ with zero derivative, albeit the exponent is smaller than $2$ and reaches this value only at $\gamma \to 3$.
 For $D_0(\omega) = \Delta_0 (\omega)/\omega$,  this yields
 \beq
 D_0(\omega) =\frac{1}{\sin{\phi_0 (\omega)}} =  1 + A
 \left(\frac{\omega_0-\omega-i0^+}{\omega_0}\right)^{2(\gamma-1)}
  \label{ar_1}
    \eeq
 where $A = O(1)$.
  The new element, compared to our approximate analysis in the previous section, is that now
Im $\phi_0 (\omega)$ develops immediately above $\omega_0$.

 \subsubsection{Density of states}

  We now show that the vanishing of ${\dot \phi} (\omega)$ at $\omega_0$  gives rise to a non-integrable singularity in the DOS at $\omega = \omega_0$.
 In explicit form, the DOS for $\phi_0 (\omega)$ from Eq.~(\ref{eq:solution_phi0}) is
\beq \label{eq:dos}
{N(\omega) \over N_0}
={\bar{g}^{\gamma}\over \omega_0} { \pi \sin [\pi (1-\gamma)]  \over 4 B_{\gamma}} { \Theta(\omega- \omega_0) \over  \left({\omega - \omega_0} \right)^{\gamma-1}},
\eeq
where $\Theta(x)$ is the unit step function.
In distinction from the toy model, there  is no gap between the macroscopically degenerate level and the continuum, but still,
$\int^{\infty}_{\omega_0} d \omega N(\omega)$ diverges at the lower limit, i.e., the DOS contains a non-integrable singularity at $\omega = \omega_0 +0$.  In practice, this implies that
the number of states within a tiny interval above $\omega_0$
 is a finite fraction of the total number of electrons $N_{tot}$ in the system.
Because $N(\omega) \propto (\gamma-2)$,  the fraction initially increases linearly with $\gamma -2$.

We show the DOS for several $\gamma$ in Fig. \ref{fig:dos}.
We see that the DOS vanishes below $\omega_0$, forms a continuum above this frequency, and displays an ``infinite'' peak at the boundary
(a non-integrable singularity).  At the highest frequency, $N(\omega)$ approaches the DOS of the normal state, $N_0$.

The non-integrability of the singularity at $\omega = \omega_0 + 0$ means that the total number of states in an arbitrary small range above $\omega_{0}$ diverges. By physical reasons, the divergence must be regularized by extending the model in a proper direction. We show below that a finite $\omega_{D}$ does not provide the regularization --- the singularity remains a non-integrable one up to some finite value of $\omega_{D}$.  One way to regularize the divergence is to keep the total number of states large but finite, by imposing the limits of integration over fermionic dispersion, and require that the total number of states in the ``infinite'' peak is a fraction, proportional to $\gamma-2$,  of the total number of states in the band.  From this perspective, the ``infinite'' peak should be viewed as a level with a macroscopic degeneracy.  A further investigation of the ``infinite'' peak is clearly called for.

We note in passing that the analysis, presented here, can be extended to smaller $\gamma$ between $1$ and $2$.  For these $\gamma$, the DOS
 still diverges at $\omega = \omega_0 +0$ with the fractional exponent $\gamma-1$, but the singularity is now integrable.

 \subsubsection{Continuity at $\gamma = 2+0$}

 \begin{figure}
\centering
\includegraphics[scale=0.6]{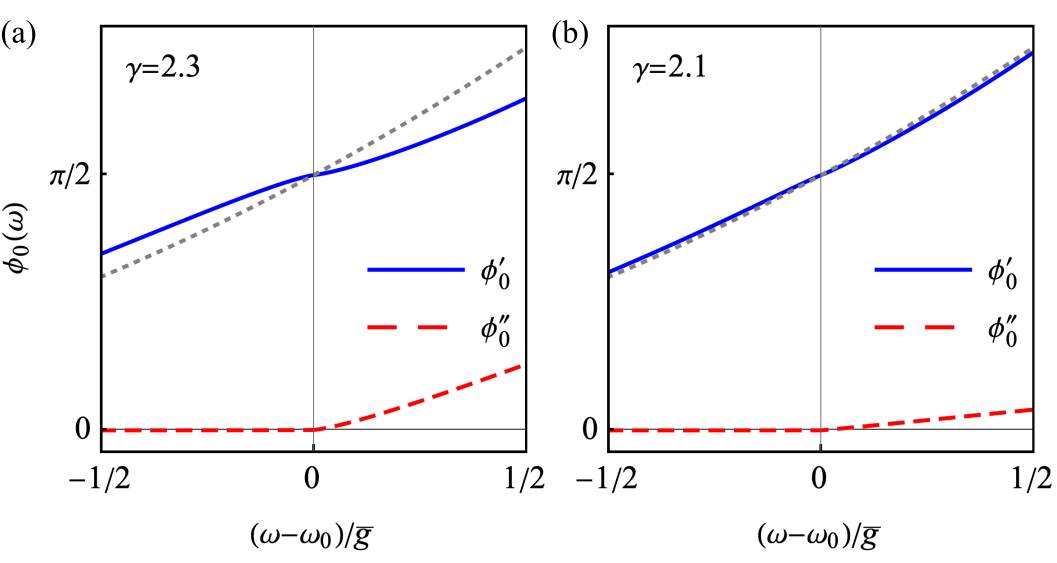}\caption{Solution of the modified gap equation Eq.~(\ref{eq:phi21}) around $\omega_0$, where (a) $\gamma=2.3$ and (b) $\gamma=2.1$.
Gray dotted line shows the solution at $\gamma=2$, i.e., $\phi_0(\omega)=\omega^2/\pi$ without imaginary part.}
\label{fig:real_axis_2}
\end{figure}

 We see from Eq. (\ref{eq:solution_phi0}) that at $\gamma = 2+0$, the frequency dependence of $\phi_0 (\omega)$ becomes $\omega^2/\pi$
 (which  corresponds to taking $\gamma \to 2$ keeping $\omega$ away from $\omega_0$), like at $\gamma =2$ (see Refs.~\cite{Karakozov_91,combescot,paper_5}).
 We show the numerical solution of Eq. (\ref{eq:phi21}) in Fig. ~\ref{fig:real_axis_2}. We see that  the behavior of $\phi_0 (\omega)$  at $\gamma >2$ continuously approaches that at $\gamma =2$:
    Im $\phi_0(\omega)$  gradually  gets smaller and   Re $\phi_0(\omega)$  approaches
$\omega^2/\pi$. 
 Simultaneously, the maxima in the continuum in the DOS get sharper and at $\gamma = 2+0$ evolve into a discrete set of $\delta$-functional peaks,
 see Fig.~\ref{fig:dos}. 
     We emphasize that the 
     continuity at $\gamma \to 2$ does not hold in our approximate treatment in the previous subsection
      and emerges only after we sum up infinite series in ${\dot \phi_0} \tan{\phi_0}$.

 On a more closer look, we find that the analysis at $\gamma \to 2+0$  needs extra care. In this limit, the series in $X$ yields
  \bea
  C(\omega) &=& \frac{\pi {\bar g}^\gamma}{2 \omega^{\gamma-2}} D(\omega) {\dot \phi}_0 \left[1 + \delta \left(X -\frac{X^2}{2} + \frac{X^3}{3} + ...\right)\right] \nonumber \\
  &=& \frac{\pi {\bar g}^\gamma}{2 \omega^{\gamma-2}} D(\omega) {\dot \phi}_0 \left[1 +\delta \log{(1+X})\right]
\label{eq:nn_1}
\eea
where, we remind, $\delta = (\gamma-2)/2$.
 Substituting into the gap equation and restricting to $ \omega \sim \omega_0$, we obtain
 \begin{align}
\dot{\phi}_0 \left[1 + \delta \log{\left(1 + \omega {\dot \phi}_0 \tan{\phi_0}\right)}\right]
=\frac{2}{\pi\bar{g}^{\gamma}} \omega^{\gamma-1}.
\label{eq:phi_2}
\end{align}
 Solving this equation, we find that $\phi_0 (\omega) \approx (2/\pi \gamma) (\omega/{\bar g})^\gamma$ up to an exponentially short distance to $\omega_0$, and within this distance
 \begin{align}
\phi_0(\omega) \simeq {\pi\over 2} + {2 \over \pi \delta}  \left( {\omega_0 \over {\bar g}} \right)^{2}
 \frac{1 -\omega/\omega_0}{\log(1 -\omega/\omega_0 - i0^+)}. \label{eq:solution_phi0b}
\end{align}
We see that $\phi_0 (\omega)$ still approaches $\pi/2$ with zero derivative, but vanishes only logarithmically.  The Im $\phi_0 (\omega)$ does develop immediately above $\omega_0$ like at larger $\gamma$, but in the immediate vicinity of $\omega_0$,
Im $\phi_0 (\omega)$ is parametrically small compared to Re $\phi_0 (\omega)$ by $1/|\log{\omega - \omega_0}|$.  If we were to neglect Im $\phi_0 (\omega)$, we would find that Re $\phi_0 (\omega)$ monotonically increases with $\omega$, as $(\omega/{\bar g})^2/\pi$,
 and just flattens in exponentially small regions near the frequencies where $\phi_0 (\omega) = \pi/2 + p \pi$.

Finally, we consider the terms with higher derivatives, like ${\ddot \phi}_0$, ${\dddot \phi}_0$, etc.
 For definiteness, let's restrict to $\gamma \geq 2$ and compare these terms with $(\gamma -2) {\dot \phi} \log X$.
Each term with a higher derivative gets renormalized by series in $X$. We evaluate the series in Appendix \ref{sec:app_expand_C}.  At large $X$, which we are interested in, the series for each term have the same asymptotic form and reduce each prefactor by $1/2$. Including these terms with rescaled prefactors, we find that the last term in
 Eq.~(\ref{eq:nn_1}) changes to
 \beq
 1 + \delta \left[ \log(1+X) + K \right]
 \label{nn_5}
 \eeq
 where
 \beq
 K = {\omega \over  \phi^{(1)}_0} \left({1\over 2!}\phi^{(2)}_0 - \frac{\omega}{3! 2} \phi^{(3)}_0 + \frac{\omega^2}{4! 2}
  \phi^{(4)}_0 + ...\right),
  \eeq
 where $\phi^{(m)}_0$ is the $m$-th derivative of $\phi_0 (\omega)$.
 We assume and then verify that at large $X$, i.e., at $\omega \approx \omega_0$,  the inclusion of the $K$ term  only changes the prefactor for  the second term in Eq.~(\ref{eq:solution_phi0b}).  To see this, we assume that at $\omega$ slightly below $\omega_0$,
 $\phi_0 (\omega) =  \pi/2 + Q (\omega_0 -\omega)/\log{(1- \omega/\omega_0)}$ with $Q=4\omega_0^{\gamma-1}/[\pi(\gamma-2) {\bar g}^{\gamma}]$,
 and compute the series for $K$ using this form of $\phi_0$.  A straightforward analysis then yields
 \beq
 K = - \frac{1}{2 \log{X}} F(X),
 \label{nn_4}
 \eeq
  where
  \bea \label{func:fx}
  F(X) &=&  2X \sum_{m=0}^\infty \frac{(-1)^m X^m}{(n+1)^2(n+2)}  \nonumber \\
  &&= -2 \left(\text{Li}_2 (-X) +  \log(1+X) (1 + 1/X) -1\right)
  \eea
  and Li$_2 (-X)$ is a polylogarithm. At large $X$, $\text{Li}_2 (-X)  \approx (-1/2) \log^2{X}$. Substituting into
  Eq.~(\ref{nn_4}),
  we obtain  $K \approx - (1/2) \log {X}$. Substituting into
  Eq.~(\ref{nn_5}),
  we see that the $\log{X}$ dependence survives, only the prefactor drops by a factor of $2$.  Then Eq.~(\ref{eq:solution_phi0b}) remains valid, with extra $2$ in the prefactor for the second term.

\subsection{
Extraction of the non-integrable singularity in the DOS directly from the integral gap equation.}
\label{sec:integral}

 We next show that the non-integrable singularity in the DOS can be obtained directly from the integral equation \eqref{el8}. For this, we first rewrite this equation in the form, which takes care of the regularization of the formal divergence of the integral for $C(\omega)$ in
 Eq.~(\ref{real_a_3_1}):
\begin{eqnarray}
&& \Delta (\omega )=
\frac{1}{2}
\int \frac{d\omega_{m}}{(-(i\omega_{m}-\omega )^{2})^{\gamma /2}}\frac{\Delta (\omega_{m})-\Delta (\omega )\frac{i\omega_{m}}{\omega }}{\sqrt{{\omega_{m}}^{2}+\Delta^{2}(\omega_{m})}}  +
\label{eq:DeltaAnalysis}\\
&&
\frac{\sin (\pi \gamma/2 )}{\gamma-2}
\left(
{\dot D} (\omega )
\int_{0}^{\omega } \frac{{\dot g}(\Omega ) d\Omega }{(\omega -\Omega)^{\gamma -2}}
-
(2-\gamma )
\int_{0}^{\omega }d\Omega \frac{D(\Omega )-D(\omega )+(\omega -\Omega ){\dot D} (\omega )}{(\omega -\Omega )^{\gamma }}g(\Omega )
\right),
\nonumber
\end{eqnarray}
where
\begin{equation}
 g(\omega )=\frac{1}{\sqrt{D^{2}(\omega )-1}}.
\end{equation}
 We take as an input the evidence from  the numerical analysis that at small $\omega$, $\Delta (\omega)$ is real, and that there exists $\omega_0$, at which
  $\Delta (\omega_{0})=\omega_{0}$. At this point,  we have
$$
\Delta (\omega_{0})=\omega_{0},\qquad D(\omega_{0})=1.
$$
Let's assume that for $\omega $ just below $\omega_{0}$ we have
$$
D(\omega )=1+A(\omega_{0}-\omega )^{\alpha }  ,\qquad \alpha >0,
$$
where $A$ is some real positive constant.

The integral over $\omega_m$ in
 Eq.~\eqref{eq:DeltaAnalysis}
 is completely regular at $\omega \rightarrow \omega_{0}$,
   the dangerous terms are the ones coming from the upper limit of integration
   over $\omega$ in the last term.
     We then
     write $\omega =\omega_{0}-\epsilon_{\omega }$, and $\Omega =\omega_{0}-\epsilon_{\omega }-\epsilon_{\Omega }$,
     assume that both $\epsilon_{\omega }$ and $\epsilon_{\Omega }$ are small, and consider
      the contribution from
      the upper limit. Then
\bea
&&g(\Omega )\approx \frac{1}{\sqrt{2A}}\frac{1}{(\epsilon_{\omega }+\epsilon_{\Omega })^{\alpha /2}},\nonumber \\
&&g'(\Omega )\approx \frac{\alpha}{2}\frac{1}{\sqrt{2A}}\frac{1}{(\epsilon_{\omega }+\epsilon_{\Omega })^{\alpha /2+1}}  .
\label{ar_2}
\eea
 Expressing
  $\epsilon_{\Omega }=x\epsilon_{\omega }$, we then obtain
the dangerous contribution to the gap equation in the form
$$
\epsilon_{\omega }^{1+\alpha/2 -\gamma } \frac{\sqrt{A}}{\sqrt{2}}
\left(
\frac{\alpha^{2}}{2}
\int_{0} \frac{ dx }{x^{\gamma -2}}\frac{1}{(1+x)^{\alpha /2+1}}
-
(2-\gamma )
\int_{0}dx
\frac{(1+x)^{\alpha }-1+ x\alpha }{x^{\gamma }(1+x)^{\alpha /2}}
\right).
$$
In order for this term to be finite, we must have
$$
\alpha \geq 2(\gamma -1)>2.
$$

 By continuity, we expect
 $\alpha =2$ at $\gamma =2$
(see previous section).
  Invoking this argument, we find
   $\alpha =2(\gamma -1)$.
    This  is exactly the same form as we obtained by summing up Taylor series, Eq.~(\ref{ar_1}).

The function $g(\Omega )$ must be analytic in the upper half plane.
 Adding $+i0$ to $\omega$ and
  substituting $\epsilon_{\omega }+\epsilon_{\Omega }=\omega_{0}-\Omega -i0$  into Eq.~(\ref{ar_2}),
   we obtain
$$
g(\Omega )\approx \frac{1}{\sqrt{2A}}\frac{1}{(\omega_{0}-\Omega -i0)^{
\gamma-1}}  .
$$
The imaginary part of $g$ is proportional to the density of states.
 We have
$$
N (\Omega )\propto  \frac{1}{\sqrt{2A}} \text{Im} \frac{1}{(\omega_{0}-\Omega -i0)^{
\gamma-1}}
  =\left\{
\begin{array}{ll}
0,&\quad \mbox{if $\Omega <\omega_{0}$}\\
\frac{1}{\sqrt{2A}}\frac{\sin (\pi
(\gamma-1))}{(\Omega -\omega_{0})^{
 \gamma-1}},&\quad \mbox{if $\Omega >\omega_{0}$}
\end{array}
\right. .
$$
This is the same expression as Eq. (\ref{eq:dos}).

\section{Finite $\omega_{D}$}
\label{sec:omega_D}

In this section, we examine whether the state with an ``infinite'' peak in the DOS is stable with respect to
perturbation imposed by a  small but finite mass of the pairing boson. On the Matsubara axis, a finite mass of the boson changes the  interaction
  to \begin{equation}
V(\Omega_{m})=\frac{\bar{g}^{\gamma}}{[\Omega_{m}^{2}+\omega_{D}^{2}]^{\gamma/2}}.
\end{equation}
A finite $\omega_D$ eliminates the solutions with large $n$, leaving only a finite number of the gap functions.
 The  number of  remaining solutions decreases  as $\omega_D$ increases, and beyond some threshold only the $n=0$ solution survives. At the same time, the form of $\Delta_0 (\omega_m)$ is only weakly affected by $\omega_D$ both for $\gamma <2$ and $\gamma >2$.

On the real axis, the effect from $\omega_D$ on the $n=0$ solution is far stronger for $\gamma >2$, but
 still it does not affect the physics qualitatively as long as $\omega_D$ remains below a finite threshold. To demonstrate this, we analyze how a finite $\omega_D$ affects the gap, re-expressed  via $\phi_0 (\omega)$ related to the gap function via $\Delta_0 (\omega) = \omega/\sin{\phi_0 (\omega)}$.   One can easily verify that in the gap equation $D(\omega) B(\omega) = A(\omega) + C(\omega)$, the terms $B(\omega)$ and $A(\omega)$ are only weakly affected by $\omega_D$, as long as $\omega_D$ remains much smaller than ${\bar g}$, but $C(\omega)$  changes
   substantially because a finite $\omega_D$ imposes a lower frequency cutoff on Im $V(\Omega) =$ Im $\bar{g}^{\gamma}/(\omega_{D}^{2} - (\Omega + i 0^+)^2)^{\gamma/2}$.

 We  analyze the effect of a finite $\omega_D$ in two steps, like in Sec.~\ref{sec:real}.
Namely, we first restrict with only ${\dot \phi}^2_0 \tan{\phi_0}$ term in the expansion of $C(\omega)$ in  the derivatives of $\phi_0 (\omega)$, and then include the series of higher-order terms.  The  equation on $\phi_0 (\omega)$ to order ${\dot \phi}^2_0 \tan{\phi_0}$ in the presence of $\omega_D$ has been derived in Paper V for $\gamma  \approx 2$.  Combining that expression with Eq.~(\ref{eq:phi}) we obtain
\begin{align}
\dot{\phi}_0  + \dot{\phi}^2_0\tan\phi_0 \times\left(\frac{\gamma-2}{2} \omega-\frac{\gamma}{4}\omega_{D}\right)
=\frac{2}{\pi\bar{g}^{\gamma}}\left(\omega^{\gamma-1}-Q_{\gamma,0}
\frac{\bar{g}^{\gamma}}{\omega^{2}}e^{i\pi\gamma/2}\sin\phi_0\right),\label{eq:eqn_phi_massive}
\end{align}
As before, we will be interested in $\omega \approx \omega_0$, where $\phi_0 (\omega)$ reaches $\pi/2$, and neglect the $Q_{\gamma,0}$ term.

\begin{figure}
    \begin{center}
      \includegraphics[width=16cm]{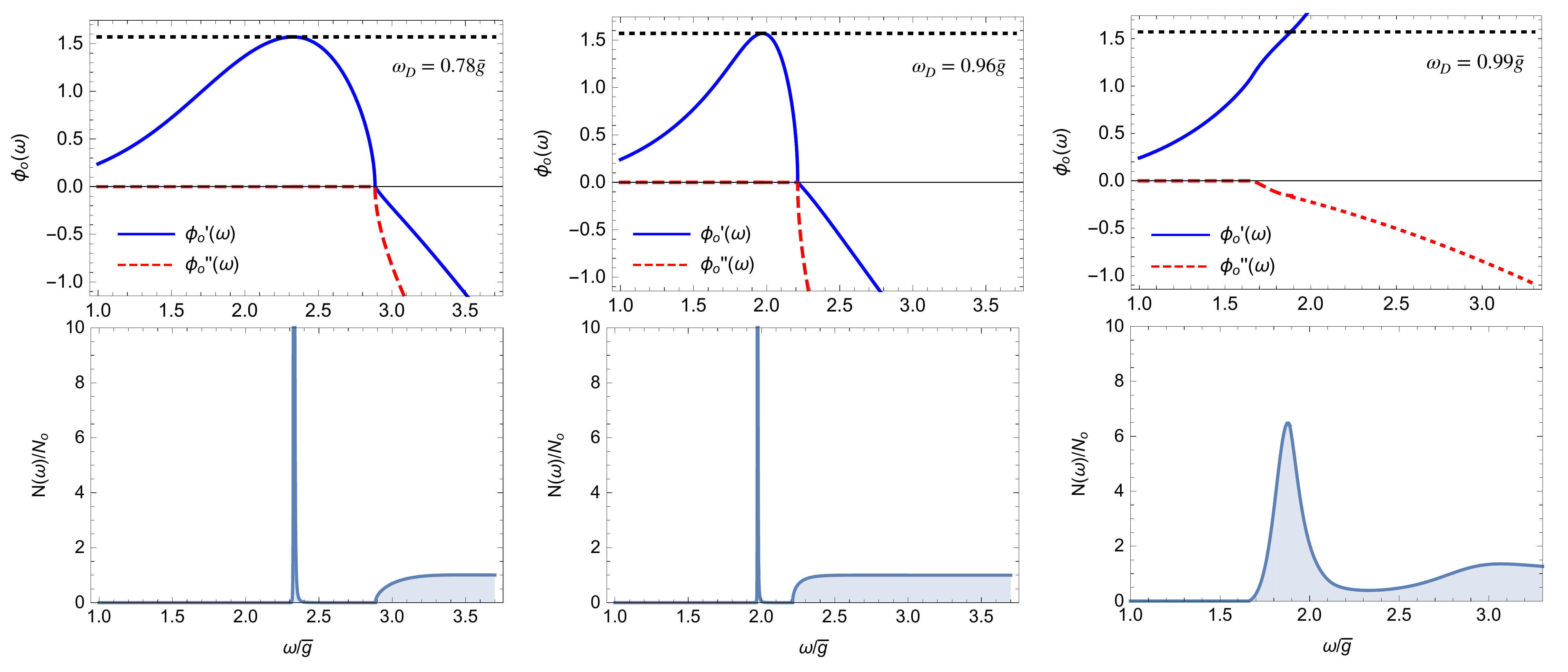}
    \end{center}
    \caption{Numerical results for $\phi_0(\omega)$ and the DOS $N(\omega)$ for a finite $\omega_D$.
      As long as $\omega_D$ is small, the ``infinite'' peak in $N(\omega)$ survives. Once $\omega_D$ exceeds a threshold value, it gets absorbed into the continuum.
      To get the DOS in panel (c), we solved Eq.~(\ref{eq:eqn_phi_massive})  near $\omega_0$, where it is valid, obtained
       Im$\phi_0 (\omega)$, and smoothly extended it to larger $\omega$.  }\label{fig:finite_omegaD}
  \end{figure}

We see from Eq.~(\ref{eq:eqn_phi_massive}) that for $\gamma >2$, the two terms in the prefactor for $\dot{\phi}^2_0\tan\phi_0$ have opposite signs and hence compete. The competition sets a
 characteristic frequency
\begin{equation}
\omega_c=\frac{\gamma}{2(\gamma-2)}\omega_{D}. 
\end{equation}
 Its relevance becomes clear once we solve Eq. (\ref{eq:eqn_phi_massive}) for ${\dot \phi}_0$ in terms of $\tan{\phi_0}$:
 \begin{equation}
\dot{\phi}_0=\frac{-1 + \sqrt{1+\frac{4(\gamma-2)}{\pi}\omega^{\gamma}(1-\frac{\omega_c}{\omega})\tan{\phi_0}}}{(\gamma-2)\omega(1-
\frac{\omega_c}{\omega})\tan{\phi_0}}
\label{eq:phidot}
\end{equation}
 At $\omega_D =0$, $\omega_c$ also vanishes. In this situation, $\phi_0 (\omega)$ remains real up to $\omega_0$, where $\phi_0 (\omega_0) = \pi/2$, and approaches this frequency quadratically, as $\pi/2 - (\omega_0 - \omega)^2 (\omega_0^{\gamma-2}/(\pi (\gamma-2) {\bar g}^\gamma))$.  This gives rise to the appearance of a macroscopically degenerate level in the DOS at $\omega = \omega_0$.  Eq.~(\ref{eq:phidot}) shows that this behavior holds as long as
 $\omega_D$ is smaller than some critical value
 $\omega_c$. At larger $\omega_D$, Im $\phi_0 (\omega)$ emerges before $\phi_0$ reaches $\pi/2$, and an
     the  bound state gets absorbed into the continuum.  We see therefore that a {\it finite}
     $\omega_D > \omega_c$ is required to destroy the ``infinite'' peak. We show this behavior in Fig.\ref{fig:finite_omegaD}.

 The $O(\omega_D)$ term in  $C(\omega)$ also contains the combination $\omega_D {\ddot \phi}_0$. As long as
  $\omega_D$ is below the threshold at
  $\omega_c$ and $\phi_0 (\omega)$ approaches $\pi/2$ quadratically,  this term only shifts $\omega_0$ by a small amount.

 We now include into $C(\omega)$ series of terms with higher powers of ${\dot \phi}_0 \tan{\phi_0}$.   We present computational details in Appendix \ref{sec:app_expand_C}
  and here quote the result.  For simplicity, we again
 focus on $\gamma$ near $2$.
  The equation on $\phi_0 (\omega)$ near $\omega_0$ becomes
 \begin{align}
\dot{\phi}_0 \left(1 - \frac{Y}{2(1+Y)} + \frac{\gamma-2}{2} \log{X}\right)
=\frac{2}{\pi\bar{g}^{\gamma}} \omega^{\gamma-1}\label{eq:eqn_phi_massive_1}
\end{align}
 where $X = \omega_0 {\dot \phi}_0 \tan{\phi_0}$ is the same as before, and $Y = \omega_D {\dot \phi}_0 \tan{\phi_0}$.
  At $\gamma =2$, the imaginary part of $\phi_0 (\omega)$ emerges at $\omega <\omega_0$, for which $Y = O(1)$.
   At $\omega_D =Y = 0$, $\phi_0 (\omega)$ remains real up to $\omega = \omega_0$ and approaches $\omega_0$ from below
   with vanishing derivative,
   which gives rise to an ``infinite'' peak in the DOS.
   When both $\omega_D$ and $\gamma-2$ are finite,  the analysis of Eq.~(\ref{eq:eqn_phi_massive_1}) shows that
     the ``infinite'' peak in the DOS survives as long as $\omega_D < \omega^*_c$, where
     \begin{equation}
     \omega^*_c  = \omega_0 e^{-\frac{2}{\gamma-2}}
     \label{nn_2}
     \end{equation}
     We see that the critical value of $\omega_D$ is still finite, although exponentially small for $\gamma$ slightly above 2.

 A finite $\omega_D$ also introduces series of terms with higher-order derivatives. The series hold in
 $\omega^{m-1}_D \phi^{(m)}_0$ ($m\geq 2$).
 We show the calculations in Appendix \ref{sec:app_expand_C}
  and here quote the results.
 The prefactors form series in $X$ and at large $X$ rescale the prefactor for each term by $1/2$.
 To understand potential relevance of these term, we use the same strategy as in the previous Section and evaluate the series in $\omega^{m-1}_D \phi^{(m)}_0$ using the solution $\phi_0 (\omega)$ at $\omega_D =0$. The series then hold in powers of $Y$ and at large $Y$ replace $- {\dot \phi}_0/2$ in Eq.~(\ref{eq:eqn_phi_massive_1}) by $- {\dot \phi}_0/2  \left(\log{(\omega_0/\omega_D)}/\log{X}\right)$.
At large $X$, the ratio of the two logarithms is small, hence summing up series of terms with higher powers of $\omega_D$ and higher derivatives of $\phi_0$ reduces the overall effect from a finite $\omega_D$. This reaffirms that the ``infinite'' peak in the DOS survives in a finite interval of $\omega_D$.

\section{Phase diagram of the $\gamma$ model}
\label{sec:gamma}

 \begin{figure}
    \begin{center}
      \includegraphics[width=16cm]{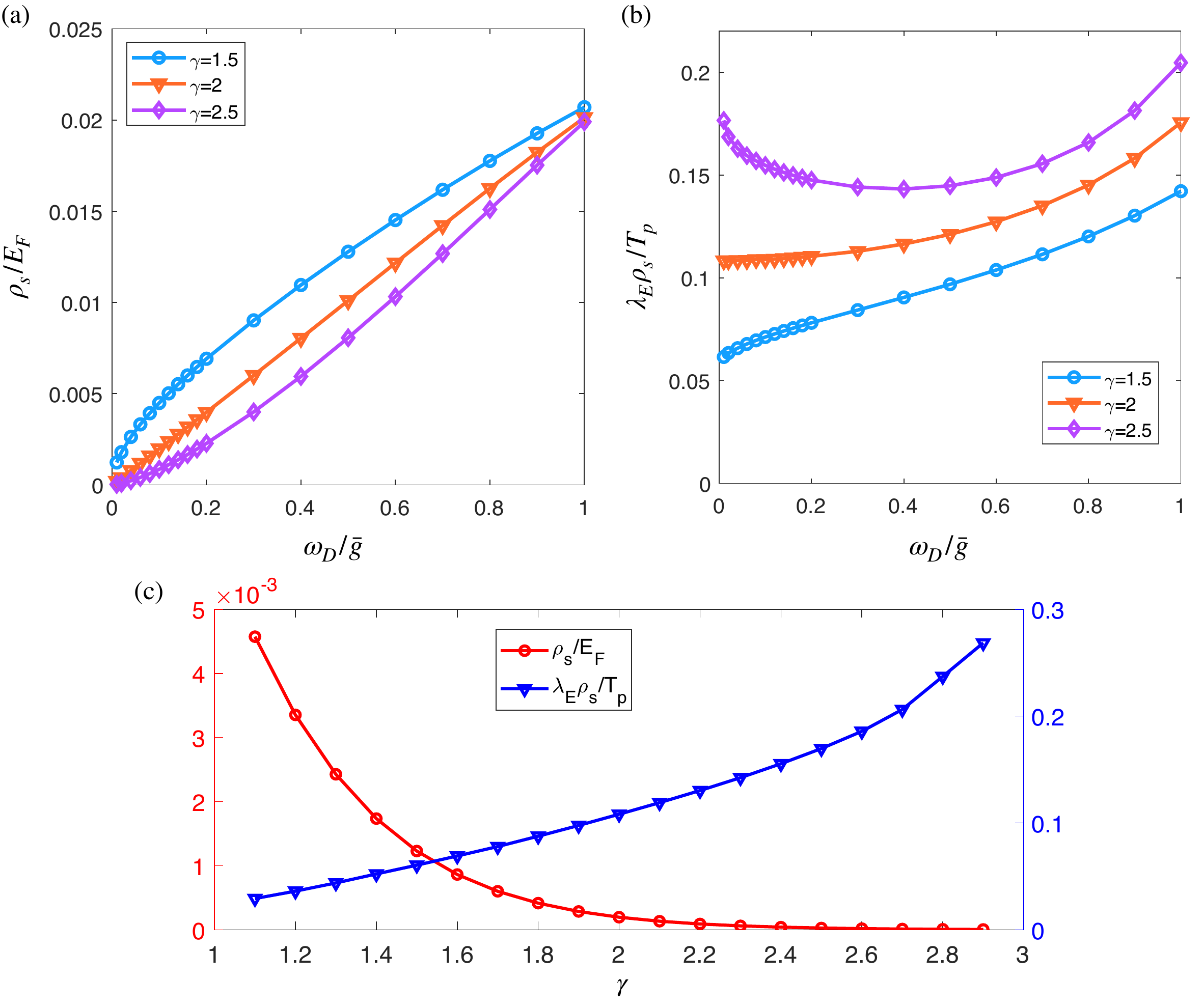}
    \end{center}
    \caption{Numerical results for the superfluid stiffness $\rho_s$  for different $\gamma$ and $\omega_D$.
     (a): $\rho_s$ in units of Fermi energy $E_F$, as a function of $\omega_D$ for different $\gamma$.
       For $\gamma >1$, the stiffness vanishes as $\omega^{\gamma-1}_D$.  We verified this dependence analytically.
     (b) The stiffness in units of $T_p/\lambda_E$, where $T_p$ is the onset temperature of the pairing and $\lambda_E = {\bar g}^\gamma/(E_F \omega^{\gamma-1}_D)$ is Eliashberg parameter, which measures the strength of corrections to side vertices in the diagrams for the self-energy and the pairing vertex (in commonly accepted language,
      $\lambda_F$ measures the strength of vertex corrections to Eliashberg theory). The parameter $\lambda_E$ has to be
       smaller than (roughly) one.  We see that $\rho_s \lambda_E/T_p$ tends to a finite value at $\omega_D =0$. This result implies that vertex corrections from only $n=0$ state  do not destroy superconducting order up to $T \leq T_p$.  (c) Comparable analysis of $\rho_s/E_F$ and $\rho_s \lambda_E/T_p$ at $\omega_D \to 0$.}       \label{fig:stiffness}
  \end{figure}

The key result of our analysis is the realization that at $T=0$,  the superconducting state at $\gamma >2$ is qualitatively different from the one at $\gamma <2$.   We label these two superconducting states as SC II and SC I,
  respectively. In both cases, the gap function in the upper half-plane of frequency is  $\Delta_0 (z)$,
   where $z = \omega' + i \omega^{''}$.

     A superconducting state  for $\gamma <2$ (SC I) does by itself evolve with $\gamma$ from BCS-like behavior for $\gamma <1$ to the novel behavior for $1<\gamma <2$, in which dynamical vortices cross, one by one, into the upper half-plane of frequency, and the phase winding of $\Delta_0 (\omega)$ along the real axis
    increases by $2\pi$ each time a new dynamical vortex moves into the upper half-plane.  From topological perspective, there is then a cascade of topological transitions at a set of  discrete $\gamma$ between $1$ and $2$.  Yet, the behavior of the DOS, $N(\omega)$, is conventional  for all $\gamma <2$ in the sense that  $N(\omega)$ is non-zero and is a continuous function of $\omega$ at frequencies above the spectral gap.
      There are no divergencies in $N(\omega)$ at $\gamma <2$,  yet a set of maxima and minima develops inside the continuum when $\gamma$ becomes close to $2$.

  The number of dynamical vortices becomes infinite at $\gamma = 2$. At this $\gamma$, several things happen: (i) an essential singularity necessarily develops at $\omega \to \infty$ as without it  an extension from an infinite array of vortex points would give a zero gap everywhere in the upper half-plane; (ii)
     an infinite number of other solutions of the gap equation, $\Delta_n (z)$, become degenerate with
    $\Delta_0 (z)$ for all $z >0$, the solutions with $n \to \infty$ form a  continuous set of $\Delta_\xi (z)$,
     (iii) $N(\omega)$ becomes the infinite set of $\delta$-functions at particular $\omega_p$.

At $\gamma >2$ (SC II), the number of dynamical vortices in $\Delta_0 (z)$ again becomes finite, and there is a cascade of topological transitions at a set of  discrete $\gamma$ between $2$ and $3$, when a vortex leaves the upper half-plane of frequency, in apparent mirror symmetry to what  happens for $1<\gamma <2$.  However, we argue that  beyond this, SC II and SC I are qualitatively different states.  Specifically, we found that
$\Delta (\omega )$
approaches $\omega_0$, where $\Delta_0 (\omega_0) = \omega_0$, with zero derivative, as $\Delta_0 (\omega_0)/\omega_0 =
1 + O(\omega_0 -\omega)^{2(\gamma-1)}$. 
At $\omega = \omega_0$, the gap function has  a branch cut, and
the density of states $N(\omega)$ develops a non-integrable singularity
(an ``infinite'' peak) 
at the lower edge of the continuum. We conjectured that if we keep the total number of states finite by imposing a finite cutoff on fermionic dispersion, the total weight under the peak will be proportional to the total number of states in the system.
  This ``infinite'' peak holds for all $\gamma >2$ (its degeneracy contains $\gamma-2$ in the prefactor) and  can be viewed as an order parameter that distinguishes between
 SC I and SC II. It is very likely that there exists another, possibly topological characteristic, which distinguishes SC II from  SC I.

 We analyzed  the effect of a finite mass of the pairing boson, $\omega_D$, and found that for $\gamma >2$ a finite $\omega_D > \omega^*_c$ is needed to transform SC II into SC I.  This analysis leads to $T=0$ phase diagram in $(\omega_D, \gamma)$ plane, shown in Fig.~\ref{fig:pd2}.
 There is a single transition line between SC I, which holds for all $\omega_D \geq 0$ at $\gamma <2$ and for
 $\omega_D > \omega^*_c$ at  $\gamma >2$, and SC II, while exists at $\gamma >2$ in the interval $ 0 \leq \omega_D \leq \omega^*_c$.

 We next consider  the phase diagram in the $(T, \gamma)$ plane at $\omega_D=0$.
 We argued in Paper V that for $\gamma =2$,
massless
``longitudinal'' fluctuations, associated with the
  continuum spectrum of condensation energies, destroy superconducting order at any finite $T$, although the onset temperature
  $T_{p} = T_{p,0}$
  for the emergence of a non-zero $\Delta_0 (z)$ is of order ${\bar g}$.
  At $\gamma <2$, longitudinal fluctuations are gapped, and it is natural to expect that $T_c$ becomes non-zero.
 We argued that $T_c$ increases gradually with $2-\gamma$, and the difference between $T_p$ and $T_c$ holds for all $\gamma <2$ and vanishes only  at $\gamma =0$.  In between $T_p$ and $T_c$, the system displays the pseudogap behavior associated with the formation of fermionic pairs without global phase coherence (preformed pairs).
  For $\gamma >2$, longitudinal fluctuations  become gapped, and it is natural to assume that $T_c$ again becomes finite and increases towards $T_p$ as $\gamma >2$ increases.

   We verified this last point by computing numerically the superconducting stiffness $\rho_s$ at $T\to 0$ (the prefactor in $F = \rho_s \int dr \nabla^2 \eta_0 (r) $, where
   $\eta_0 (r)$ 
   is the phase of the order parameter $\Delta_0 (r) = \Delta_0 e^{i\eta_0(r)} $.
   We show the results for different $\gamma$ and $\omega_D$ in Fig. \ref{fig:stiffness}. In the calculations, we only included the $n=0$ solution, i.e., we neglected fluctuation corrections from the solutions with other $n$.

   At small $\omega_D$,  the stiffness,  expressed in units of $E_F$, rapidly decreases with increasing $\gamma$. Taken at a face value, this would imply that the strength of phase fluctuations rapidly increases with $\gamma$. One has to be careful here, however, because our analysis, based on the analysis of modified Eliasberg equations, is valid as long as
     corrections to Eliashberg theory are small.  These corrections come from the renormalizations of side vertices in the diagrams for fermionic self-energy and the pairing vertex and  hold in powers of the Eliashberg parameter $\lambda_E$.   For an electron phonon problem (the case $\gamma =2$ in our notations), the Eliashberg parameter is $\lambda_E = a_2 {\bar g}^2/(E_F \omega_D)$, where $a_2 = O(1)$  (see, e.g., Paper V and Ref.\cite{Chubukov_2020b}). To keep it small at small $\omega_D/{\bar g}$, one need to simultaneously increase $E_F$.  The stiffness, re-expressed in units of the onset temperature for the pairing, $T_p$, scales as $\rho_s/T_p \sim 1/\lambda_E$.  Then, as long as
      $\lambda_E \leq 1$, the ratio $\rho_s/T_p$ does not become small at small $\omega_D$, which implies that phase fluctuations from the $n=0$ solution alone cannot substantially  reduce the actual $T_c$ compared to $T_p$.

      For $\gamma \neq 2$, the Eliashberg parameter is,
      up to a prefactor,  $\lambda_E = a_\gamma {\bar g}^\gamma/(E_F \omega^{\gamma-1}_D)$. In panels (b) and (c) of Fig. \ref{fig:stiffness} we plot $\rho_s$ in units of $T_p/\lambda_E$, with $T_p$ taken from~\cite{Wang2016}.
         We see that this ratio remains finite at $\omega_D \to 0$ for all $\gamma >1$, where one needs to adjust $E_F$ at small $\omega_D$ to keep $\lambda_E$ small.  This implies that within Eliashberg theory, phase fluctuations from the $n=0$ state  do not destroy superconducting order even at $\omega_D \to 0$.   Moreover, if
       the prefactor $a_\gamma$ weakly depends on $\gamma$,  the ratio  $\rho_s/(T_p/\lambda_E) = \rho_s \lambda_E/T_p$  actually increases with increasing $\gamma$, i.e., for   $\lambda_E =1$, which is  at the boundary of applicability of the Eliashberg theory, the ratio $\rho_s/T_p$  actually increases with $\gamma$, i.e., phase fluctuations become weaker.

        A more subtle question is whether for $\gamma >2$,  the order below $T_c$ is SC II.
  In our approximate analysis, Eq.~(\ref{eq:phi}), the order remains SC II up to some critical temperature, which is natural to be associated with $T_c$.
  Indeed, at a finite $T$ and $\gamma \approx 2$, the prefactor for the ${\dot \phi}^2_0 \tan{\phi_0}$ term, which plays the crucial role in distinguishing between
   SC I and SC II, is
    \beq
    \frac{\gamma-2}{2} \omega - T,
    \label{nn_6}
    \eeq
     see Ref. \cite{combescot}.
    The SC II state then holds as long as this coefficient is positive
    for $\omega \approx \omega_0$,
    which holds at $T < ((\gamma-2)/2) \omega_0$.  In this respect, $T$ plays the same role as a finite mass of a boson field, $\omega_D$.
     When we include infinite series in the derivatives of $\phi_0$, the analysis becomes more involved, but we still
      can  identify a characteristic  temperature $T^*_c$, which separates the behavior at higher $T$, when Im $\phi_0 (\omega)$ develops before it would flatten due to $(\gamma-2) \log{X}$ term in (\ref{eq:nn_1}),
        and at smaller $T$, when $\phi_0$ flattens up before temperature effects become relevant.  This scale is
       \begin{equation}
     T^*_c  \sim \omega_0 e^{-\frac{2}{\gamma-2}},
     \label{nn_2_1}
     \end{equation}
       It is similar to $\omega^*_c$ in (\ref{nn_2}). This $T_c^*$ is exponentially small, but finite, hence,
        the order SC II survives in a finite range of $T$. Whether $T^*_c$ coincides with the actual $T_c$, is beyond the scope of our analysis.  Assuming that it does, we
         arrive at the ``symmetric'' phase diagram shown in Fig.~\ref{fig:pd1}, with two distinct ordered phases SC I and  SC II, and the pseudogap phase in between.

       There is one
       caveat
       that needs to be addressed in further studies.
        In the discussion above we assumed
         that $\Delta (\omega )$ does not acquire an imaginary part for frequencies below $\omega_{0}$.
At a finite $T$, one generally expects that the DOS becomes non-zero for all $\omega$,
in which case the non-integrable singularity in $N(\omega)$ gets regularized.
A more careful
extension of the present approach to finite $T$ is needed to address this issue.

\section{Conclusions}
\label{sec:conclusions}

In this paper, the sixth in the series, we analyzed the interplay between non-Fermi liquid and pairing in the effective low-energy model of fermions with singular dynamical interaction $V(\Omega_m) = {\bar g}^\gamma/|\Omega_m|^\gamma$
 (the $\gamma$ model). The model describes low-energy physics of various quantum-critical metallic systems at the verge of an instability towards density or spin order as well as pairing of fermions at the half-filled Landau level, color superconductivity, and pairing in SYK-type models (see Paper I for the list of microscopic models).
  In previous publications, Paper I-V, we analyzed the physics of the model with $\gamma \leq 2$.  The key outcome
   of those studies was that a peculiar quantum-critical behavior   develops within this space of critical models
     as the exponent $\gamma$ approaches $\gamma =2$. The critical behavior is with a topological twist, as the number of  dynamical vortices in the upper half-plane of frequency tends to infinity at $\gamma =2$. In this paper we consider the $\gamma$-model with  exponents $2< \gamma <3$ and address the issue what happens on the other side of the quantum transition. We argue that the system moves away from criticality, e.g., the number of dynamical vortices becomes finite and decreases with increasing $\gamma$. This is similar to what happens when $\gamma$  decreases from $\gamma =2$.  Our key result, however, is the discovery that  superconducting order for $\gamma >2$ is qualitatively different from that for $\gamma <2$ (we labeled these states as SC II and SC I, respectively).   Specifically, we found that for $\gamma >2$, the DOS has a non-integrable singularity at the lower edge of the gapped continuum. In physical terms, this implies that the spectrum of excited states contains a level with  macroscopic degeneracy proportional to the total number of states in the system. We obtained the phase diagram at $T=0$ in variables $(\omega_D, \gamma)$, where $\omega_D$ is the mass of a pairing boson, and argued that for $\gamma >2$, the SC II state exists in a finite range of $\omega_D$ (Fig.~\ref{fig:pd2}).
     We conjectured  that  SC II  state survives at a finite $T$ and  rationalized the phase diagram in Fig.~\ref{fig:pd1}
       in variables $(T,\gamma)$ for $\omega_D =0$. The phase diagram contains two distinct superconducting phases SC I and SC II and an intermediate state with preformed pairs and pseudogap behavior of observables.

 From physics perspective, the appearance of
          an ``infinite'' peak
          can be understood using the same reasoning as in Ref. \cite{combescot}, as a bound state  between an excitation and an off-diagonal pairing field that this excitation can modify via the self-energy.  Indeed, we find that the self-energy $\Sigma (\omega)$ becomes singular
          at the lower end of the continuum, where $\Delta (\omega) = \omega$,
           i.e., at this frequency the effective potential, acting on a fermion in a superconductor, is infinite. A fermion in an infinite potential undergoes a self-trapping that generally
         leads to bound states.
          This argument however, does not immediately explains why we get a non-integrable singularity.

   The emergence of the non-integrable singularity may be related to the fact that for $\gamma >2$,
            the gap equation on the real axis contains a formally divergent contribution,
             which needs to be  regularized.
            The  divergence comes from the interaction $V(\Omega)$
             in the limit of zero frequency transfer $\Omega \to 0$.
               The interaction $V(\Omega \to 0)$ scatters with vanishingly small  frequency transfer and in this respect acts on electrons in the same way as impurities.
               The contribution from $V(0)$ that cancels out without regularization,
               is analogous to the
               contribution from non-magnetic impurities, while the one, which  cancels out only after regularization, is analogous to the contribution from  magnetic impurities.  In this respect, there may be  a similarity between our bound state and Yu-Shiba-Rusinov in-gap bound state in the DOS of a superconductor in the presence of magnetic impurities~\cite{Yu_65,Shiba_68,Rusinov_69}.

               Finally, the very fact that the leading order in the expansion in $X = \omega {\dot \phi}_0 \tan{\phi_0}$   captures the
divergence in the DOS, but does not capture the power-law singularity at the edge of the continuum, is similar
to the situation in the X-ray Fermi edge and Kondo problems (see e.g., \cite{Nozieres,Khvesh,Guinea,Affleck,Mahan} and references therein).  From this perspective, one might think that effects similar to the orthogonality catastrophe~\cite{Anderson} are also at play in the $\gamma$-model despite that this model is for a clean system.

 We call for more efforts to establish physical interpretation of the non-integrable singularity in the DOS for $\gamma >2$.

\acknowledgements We thank I. Aleiner, B. Altshuler, E. Berg, D.
Chowdhury, L. Classen, R. Combescot, K. Efetov, R. Fernandes, A. Finkelstein, E.
Fradkin, A. Georges, S. Hartnol, S. Karchu, S. Kivelson, I. Klebanov,
A. Klein, R. Laughlin, S-S. Lee, G. Lonzarich, D. Maslov, F. Marsiglio,
I. Mazin, M. Metlitski, W. Metzner, A. Millis, D. Mozyrsky, C. Pepan,
V. Pokrovsky, N. Prokofiev, S. Raghu, S. Sachdev, T. Senthil, D. Scalapino,
Y. Schattner, J. Schmalian, D. Son, G. Tarnopolsky, A-M Tremblay,
A. Tsvelik, G. Torroba, Y. Wang  E. Yuzbashyan, and J. Zaanen for useful discussions of this and previous works (Papers I-V).  The work by
 Y.M.W., S.-S.Z, and A.V. C.  was supported by the NSF DMR-1834856.
  Y.-M.W, S.-S.Z.,and A.V.C also acknowledge the hospitality of KITP at UCSB, where part of the
work has been conducted. The research at KITP is supported by the
National Science Foundation under Grant No. NSF PHY-1748958.

\appendix

\section{KK transformation for the interaction}
\label{sec:KK}

In this Appendix, we discuss the subtlety with
 expressing the gap equation on the real axis, Eq.~(\ref{el8}),  in terms of $C(\omega)$,  given by
 Eq. (\ref{real_a_3_1}). Taken at a face value, the integral in the r.h.s. of (\ref{real_a_3_1}) contains the piece
 \beq
 -i {\bar g}^\gamma \sin{\frac{\pi \gamma}{2}} \frac{\frac{dD(\omega)}{d\omega}}{\sqrt{1-D^2 (\omega)}}
  \int^\omega_{0+} d\Omega \frac{1}{\Omega^{\gamma-1}},
   \eeq
    For $\gamma >2$, the integral formally diverges and has to be properly regularized.

 We went back to the computational steps, involved in the derivation of the gap equation on the real axis, and traced the divergence in the integral for $C(\omega)$ to the divergence in the KK relation for the interaction
  on the real axis.  Specifically, on the real axis,
 \beq
 V(\Omega) =  \left(\frac{{\bar g}}{|\Omega|}\right)^{\gamma} \left(\cos{\frac{\pi \gamma}{2}} + i \sin{\frac{\pi \gamma}{2}} \text{sgn}\Omega\right)
\label{app_1}
\eeq
The derivation of $C(\Omega)$ uses the KK relation expressing $V^{'}(\Omega)$ in terms of $V^{''} (\Omega)$:
\beq
V' (\omega) = \frac{1}{\pi} P\int_{-\infty}^\infty \frac{V^{''} (x)}{x-\Omega} = \frac{2}{\pi} P\int_{0}^\infty \frac{V^{''} (x) x}{x^2-\Omega^2}
\label{app_1_1}
\eeq
 where $P$ stands for principle value.
 Rescaling $x$ by $\Omega$ we find that for $V(\Omega)$ from (\ref{app_1}), this relation is satisfied if
 \beq
 \frac{2}{\pi}\int_0^\infty \frac{dy}{y^{\gamma-1}} \frac{1}{y^2-1} = \cot{\frac{\pi \gamma}{2}}
  \label{app_3}
\eeq
 For $\gamma <2$, this relation holds,
 as one can easily verify, but for  $\gamma >2$, the integral in the l.h.s. of (\ref{app_3}) diverges.

 We argue that to  avoid the divergence and satisfy the KK relation for all $\gamma$,  one has to modify the integration contour to the one shown in Fig.~\ref{fig:contour},
 which by-passes $y=0$ by moving slightly into the upper half-plane of frequency. Indeed,
 integrating over the contour by standard means, we find
  that the integral in the l.h.s. of Eqn. (\ref{app_3}) gets modified to
   \beq
  \frac{2}{\pi}\left[ \int_{\epsilon/\omega}^\infty \frac{dy}{y^{\gamma-1}} \frac{1}{y^2-1}  + \left(\frac{\omega}{\epsilon}\right)^{\gamma-2} \frac{1}{\gamma -2}\right]
   \label{app_3_1}
   \eeq
  for $2<\gamma <4$.  The remaining integral is
    \bea
  && \int_{\epsilon/\omega}^\infty \frac{dy}{y^{\gamma-1}} \frac{1}{y^2-1} =  -\int_{\epsilon/\omega}^\infty \frac{dy}{y^{\gamma-1}} + \int_{0}^\infty \frac{dy y^(3-\gamma)}{y^2-1} \nonumber \\
  && = - \left(\frac{\omega}{\epsilon}\right)^{\gamma-2} \frac{1}{\gamma-1} - \tan{\frac{ (\gamma-1)\pi}{2}}
  \label{app_5}
  \eea
   Substituting into (\ref{app_3_1}), we find that the divergent term cancels out, and the KK relation is
    satisfied.
   For $\gamma >4$, the subleading term in (\ref{app_5}) also diverges, and the integral over a half-circle near
   $z=0$ has to be computed by including $(\epsilon/\omega)^2$ terms (and higher powers for even larger $\gamma>6$).
    We verified that the subleading  divergent terms also cancel out, i.e., integrating over the modified contour one does satisfy the KK relation (\ref{app_1_1}) for all $\gamma$.
     One can also check that the  other KK relation
     \beq
      V^{''} (\omega) = -\frac{1}{\pi} P\int_{-\infty}^\infty \frac{V^{'} (x)}{x-\Omega} =- \frac{2\Omega}{\pi} P\int_0^\infty \frac{V^{'} (x)}{x^2-\Omega^2}
\label{app_2}
\eeq
 is also satisfied for all $\gamma$, despite that the integral in the r.h.s. of (\ref{app_2}) formally diverges
  for $\gamma >1$.  The Cauchy relation between
  $V(\Omega_m)$ and $V^{''} (\Omega)$:  $V(\Omega_m) = (1/\pi) \int_0^\infty dx V^{''} (x) x/(x^2 + \Omega^2_m)$  is also satisfied for  the integration contour as in Fig.~\ref{fig:contour}.

  In practical terms, bending of the integration contour to by-pass the $z=0$ point is equivalent to just cancelling out the divergent terms in the KK transformation.  For $C(\omega)$,
   this implies that $ \int_{0^+}^\omega d\Omega/\Omega^{\gamma-1}$ has to be evaluated as
      \beq
      \int_{\epsilon}^\omega \frac{d\Omega}{\Omega^{\gamma-1}} - \frac{1}{\gamma-2}  \frac{1}{\epsilon^{\gamma-2}} = - \frac{1}{\gamma-2} \frac{1}{\omega^{\gamma-2}}
      \label{app_6}
      \eeq
  Using this procedure, one obtains that the prefactor  for the ${\dot \phi}_0$ term in the gap equation  evolves smoothly through $\gamma =2$.

\section{The gap function along the real axis}

\label{sec:app_gap_func}

When the critical boson becomes massive, the Eliashberg equantion
along the Matsubara axis takes the following form
\begin{equation}
\Delta(\omega_{m})=\bar{g}^{\gamma}\pi T\sum_{\omega_{m}^{\prime}}\frac{\Delta(\omega_{m}^{\prime})-\Delta(\omega_{m})\frac{\omega_{m}^{\prime}}{\omega_{m}}}{\sqrt{(\omega_{m}^{\prime})^{2}+\Delta^{2}(\omega_{m}^{\prime})}}\frac{1}{\left[(\omega_{m}^{\prime}-\omega_{m})^{2}+\omega_{D}^{2}\right]^{\gamma/2}},
\end{equation}
where $\omega_{D}>0$ is the mass of the intermediate boson. In this
section, we make the analytic continuation of the above equation to
the real axis.

To that end, we use the spectral representation of the interaction
$\chi(\omega_{m})=(1/\pi)\int d\omega\chi^{\prime\prime}(\omega)/(\omega-i\omega_{m})$,
where $\chi^{\prime\prime}(\omega)$ is the imaginary part of the
interaction along the real axis
\begin{equation}
\chi(\omega)=\frac{\bar{g}^{\gamma}}{(\omega_{D}-\omega-i\delta)^{\gamma/2}(\omega_{D}+\omega+i\delta)^{\gamma/2}},
\end{equation}
where $\delta$ is an infinitesimal positive number. Noting that $\text{Arg}[(\omega_{D}-\omega-i\delta)(\omega_{D}+\omega+i\delta)]=-\pi\text{sign}\omega\Theta(\rvert\omega\rvert-\omega_{D})$,
we have
\begin{align}
\chi^{\prime}(\omega) & =\frac{\bar{g}^{\gamma}}{\left(\rvert\omega\rvert^{2}-\omega_{D}^{2}\right)^{\gamma/2}}\Theta(\omega_{D}-\rvert\omega\rvert)+\frac{\bar{g}^{\gamma}}{\left(\rvert\omega\rvert^{2}-\omega_{D}^{2}\right)^{\gamma/2}}\cos(\frac{\pi\gamma}{2})\Theta(\rvert\omega\rvert-\omega_{D}),
\end{align}
\begin{align}
\chi^{\prime\prime}(\omega) & =\frac{\bar{g}^{\gamma}}{\left(\rvert\omega\rvert^{2}-\omega_{D}^{2}\right)^{\gamma/2}}\sin(\frac{\pi\gamma}{2})\text{sign}\omega\Theta(\rvert\omega\rvert-\omega_{D}).
\end{align}
With this representation, the gap equation can be rewritten as
\begin{equation}
\Delta(\omega_{m})=\int_{-\infty}^{\infty}d\omega\chi^{\prime\prime}(\omega)\left\{ T\sum_{\omega_{m}^{\prime}}\frac{\Delta(\omega_{m}^{\prime})-\Delta(\omega_{m})\frac{\omega_{m}^{\prime}}{\omega_{m}}}{\sqrt{(\omega_{m}^{\prime})^{2}+\Delta^{2}(\omega_{m}^{\prime})}}\frac{1}{\omega-i(\omega_{m}-\omega_{m}^{\prime})}\right\} .
\end{equation}
Now we make the analytic continuation $i\omega_{m}\rightarrow z$,
while keeping the terms within the brace bracket analytic on the upper
complex plane:
\begin{align}
 & T\sum_{\omega_{m}^{\prime}}\frac{\Delta(\omega_{m}^{\prime})-\Delta(\omega_{m})\frac{\omega_{m}^{\prime}}{\omega_{m}}}{\sqrt{(\omega_{m}^{\prime})^{2}+\Delta^{2}(\omega_{m}^{\prime})}}\frac{1}{\omega-i(\omega_{m}-\omega_{m}^{\prime})} \nonumber \\
\rightarrow & T\sum_{\omega_{m}^{\prime}}\frac{\Delta(\omega_{m}^{\prime})}{\sqrt{(\omega_{m}^{\prime})^{2}+\Delta^{2}(\omega_{m}^{\prime})}}\frac{1}{\omega-z+i\omega_{m}^{\prime}}  -\frac{\Delta(\omega_{m})}{\omega_{m}}T\sum_{\omega_{m}^{\prime}}\frac{\omega_{m}^{\prime}}{\sqrt{(\omega_{m}^{\prime})^{2}+\Delta^{2}(\omega_{m}^{\prime})}}\frac{1}{\omega-z+i\omega_{m}^{\prime}}
\nonumber \\
 & -\frac{1}{2}\frac{\Delta(z-\omega)}{\sqrt{-(z-\omega)^{2}+\Delta^{2}(z-\omega)}}\left( \tanh\frac{\omega-z}{2T}-\coth\frac{\omega}{2T}\right)  \nonumber \\
 &+\frac{1}{2}\frac{\Delta(z)}{z}\frac{z-\omega}{\sqrt{-(z-\omega)^{2}+\Delta^{2}(z-\omega)}}\left( \tanh\frac{\omega-z}{2T}-\coth\frac{\omega}{2T}\right).
\end{align}
The additional terms except that from the replacement $i\omega_{m}\rightarrow z$
ensure that the extended function of $z$ gets rid of the pole at
$z=\omega+i\omega_{m}^{\prime}$ ($\rvert\omega\rvert>\omega_{D}$).
The gap equation on the upper complex plane takes the form
\begin{align}
zD(z) & =\pi T\sum_{\omega_{m}^{\prime}}\frac{\Delta(\omega_{m}^{\prime})}{\sqrt{(\omega_{m}^{\prime})^{2}+\Delta^{2}(\omega_{m}^{\prime})}}\chi(\omega_{m}^{\prime}+iz)-iD(z)\pi T\sum_{\omega_{m}^{\prime}}\frac{\omega_{m}^{\prime}}{\sqrt{(\omega_{m}^{\prime})^{2}+\Delta^{2}(\omega_{m}^{\prime})}}\chi(\omega_{m}^{\prime}+iz)\nonumber \\
 & -\frac{1}{2}\int_{-\infty}^{\infty}d\omega\chi^{\prime\prime}(\omega)\frac{\Delta(z-\omega)-(z-\omega)D(z)}{\sqrt{-(z-\omega)^{2}+\Delta^{2}(z-\omega)}}\left\{ \tanh\frac{\omega-z}{2T}-\coth\frac{\omega}{2T}\right\} ,
\end{align}
where $D(z)=\Delta(z)/z$ and $V(-iz)=\left(\bar{g}^{2}/(\omega_{D}^{2}-z^{2})\right)^{\gamma/2}$.
In a compact form, we have
\begin{align*}
zD(z)B(z) & =A(z)+C(z),
\end{align*}
where
\begin{align}
A(z) & =\pi T\sum_{\omega_{m}^{\prime}>0}\frac{D(\omega_{m}^{\prime})}{\sqrt{1+D^{2}(\omega_{m}^{\prime})}}\left(\chi(\omega_{m}^{\prime}+iz)+\chi(\omega_{m}^{\prime}-iz)\right),\\
B(z) & =1+i\frac{\pi T}{z}\sum_{\omega_{m}^{\prime}>0}\frac{1}{\sqrt{1+D^{2}(\omega_{m}^{\prime})}}\left(\chi(\omega_{m}^{\prime}+iz)-\chi(\omega_{m}^{\prime}-iz)\right),\\
C(z) & =-\frac{1}{2}\int_{-\infty}^{\infty}d\omega\chi^{\prime\prime}(\omega)\frac{\Delta(z-\omega)-(z-\omega)D(z)}{\sqrt{-(z-\omega)^{2}+\Delta^{2}(z-\omega)}}\left\{ \tanh\frac{\omega-z}{2T}-\coth\frac{\omega}{2T}\right\} .
\end{align}
Below we consider the real axis where we replace $z$ by $\omega+i\delta$.
At zero temperature, using the spectral representation of the interaction
$\chi(\omega)$, the above functions reduce to
\begin{align}
A(\omega) & =\frac{1}{2}\int_{0}^{\infty}d\omega_{m}\frac{D(\omega_{m})}{\sqrt{1+D^{2}(\omega_{m})}}\left(\chi(\omega_{m}+i\omega)+\chi(\omega_{m}-i\omega)\right)\\
B(\omega) & =1+\frac{i}{2z}\int_{0}^{\infty}d\omega_{m}\frac{1}{\sqrt{1+D^{2}(\omega_{m})}}\left(\chi(\omega_{m}+i\omega)-\chi(\omega_{m}-i\omega)\right)\\
C(\omega) & =\frac{i}{2}\int_{0}^{\rvert\omega\rvert}d\Omega\chi^{\prime\prime}(\Omega)\frac{D(\omega-\Omega)-D(\omega)}{\sqrt{1-D^{2}(\omega-\Omega)}}.
\end{align}
Once we obtained $D(\omega_{m})$ by solving the Eliashberg equation
along the Matsubara axis, $A(\omega)$ and $B(\omega)$ are known
functions.

\section{Expansion of $C(\omega)$}
\label{sec:app_expand_C}
We evaluate the integral for $C(\omega)$ in (\ref{real_a_3_1})
 by Taylor-expanding the integrand in powers of internal  $\Omega$,  integrating each term in the expansion, and summing up the series. This procedure is inspired by the fact that only one term in the series survives at $\gamma =2$. However,
  away from this $\gamma$, an infinite number of terms appear with the same prefactor $(\gamma-2)$, and one has to sum up infinite series.

\subsection{At a QCP}
We first consider the case at a QCP and perform the integral over $\Omega$ at each order of the expansion:
\begin{align}
\int_0^{\omega} {d\Omega \over \Omega^{\gamma}} \Omega^n = {\omega^{n+1-\gamma} \over n+1-\gamma}, n=1,2,....
\end{align}
The infrared divergence for $n=1$ is avoided using the trick discussed in Appendix~\ref{sec:KK}.
The expansion of $C(\omega)$ is then given by a differential form
\begin{align}
C(\omega) = & {\bar{g}^{\gamma} \over \omega^{\gamma-2}}
{ \sin\frac{\pi\gamma}{2} \over 2-\gamma} D(\omega) \Bigg\{  \dot{\phi}  + \frac{\gamma-2}{2(3-\gamma)} \omega \left[\tan\phi\dot{\phi}^{2}+\ddot{\phi}\right]  \nonumber \\
& - \frac{\gamma-2}{6(4-\gamma)} \omega^{2} \left[ (2+3\tan^{2}\phi)\dot{\phi}^{3}+3\tan\phi\dot{\phi}\ddot{\phi}+\dddot{\phi}\right]  \nonumber \\
 & + {\gamma-2 \over 24 (5-\gamma)} \omega^{3} \left[\left(11\tan\phi+12\tan^{3}\phi\right)\dot{\phi}^{4}+\left(12+18\tan^{2}\phi\right)\dot{\phi}^{2}\ddot{\phi}  +3\tan\phi\ddot{\phi}^{2}\right. \nonumber \\
 & \left.\ \ \ \ \ \ \ \ \ \ \ \ \ \ \ \ \ \ \ \ \ \ +4\tan\phi(\dot{\phi})(\dddot{\phi})+\phi^{(4)}\right] \nonumber \\
 & - \frac{\gamma-2}{120(6-\gamma )} \omega^{4} \Bigg[\left(16+75\tan^{2}\phi+60\tan^{4}\phi\right)\dot{\phi}^{5} +\left(110\tan\phi+120\tan^{3}\phi\right)\dot{\phi}^{3}\ddot{\phi} \nonumber \\
 &   \ \ \ \ \ \ \ \ \ \ \ \ \ \ \ \ \ \ \ \ \ \  +\left(20+30\tan^{2}\phi\right)\dot{\phi}^{2}\dddot{\phi} +5\left(6+9\tan^{2}\phi\right)\dot{\phi}\ddot{\phi}^{2}+5\tan\phi\dot{\phi}\phi^{(4)} \nonumber \\
 &   \ \ \ \ \ \ \ \ \ \ \ \ \ \ \ \ \ \ \ \ \ \  +10\tan\phi(\ddot{\phi})(\dddot{\phi})+\phi^{(5)}\Bigg] + ... \Bigg\}.
\label{eq:C2}
\end{align}
The order of this expansion is equal to the number of derivatives with respect to $\omega$ (denoted as $M$).
The leading order $M=1$ survives at $\gamma-2$. All the higher order terms are proportional to the small parameter $\gamma-2$.
Clearly, the small-$\Omega$ expansion is not equivalent to a small-$(\gamma-2)$ expansion.

As we are mainly interested in the gap function around $\omega_0$, where $\phi=\pi/2$ and $\tan \phi=\infty$, we choose the highest power of $\tan\phi$ in the coefficients of each differential term in Eq.~(\ref{eq:C2}).
Keeping only the first derivative terms gives rise to
\bea
 \label{eq:nn_11b}
 C(\omega) &=& \frac{{\bar g}^\gamma}{\omega^{\gamma-2}} \frac{\sin{\frac{\pi \gamma}{2}}}{2-\gamma} D(\omega) {\dot \phi} \left[1 +  \frac{\gamma-2}{2(3-\gamma)} X -\frac{\gamma-2}{2(4-\gamma)} X^2 + \frac{\gamma-2}{2(5-\gamma)} X^3 +...\right],
\eea
namely Eq.~(\ref{eq:nn_11}) in the main text, where $X=\omega \tan \phi \dot{\phi}$.
This leads to the gap equation in Eq.~(\ref{eq:phi21}), which has been analyzed in Sec.~\ref{sec:real_sub1}.

Now we examine the effect of terms with higher derivatives (e.g., ${\ddot \phi}, \dddot{\phi}$, etc.) on the solution around $\omega_0$ by evaluating these terms using the above approximate solution.
To simplify the discussion, we consider the case $\gamma=2+0^+$ as an example, where the solution at $\omega$ slightly below $\omega_0$ reads $\phi = \pi/2 + Q (\omega_0-\omega)/\log(1-\omega/\omega_0)$ with $Q= 4 \omega_0^{\gamma-1} / [\pi (\gamma-2){\bar g})^{\gamma}]$ (see Eq.~(\ref{eq:solution_phi0b}).
For practical reasons, we consider a subset whose contribution to $C(\omega)$ is
\begin{align}
&- \frac{{\bar g}^\gamma \sin{\frac{\pi \gamma}{2}} }{\omega^{\gamma-2}} D_0(\omega) \Bigg[   \nonumber \\
&  {1\over 2!}     \omega \ddot{\phi} \left( 1-{1\over 2}X + {1\over 2}X^2 - {1\over 2}X^3 + ... \right)  \nonumber \\
& -{1\over 3! 2}  \omega^2 \dddot{\phi} \left( 1-{2\over 3}X + {3\over 4}X^2 - {4\over 5}X^3 + ... \right)  \nonumber \\
&  + {1\over 4! 3}  \omega^3 \ddddot{\phi} \left( 1-{3\over 4} X(1 - {6\over 5}X + {8\over 6}X^2 + ...) \right),    \nonumber \\
&  -... \Bigg],
\label{eq:series2}
\end{align}
in which the coefficients of the $n$-th derivative $\phi^{(n)}$ are formed by series in $X$.
In the limit of $X\to \infty$, which corresponds to $\omega \to \omega_0$, the series in $X$ in each term sums up to $1/2$.
Evaluating the differentials, $\dot{\phi} = Q/\log X$, $\omega_0^{n-1} \phi^{(n\geq 2)} = - (n-2)! Q X^{n-1}/\log^2 X$, one obtains the sum
\begin{align}
&{1\over 2!}     \omega \ddot{\phi} -{1\over 3! 2}  \omega^2 \dddot{\phi} + {1\over 4! 3}  \omega^3 \ddddot{\phi} - ...  \nonumber \\
=& -{Q\over \log^2 X}  \sum_{m=0}^{\infty} {(-1)^{m}X^{m+1} \over (m+1)^2(m+2)}.
\end{align}
Including these contributions to $C(\omega)$, we obtain the modified gap equation
\bea
{\dot \phi} \left[1 +  \frac{\gamma-2}{2} [\log(1+ X) + K] \right] = {2 \omega^{\gamma-1}\over \pi \bar{g}^{\gamma}},
\eea
where $K$ and $F(X)$ are the same functions defined in Eqs.~(\ref{nn_4}), (\ref{func:fx}) of the main text.
At large $X$, $F(X)\approx \log^2{X}$ and $K \approx - (1/2) \log {X}$.
Therefore, the leading order term $\log(X)$ near $\omega_0$ drops by a factor of $2$. The only change of the functional form of $\phi(\omega)$ near $\pi/2$
is an extra factor of $1/2$.

\subsection{Away from a QCP}
Next, we consider the effect of a finite but small mass ($\omega_D>0$) of the critical boson.
We redo the integral over $\Omega$ in the presence of a finite $\omega_D$:
\begin{align}
\int_{\omega_D}^{\omega} {d\Omega \over (\Omega^2-\omega_D^2)^{\gamma/2}} \Omega^n = & {\omega_D^{n+1-\gamma} \over 2} B_{1-({\omega_D\over \omega})^2}\left( 1-{\gamma \over 2}, {\gamma - n - 1 \over 2} \right),
\end{align}
where $B_z(a,b)$ refers to the incomplete Beta function. The divergence at $\Omega = \omega_D$ at $\gamma>2$ is again avoided using the trick discussed in Appendix~\ref{sec:KK}.
Near $\gamma=2$, this integral depends on the ratio between $\omega_D^{n-1}$ and $\gamma-2$, i.e.,
\begin{align}
\int_{\omega_D}^{\omega} {d\Omega \over (\Omega^2-\omega_D^2)^{\gamma/2}} \Omega^n
= {1 \over 2-\gamma} \omega_D^{n-1}  + {\cal O}((2-\gamma)^0).
\end{align}
The function $C(\omega)$, however, is regular because $1/(2-\gamma)$ is cancelled out by the small factor $\sin( \pi\gamma/2)$ from the interaction function.
Subtracting the contribution at $\omega_D=0$ and keeping only the leading order in $\gamma-2$, we obtain the modification to $C(\omega)$ due to a finite mass
\begin{align}\label{eq:CY}
 & {{\bar g}^{\gamma} \over \omega^{\gamma-2}} { \sin {\pi \gamma \over 2} \over 2-\gamma} D(\omega) \Bigg[  \nonumber \\
 & -{1\over 2}\dot{\phi} \left( Y - Y^2 + Y^3 - Y^4  + ... \right)  \nonumber \\
 & -  \left(  {1\over 2!} \omega_D \ddot{\phi} - {1\over 3!}\omega_D^2 \dddot{\phi} +  {1\over 4!}\omega_D^3 \ddddot{\phi}  + ...  \right) \nonumber \\
 & \times  \left( 1-Y+{3\over 2}Y^2-2Y^3 + {5\over 2}Y^4 + ...   \right)\Bigg],
\end{align}
where $Y=\omega_D \tan \phi \dot{\phi}$.

Ignoring the second and higher order derivatives, we obtain the gap function Eq.~(\ref{eq:eqn_phi_massive_1}) for a finite mass of the boson.
Its effect on the gap function has been analyzed in Sec.~\ref{sec:omega_D} of the main text.

Then, we verify that these neglected terms do not affect the solution around $\omega_0$ using the same strategy for a QCP case, namely, by evaluating them explicitly using the solution Eq.~(\ref{eq:solution_phi0b}) near $\omega_0$.
The series formed by $Y$ sums up to $1/2$ when $Y\to \infty$ at $\omega_0$.
The sum over terms including higher-order differentials reads
\begin{align}
& {1\over 2!} \omega_D \ddot{\phi} - {1\over 3!}\omega_D^2 \dddot{\phi} +  {1\over 4!}\omega_D^3 \ddddot{\phi}  + ... \nonumber \\
 = & - {Q\over \log^2 X} Y \sum_{m=0}^{\infty} {(-1)^m \over (m+1)(m+2)} Y^{m} \nonumber \\
 = & - {Q\over \log^2 X}  \left( {1 + Y \over Y} \log (1+Y) - {1\over Y} \right).
\end{align}
Near $\omega_0$, where $X,Y\to \infty$, the above sum reduces to $-Q\log Y/\log X$ asymptotically.
Adding their contributions to $C(\omega)$, the modified gap equation takes the form
\begin{align}
 {\dot \phi} \left[1 - {\log (\omega_0/\omega_D) \over 2 \log X }  +  \frac{\gamma-2}{2} \log(X) \right] = {2\over \pi \bar{g}^{\gamma}}\omega^{\gamma-1}.
\end{align}
Without including these higher order terms, the second term $ {\log (\omega_0/\omega_D) }/(2 \log X)$ becomes a constant $1/2$.
At large $X$, the ratio of the two logarithms is small, hence summing up series of terms with higher powers of $\omega_D$ and higher derivatives of $\phi_0$ reduces the overall effect from a finite $\omega_D$.

\section{The $n=\infty$ solution on the upper complex plane}
\label{sec:app_exact}

The $n=\infty$ solution along the Matsubara axis is given analytically by the same expression as for $\gamma \leq 2$, and we refer to Refs.~\cite{paper_1,paper_4,paper_5} for details.
Its analytic continuation towards to the upper complex plane of frequency is obtained by a rotation of frequency axis, $i\omega_m \to z = \omega^{\prime} + i \omega^{\prime\prime} = |z|e^{i\psi}$, which gives rise to
\beq
 \Delta_\infty ( z ) = \int_{-\infty}^\infty dk  \frac{e^{-\theta k} e^{-i I_k - i k \log y_z}}{ \sqrt{ \cosh(\pi (k-\beta ))\cosh(\pi (k+\beta)) } } .
 \label{gap_z}
 \eeq
 where $y_z = (\rvert z\rvert/{\bar g})^{\gamma}$, $\theta = (\pi/2 - \psi) \gamma$, and
\beq
  b_k = \frac{e^{-i (I_k + k\log{(\gamma-1)})}}{\left[\cosh(\pi (k-\beta))\cosh(\pi (k+\beta))\right]^{1/2}}.
  \label{nn_2_1_1}
  \eeq
 Here
\beq
  I_k = \frac{1}{2} \int_{-\infty}^\infty dk' \log{|\epsilon_{k'} -1|} \tanh{\pi (k'-k)},
  \label{nn_2_2}
  \eeq
\beq
\epsilon_{k'} = \frac{1-\gamma}{2} \frac{\Gamma\left(\frac{\gamma}{2}\left(1 + 2i k'\right)\right)\Gamma\left(\frac{\gamma}{2}\left(1 - 2i k'\right)\right)}{\Gamma(\gamma)} \left(1+ \frac{\cosh{\pi  \gamma k'}}{\cos{\pi \gamma/2}}\right),
\label{nn_3}
\eeq
and $\beta>0$ is the solution of $\epsilon_{\beta} =1$.
This extension is limited to the region $- \pi/\gamma < \psi - \pi/2 < \pi/\gamma $ where the integral giving rise to $\Delta_{\infty}(z)$ is convergent.
The critical axis $\psi = \pi/2 \pm \pi/\gamma$ is on the lower complex plane when $\gamma<2$, and rotates to the upper plane when $\gamma>2$.
Along the critical axis, the behavior of $\Delta_{\infty}(z)$ is very similar to that along the real axis at $\gamma=2$, where
the phase $\eta_{\infty} (z) = \text{Arg}(D_{\infty}(z))$ winds up to infinity as $|z| \to \infty$, while the amplitude follows a power-law increase $\sim |z|^{\gamma/2/(\gamma-1)}$.
The phase winding is attributed to the existence of an array of infinite vortices that line up along the critical axis as $|z|\to \infty$.
Consequently, there are only a finite number of vortices on the upper half-plane when $\gamma<2$ but infinite number of vortices when $\gamma\geq2$.
This evolution has been shown schematically by Fig.~\ref{fig:vortex} in the main text. Representative examples are provided in Fig.~\ref{fig:vortex_1} (b).

\section{A discrete set of solutions of the non-linear gap equation}
\label{app:discrete_set}

 Here we present the details of the analysis of a discrete set of solutions $\Delta_n (\omega_m)$.
 We depart from the solution of the linearized gap equation and expand the solution of the full non-linear gap equation in powers of $\Delta$
  as
 \begin{equation}
\Delta(\omega_{m})=\sum_{j=0}^{\infty} \epsilon^{2j+1} \Delta^{(2j+1)}(\omega_{m}),\label{exx1_aa}
\end{equation}
where $\Delta^{(1)}(\omega_{m}) = \Delta_{\infty} (\omega_m)$. We then solve iteratively for $\Delta^{(2j+1)}$  in terms
 of $\Delta^{(2j'+1)}$  and $j' <j$.

The gap equation must be satisfied at each order of $\epsilon$, which imposes the following equation
\begin{align}
\omega_{m}D^{(2j+1)}(\omega_{m})-\frac{\bar{g}^{\gamma}}{2}\int_{-\infty}^{\infty}d\omega_{m}^{\prime}\left(D^{(2j+1)}(\omega_{m}^{\prime})-D^{(2j+1)}(\omega_{m})\right)\frac{\text{sign}(\omega_{m}^{\prime})}{\rvert\omega_{m}^{\prime}-\omega_{m}\rvert^{\gamma}} & =K^{(2j+1)}(\omega_{m}),
\end{align}
with $j=0,1,2,...$.
The source term $K^{(2j+1)}(\omega_{m})$ is built from the gap function of a lower order $1\leq j^{\prime}<j$.
For example, the first two orders are given by
\begin{align}
K^{(0)}(\omega_{m}) & =0,\\
K^{(3)}(\omega_{m}) & =-\frac{\bar{g}^{\gamma}}{4}\int_{-\infty}^{\infty}d\omega_{m}^{\prime} \frac{\text{sgn}(\omega_{m}^{\prime})}{\rvert\omega_{m}^{\prime}-\omega_{m}\rvert^{\gamma}} \left(D^{(1)}(\omega_{m}^{\prime})-D^{(1)}(\omega_{m})\right)D^{(1)2}(\omega_{m}^{\prime}).
\end{align}
Since $K^{(0)}(\omega_{m}) =0$, the leading order is given by the solution of the linearized gap equation
\begin{equation}
D^{(1)}(\omega_{m})=D_{\infty}(\omega_{m}).
\end{equation}
At $\omega\ll\bar{g}$, there is
\begin{equation}
D^{(1)}(\omega_{m})\rightarrow 2 \text{sgn}(\omega_{m})\left(\frac{\rvert\omega_{m}\rvert}{\bar{g}}\right)^{\delta}\cos f(\omega_{m}),\label{eq:asymptotic}
\end{equation}
where $\delta={(\gamma-2)}/{2}$ and
\begin{equation}
f(\omega_{m})=\beta\log\frac{\rvert\omega_{m}\rvert^{\gamma}}{\bar{g}^{\gamma}}+\phi.
\end{equation}
We note that the term $\omega_{m}D^{(0)}(\omega_{m})$ in the gap equation is irrelevant for the small frequency behavior.
This holds true for each subleading order to be discussed.

Provided the leading order $j=0$ solved, one can compute the source term at the next order, $K^{(3)}(\omega)$,
and then search for the induced solution $D^{(3)}(\omega_{m})$. For the smallest frequency, $K^{(3)}(\omega)$ is free from the ultra-violet details, and thus fully determined by the asymptotic form of $D^{(1)}(\omega)$ in Eq. (\ref{eq:asymptotic}). One can continue this process to higher orders, which is summarized as a two-step iterative procedure.
(1) Once we solved the solution at orders $j'<j$, we first compute the source term at order $j$:
\begin{equation}\label{source_Kj}
K^{(j)}(\omega_{m})=\bar{g}\left(\frac{\rvert\omega_{m}\rvert}{\bar{g}}\right)^{(j-\frac{1}{2})(\gamma-2)-1}\sum_{r=0}^{j}e^{i(2r+1)f(\omega_{m})}I_{2r+1}^{(2j+1)}+c.c.,
\end{equation}
where $I_{2r+1}^{(2j+1)}$ is determined from the lower-order solutions. For $j=1$, we use the $j=0$ solution and obtain
\begin{align}
I_{1}^{(3)} & =-\frac{1}{4}I(3\delta+i\beta\gamma,2\delta+2i\beta\gamma)-\frac{1}{2}I(3\delta+i\beta\gamma,2\delta),\\
I_{3}^{(3)} & =-\frac{1}{4}I(3\delta+3i\beta\gamma,2\delta+2i\beta\gamma).
\end{align}
Here we have defined the integrals
\bea
I(a,b) & = &\int_{-\infty}^{\infty}\frac{dx}{\rvert x-1\rvert^{\gamma}}\left(\rvert x\rvert^{a}-\text{sign}(x)\rvert x\rvert^{b}\right) = B(\gamma-1-a,1+a)+B(\gamma-1-b,1+b)
\nonumber \\
 & + &\frac{\pi\csc(\pi\gamma)}{\Gamma(\gamma)}\left(\frac{\Gamma(1+a)}{\Gamma(2-\gamma+a)}-\frac{\Gamma(1+b)}{\Gamma(2-\gamma+b)} +
 \frac{\Gamma(\gamma-1-a)}{\Gamma(-a)}-\frac{\Gamma(\gamma-1-b)}{\Gamma(-b)}\right),
\eea
where $B(x,y)$ is the Beta function.
The convergence of this integral requires $\gamma<3$ and $-1<\text{Re}[a],\text{Re}[b]<\gamma-1$.
On order $j=1$, it requires $\gamma<3$;
on an arbitrary order $j>1$, it requires $\gamma<2+{2}/{(2j-1)}$.

(2) The source term in Eq.~(\ref{source_Kj}) leads to the induced solution at order $j$:
\begin{align}
D^{(j)}(\omega_{m}) & \simeq 2 \text{sgn}(\omega_{m})
\left(\frac{\rvert\omega_{m}\rvert}{\bar{g}}\right)^{(2j+1)\delta}
\sum_{r=0}^{j}Q_{2r+1}^{(2j+1)}\cos\left((2r+1)f(\omega_{m})+\phi_{2r+1}^{(2j+1)}\right).
\end{align}
where
\begin{equation}
Q_{2r+1}^{(2j+1)}\exp[{i\phi_{2r+1}^{(2j+1)}}] = - 2{I_{2r+1}^{(2j+1)}}/{J_{2r+1}^{(2j+1)}},\;r=0,1,...,j.\label{eq:Qj}
\end{equation}
and
\begin{equation}
J_{2r+1}^{(2j+1)}=I\left( (2j+1)\delta + i\beta\gamma(2r+1),0\right).
\end{equation}
The integrals $J_{2r+1}^{(2j+1)}$ is convergent under the same condition as $I_{2r+1}^{(2j+1)}$.

To apply the above iterative procedure for any given $\gamma>2$, however, we must stop at a finite order $j\sim1/(\gamma-2)$, above which, the gap function cannot be satisfied
because the divergence in both integrals $I_{2r+1}^{(2j+1)}$ and $J_{2r+1}^{(2j+1)}$ cannot be cancelled out from the equation.
The divergence indicates the gap equation at the low-frequency limit depends on the gap function at the higher frequency, which in turns depends on the parameter $\epsilon$.
In other words, $\epsilon$ enters the gap equation at each order by renormalizing the divergence.
To satisfy the gap equation, only a discretized set of $\epsilon$ is possible, indicating that the solutions form an infinite and discrete set.

\section{Behavior of $\Delta_0 (\omega_m)$ in the extended $\gamma$-model at $M \to 0$}
\label{sec:app_extended_model}

 The numerical solution in Fig.~\ref{fig:nleqM} (b) for $\gamma>2$ shows that
  $\Delta_0 (\bo_m)$, where $\bo_m$ is a properly normalized frequency,  vanishes at $M=0$ in a rather peculiar way:
the gap function at zero frequency, $\Delta_0(0)$, gradually  decreases as $M$  gets smaller and vanishes at $M =0$,
 however
 the full function $\Delta_0(\bo_m)$ remains finite at $M =0+$ and scales as $\bo$ at small frequencies.

In this section, we analyze the behavior of $\Delta_0 (\bo_m)$ analytically and argue that at $M =0+$,
 there exists a one-parameter  continuous set $\Delta_{0,\varepsilon} (\bo_m)$, specified by a parameter $\varepsilon$,
  which runs between $\varepsilon_{\text{min}} = 0+$ and a finite $\varepsilon_{\text{max}}$.  All $\Delta_{0,\varepsilon} (\bo_m)$ vanish at $\bo_m =0$
   and scale linearly with $\bo_m$ at small frequencies, but the slope is proportional to $\varepsilon$. As $M$ approaches zero from the positive side, the gap function $\Delta_0 (\bo_m)$ approaches  $\Delta_{0,\varepsilon_{\text{max}}} (\bo_m)$, while as $M$ approaches zero from the negative side,
 the gap function is infinitesimally small and approaches $\Delta_{0,\varepsilon_{\text{min}}} (\bo_m)$,

The gap function with $\varepsilon_{\text{min}}$ is the solution of the linearized gap equation.
At  small frequencies, $\Delta_{0,\varepsilon_{\text{min}}} (\bo_m)$  is the sum of two power-laws $(\bo_m)^{a_{1,2}}$.
At $M \to 0$,  $a_1$ approaches $1$ and $a_2$ approaches $\gamma-1>1$, hence $(\bo_m)^{a_{1}}$ is much larger,
 hence $\Delta_{0,\varepsilon_{\text{min}}} (\bo_m)$  is linear in $\bo_m$ at small frequencies.
   Like we said, the numerical  solution of the non-linear
   gap equation at $M \to 0$ also shows linear dependence of the gap function on frequency at small $\bo_m$.
 Based on this analogy, we assume that at $M \to 0$, there is a set of gap functions $\Delta_{0,\varepsilon} (\bo_m)$, which at small $\bo_m$ are all linear in $\bo_m$ at $M = 0+$ and at vanishingly small but finite $M$ behave as
   $\Delta_{0,\varepsilon} (\bo_m) = \varepsilon|\bo_m|^{1+\delta} \text{sign}(\bo_m)$, where $\delta$ scales with $M$.

To determine the two parameters $\varepsilon$ and $\delta$, we substitute this trial function into the modified gap equation in Eq.~(\ref{3_12b}).
In the infrared limit, the bare $\bo_m$ term in the l.h.s. is irrelevant, and ignoring it
we rewrite Eq.~(\ref{3_12b}) as
\bea
&& \int {d \bo_m^{\prime} \over |\bo_m-\bo'_m|^{\gamma} } \left( {|D(\bo_m)| \over \sqrt{1+D^2(\bo_m)} } - {|D(\bo'_m)| \over \sqrt{1+D^2(\bo'_m)} } \right)
\nonumber \\
& = & M D(\bo_m) \int ~ {d \bo_m^{\prime} \over |\bo_m-\bo'_m|^{\gamma} } \left( {{\mbox{\sign}}(\bo_m) \over \sqrt{1+D^2(\bo_m)}} - { {\mbox{\sign}} (\bo'_m) \over \sqrt{1+D^2(\bo'_m)} } \right).
\eea
Substituting the trial function into
 this equation, expanding to order $\epsilon^3$, and evaluating the integrals which turn out to be convergent in in both infrared and ultra-violet limits, we obtain at vanishing $\delta$
\bea
{\delta \over 2 } I(\gamma) \left(1-\varepsilon^2
  +  {\cal O}(\varepsilon^4)\right) = {M \over \gamma -1 } + {\cal O}(M \varepsilon^2),
\label{eq:F}
\eea
where
\bea
I(\gamma)&=& -\int_0^\infty dx \log{x} \left(\frac{1}{|1-x|^\gamma} + \frac{1}{(1+x)^\gamma}\right) \nonumber \\
 &=& {1\over \gamma-1}  \left(H(\gamma-2)-H(1-\gamma) + {\pi \over \sin \pi \gamma} \right).
\eea
and $H(x)$ is the  Harmonic number,  analytically continued from  $H(n)=\sum_{k=1}^{n} 1/k$.
In the two limits
$I(\gamma) \simeq \pi^2(\gamma-2)/2$ near $\gamma =2$  and
$I(\gamma) \simeq 1/(3-\gamma)$ near $\gamma=3$.

We see that $\delta \propto M$,
 as we
 anticipated.
Eq. (\ref{eq:F}) sets one condition on two parameters, $\delta$ and $\varepsilon$ and therefore allows for a continuous set of solutions.
Taking the limit $M\to 0$
 and  keeping $\delta/M = \alpha >0 $ as a constant, we obtain  $\varepsilon$ as a function of $\alpha$:
\bea
\varepsilon  = \sqrt{
1 - { 2 \over (\gamma -1 )I(\gamma)}{1\over \alpha}
}
\eea
As $\alpha$ varies between
$(\gamma-1)I(\gamma)/2$ and
 $\infty$, the amplitude parameter $\varepsilon$ changes continuously from
  $\varepsilon_{\text{min}} = 0+$ to $\varepsilon_{\text{max}} = 1
  $.
The gap function with $\varepsilon_{\text{min}}$ is the solution of the linearized gap equation, which is also the only solution one can obtain by approaching $M=0$ from negative $M$,  while
 the solution with  $\varepsilon_{\text{max}}$ is $\Delta_0 (\bo_m)$ that we obtained numerically by solving the non-linear gap equation at $M \to 0$ coming from positive $M$.

\bibliography{gamma_larger2}

\end{document}